\documentclass[reprint,
superscriptaddress,
 amsmath,amssymb,
 aps,
 prl,
]{revtex4-2}

\usepackage{graphicx}
\usepackage{dcolumn}
\usepackage{bm}
\usepackage{xcolor}
\usepackage{upgreek}
\usepackage{orcidlink}
\usepackage{hyperref}
\usepackage{wrapfig}
\usepackage{soul}
\usepackage[symbol]{footmisc}

\begin{document}

\preprint{APS/123-QED}

\title{Polymer extension at stagnation points \\governs flow thickening of polymer solutions in ordered porous media}

\author{Emily Y. Chen \orcidlink{0009-0004-6007-3460}}
\affiliation{%
 Department of Chemical and Biological Engineering, Princeton University, Princeton, NJ 08544, USA
}%

\author{Simon J. Haward \orcidlink{0000-0002-1884-4100}}
\author{Amy Q. Shen \orcidlink{0000-0002-1222-6264}}
\affiliation{
 Micro/Bio/Nanofluidics Unit, Okinawa Institute of Science and Technology Graduate University, Onna-son, Okinawa 904-0495, Japan
}%

\author{Sujit S. Datta \orcidlink{0000-0003-2400-1561}}%
 \email{ssdatta@caltech.edu}
\affiliation{%
 Division of Chemistry and Chemical Engineering, California Institute of Technology, Pasadena, CA 91125, USA
 }%
 \affiliation{%
 Department of Chemical and Biological Engineering, Princeton University, Princeton, NJ 08544, USA
 }%

\date{\today}
\begin{abstract}
\noindent 
Polymer solutions exhibit anomalous flow thickening---marked by an abrupt increase in the macroscopic flow resistance---above a threshold flow rate in a porous medium, but not in bulk solution. This phenomenon has evaded a mechanistic description for over half a century. Here, we develop a model that quantitatively links pore-scale flow fields and fluid rheology to macroscopic flow thickening, and validate it in experiments in two- and three-dimensional (2D and 3D) porous media. We find that flow thickening in ordered media is governed by polymer extension at stagnation points---in contrast to disordered media, where viscous dissipation by unsteady flow fluctuations also contributes substantially. Our results provide a foundation to predict and control such flows in energy, environmental, industrial, and microfluidic applications.

\end{abstract}

\maketitle
The flow of viscoelastic polymer solutions through porous media at low Reynolds number (Re $\ll 1$) underlies diverse chemical production~\cite{petrie_instabilities_1976,denn_polymer_2008,turner_review_2014,elbadawi_polymeric_2018}, additive manufacturing~\cite{agassant_polymer_2006,das_importance_2021}, separation~\cite{kozicki_filtration_1994,luo_high_1996,bourgeat_filtration_2003,gritti_harnessing_2024}, energy production~\cite{durst_flows_1981,okandan_heavy_1984,seright_new_2010,sorbie_polymer-improved_2013,clarke_how_2016,pogaku_polymer_2018,mirzaie_yegane_fundamentals_2022,di_dato_impact_2022}, and environmental remediation~\cite{roote_technology_1998,smith_compatibility_2008,huo_surfactant-enhanced_2020,hartmann_risk_2021} processes. And yet, predicting how pore-scale dynamics set the macroscopic flow resistance, quantified by the ``apparent viscosity'' $\eta_{\mathrm{app}}\equiv\frac{\Delta P}{L}\frac{k}{Q/A}$, remains an open problem; here, Darcy's law relates the total pressure drop $\Delta P$ across a length $L$ of the medium, the volumetric flow rate $Q$ through the cross section $A$, and the permeability $k$~\cite{whitaker_flow_1986}. Indeed, despite the typical shear-thinning rheology of these fluids, they anomalously `flow thicken' in a porous medium: $\eta_{\mathrm{app}}$ \emph{increases}, not decreases, above a threshold flow rate~\cite{pye_improved_1964,sandiford_laboratory_1964,jones_flow_1966,dauben_flow_1967,marshall_flow_1967,savins_shear_1968,savins_non-newtonian_1969,gogarty_viscoelastic_1972,kulicke_flow_1984}. The underlying reason has been a mystery for over half a century. Various mechanisms have been proposed~\cite{burcik_note_1965,sadowski_non-newtonian_1965-1,mungan_aspects_1966,gogarty_viscoelastic_1972,hirasaki_analysis_1974,dominguez_retention_1977,jones_flow_1979,bagassi_behavior_1989,tam_remarks_1991,zitha_unsteady-state_2001,sadowski_non-newtonian_1965,james_laminar_1975,moan_entrance_1979,jones_flow_1979,chauveteau_onset_1981,durst_flows_1981,haas_viscoelastic_1982,chauveteau_thickening_1984,gupta_viscoelastic_1985,magueur_effect_1985,ghoniem_extensional_1985,durst_nature_1987,tam_remarks_1991,chmielewski_effect_1992,skartsis_polymeric_1992,kozicki_viscoelastic_2001,rothstein_axisymmetric_2001,kozicki_flow_2002,haward_viscosity_2003,odell_viscosity_2006,walters_competing_2009,zamani_effect_2015,skauge_polymer_2018,ibezim_viscoelastic_2021,mokhtari_modified_2022,deiber_modeling_1981,vorwerk_shearing_1994,helmreich_non-viscous_1995,rothstein_axisymmetric_2001,walters_competing_2009,james_slow_2012,james_n1_2016,de_viscoelastic_2017,de_viscoelastic_2017-1,liu_flow_2017,savins_shear_1968,odell_non-newtonian_1988,rodriguez_flow_1992,saez_flow_1994,kauser_flow_1999,chmielewski_elastic_1993,galindo-rosales_microfluidic_2012,clarke_mechanism_2015,machado_extra_2016,mitchell_viscoelastic_2016,kawale_elastic_2017,hemingway_thickening_2018,qin_flow_2019,ekanem_signature_2020,browne_elastic_2021,browne_homogenizing_2023}, but quantifying their relative contributions to flow thickening has been challenging due to the opacity of typical media. Instead, prior work has relied on numerical simulations to characterize pore-scale flow dynamics~\cite{chilcott_creeping_1988,talwar_flow_1995,khomami_stability_1997,de_viscoelastic_2017,de_viscoelastic_2017-1,liu_flow_2017,mokhtari_birefringent_2022,mokhtari_modified_2022} or on semi-empirical models with adjustable fitting parameters ~\cite{wissler_viscoelastic_1971,vossoughi_pressure_1974,dharmadhikari_flow_1985,bendova_pressure_2009,sobti_creeping_2014}.

Recent work has identified the central roles of two key mechanisms. Direct flow visualization within \emph{disordered} 3D sphere packings~\cite{browne_elastic_2021,browne_harnessing_2024} demonstrated that the onset of an elastic flow instability---which generates unsteady pore-scale flow fluctuations~\cite{datta2022perspectives,groisman_elastic_2000,groisman_elastic_2004,zilz_geometric_2012,casanellas_stabilizing_2016,qin_characterizing_2017,qin_flow_2019,van_buel_characterizing_2022,oztekin1997stability,arratia_elastic_2006,poole2007purely,shi_growth_2016,zhao_flow_2016,qin_upstream_2019,haward2019flow,haward_asymmetric_2020,haward_stagnation_2021,hopkins_upstream_2022,hopkins_effect_2022}---coincides with the onset of flow thickening, and that the resulting enhancement of viscous dissipation accounts for much of the increase in $\eta_{\mathrm{app}}$ near this onset. However, the elastic instability contribution alone progressively underpredicts flow thickening at higher flow rates---an effect that is particularly pronounced in \emph{ordered} 3D sphere packings, where polymeric stresses generated by extensional flow have been suggested to govern flow thickening instead~\cite{chen_stagnation_2024}, as corroborated by independent 2D experiments~\cite{haward2026effects}. Yet a quantitative way to link extension at the pore scale to macroscopic flow thickening in porous media has remained lacking. Here, we develop this link from a mechanical energy balance and validate it experimentally in 2D and 3D porous media spanning a broad range of geometries.

We consider incompressible, inertia-free, viscoelastic flow through a porous medium~\cite{snoeijer_relationship_2020,browne_elastic_2021,supplementary_ref}. The rate of work done by the imposed pressure drop, per unit volume of the medium, $\frac{\Delta P}{L}\cdot\frac{Q}{A}$, is balanced by the time ($t$)-and volume ($V$)-averaged rate of viscous dissipation,  $\langle \boldsymbol\tau : \nabla \mathbf{u} \rangle_{t,V}$; here, $\boldsymbol\tau$ and $\mathbf{u}$ are the fluid stress tensor and velocity, respectively. 
Previous work~\cite{browne_elastic_2021} applied a Reynolds decomposition that separates this dissipation into mean and fluctuating components, $(\boldsymbol\tau : \nabla \mathbf{u})_0$ and $\langle \chi \rangle_{t,V} \sim \langle \mathbf{s'}:\mathbf{s'}\rangle_{t,V}$, respectively, where $\mathbf{s}$ is the rate-of-strain tensor and a prime denotes the fluctuating part. The fluctuating component, $\langle \chi \rangle_{t,V}$, captures the added viscous dissipation arising from unsteady flow fluctuations generated by the elastic instability. That decomposition, however, used a generalized-Newtonian (power-law) constitutive model of the fluid that neglects flow history-dependent polymeric stresses, and therefore omits the extensional contribution that we expect is required to fully capture flow thickening. 

Instead, we extend this decomposition by separating the mean dissipation into its shear ($\tau_{ij, i \neq j}$) and extensional ($\tau_{ij, i = j}$) components. Rewriting the extensional stresses in terms of the extensional viscosity $\eta_E = \frac{\tau_{xx}-\tau_{yy}}{\dot\epsilon}$ and the extensional strain rate $\dot \epsilon = \frac{\partial u}{\partial x}$ yields an additional dissipation contribution $\langle \eta_E \dot\epsilon^2 \rangle_{t,V}$, which captures the energy cost of stretching polymer chains under flow. Combining all three contributions in Darcy's law and normalizing by the bulk shear viscosity $\eta_I$ evaluated at the characteristic pore-scale shear rate $\dot\gamma$ then gives our first main result:
\begin{equation}
\label{Eq:CaberPB}
    \frac{\eta_\mathrm{app}}{\eta_I} = 1 + \frac{k\langle \chi \rangle_{t,V}}{(Q/A)^2 \eta_I} + \frac{k \ \langle\mathrm{Tr}\rangle_{t,V} \ \eta_0 \ \zeta^2}{\phi^2 D_p^2 \eta_I},
\end{equation} 
where $\phi$ is the medium porosity and $D_p$ is the mean diameter of the pillars (2D) or grains (3D) composing the solid phase of the medium. The first term on the right-hand side of Eq.~\eqref{Eq:CaberPB} represents the flow resistance from the bulk shear viscosity of the fluid. The second term reflects the viscous dissipation generated by unsteady flow fluctuations arising from an elastic instability when the Weissenberg number comparing elastic to viscous stresses in the fluid, $\mathrm{Wi}\equiv N_1(\dot\gamma)/2\sigma(\dot\gamma)$, exceeds a threshold value $\mathrm{Wi}_c$~\cite{browne_elastic_2021}; here, $N_1$ and $\sigma$ are the first normal stress difference and shear stress, respectively. The third term is the new extensional contribution, where the Trouton ratio $\mathrm{Tr}\equiv\eta_E/\eta_0$~\cite{trouton_coefficient_1906,anna_elasto-capillary_2001} compares the $\mathrm{Wi}$-dependent transient extensional viscosity to the zero-shear viscosity, $\eta_0$, and the characteristic extensional strain rate is $\dot\epsilon = \frac{Q\zeta}{\phi A D_p}$, where $Q/(\phi A)$ is the average flow speed through the pore space and $D_p/\zeta$ is the effective length scale over which that velocity changes as the fluid is squeezed past each obstacle; the value of the numerical factor $\zeta$ is determined from the unit-cell geometry~\cite{supplementary_ref}.

\begin{figure}
\includegraphics{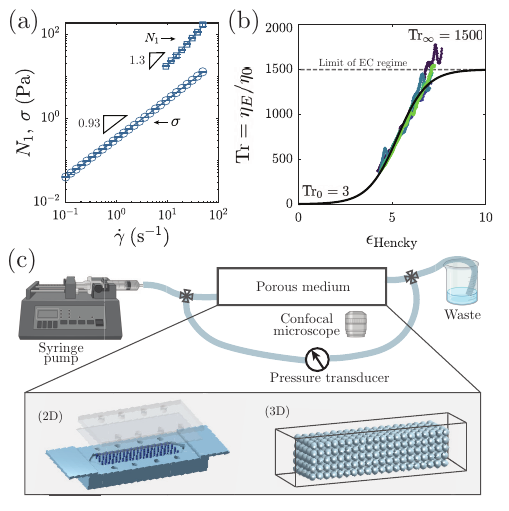}
\caption{\label{fig:exptl} Fluid rheology and experimental setup. (a) Steady shear rheology for the polymer solution. Solid curves are power law fits. Error bars correspond to one standard deviation from 3 replicate measurements.  (b) Transient extensional rheology from capillary break-up rheometry showing the Trouton ratio $\mathrm{Tr}$ versus accumulated Hencky strain $\epsilon_{\mathrm{Hencky}}$. The solid curve is a logistic fit to the elastocapillary thinning regime. Different colors show different experimental replicates. (c) Schematic of experimental imaging and pressure drop measurements. Inset: 2D hexagonal pillar array device (left) and 3D consolidated sphere packing (right).}
\end{figure}

To evaluate the accuracy of Eq.~\eqref{Eq:CaberPB}, we perform experiments using a polymer solution flowing through porous media of controlled ordered geometries [Figure~\ref{fig:exptl}]. The solution is composed of $c=300$~ppm ($\approx0.5c^*$, the critical overlap concentration) partially hydrolyzed polyacrylamide (18 MDa, 30\% hydrolyzed) in a viscous solvent composed of 82.6\% glycerol, 10.4\% dimethyl sulfoxide, 6\% ultrapure Millipore water, and 1\% sodium chloride (by weight). This fluid is weakly shear-thinning and highly elastic, as quantified by shear rheology in Fig.~\ref{fig:exptl}(a). We characterize the fluid's extensional response using capillary break-up rheometry (CaBER); the Trouton ratio $\mathrm{Tr}$ grows with accumulated Hencky strain $\epsilon_{\mathrm{Hencky}}$ from the Newtonian value $\mathrm{Tr}_0 = 3$, saturating at $\mathrm{Tr}_\infty = 1500$ as the stretching polymer chains approach their finite extensibility [Fig.~\ref{fig:exptl}(b)] ~\cite{anna_elasto-capillary_2001}.

We study flow through two classes of ordered porous medium: (i) 2D millifluidic arrays of hexagonally arranged pillars, fabricated by stereolithographic 3D printing, in two configurations (`staggered' and `aligned' relative to the mean flow direction), and (ii) 3D consolidated sphere packings, fabricated by selective laser-induced etching of fused silica, in two configurations (simple cubic, SC, and body-centered cuboid, BC) [Fig.~\ref{fig:exptl}(c)]. The 3D packings are rendered transparent by refractive-index matching of the solvent to the fused-silica matrix~\cite{datta2013spatial,supplementary_ref}, enabling direct flow visualization at the pore scale by confocal microscopy. We seed the polymer solution with 1~$\upmu$m fluorescent tracer particles and measure the time-resolved velocity field $\mathbf{u}$ by particle-image velocimetry~\cite{thielicke_pivlab_2014} in each unit cell across a range of imposed Weissenberg numbers. Simultaneously, we use a pressure transducer connected in parallel to measure the pressure drop across the medium~\cite{supplementary_ref}, which we convert to $\eta_{\mathrm{app}}$ via Darcy's law.

\begin{figure}
\includegraphics{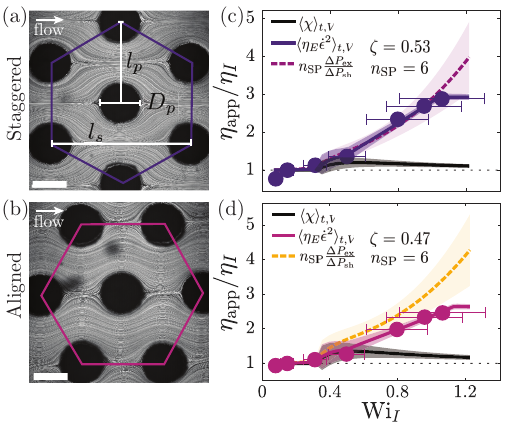}
\caption{\label{fig:2darrays} Experiments in 2D hexagonal arrays. (a-b) Pathline images of steady flow ($\mathrm{Wi} < \mathrm{Wi}_c$) in the staggered and aligned arrays with pillar diameter $D_p$, streamwise pillar spacing $l_s$, and pillar center-to-center distance $l_p$. Flow is from left to right. (c-d) Normalized apparent viscosity $\eta_{\mathrm{app}}/\eta_I$ versus Weissenberg number. Points: bulk pressure drop measurements. Black curves: elastic instability contribution alone [second term of Eq.~\eqref{Eq:CaberPB}]. Solid colored curves: full Eq.~\eqref{Eq:CaberPB} with the extensional contribution, with $\zeta \approx 0.5$ corresponding to an effective extensional length scale $D_p/\zeta\approx2D_p$ between the pillar diameter and the streamwise pillar spacing $l_s$~\cite{supplementary_ref}. Dashed colored curves: Eq.~\eqref{eq:oscerPB} with $n_{\mathrm{SP}}=6$ counted directly from the unit-cell geometry. The slight overprediction at higher $\mathrm{Wi}$ in the aligned array likely reflects partial screening of stagnation points~\cite{haward_stagnation_2021}.  Shaded regions reflect error propagation from the experimental measurements.
}
\end{figure}

We first examine flow through the 2D hexagonal pillar arrays [Fig.~\ref{fig:2darrays}(a,b)]. Bulk pressure drop measurements show that flow thickening sets in above $\mathrm{Wi}_c \approx 0.3$ in both configurations [points in Fig.~\ref{fig:2darrays}(c-d)], coincident with the onset of the elastic instability~\cite{supplementary_ref}. However, the elastic instability contribution to the flow resistance [second term in Eq.~\eqref{Eq:CaberPB}] falls far short of the measured resistance, as shown by the black curves in Fig.~\ref{fig:2darrays}(c,d), indicating that the extensional contribution is required. We evaluate this new contribution [third term in Eq.~\eqref{Eq:CaberPB}] by extracting, from the measured pore-scale velocity fields, the probability density $p(\epsilon_{\mathrm{Hencky}})$ of accumulated Hencky strain experienced by fluid elements along Lagrangian trajectories integrated over one polymer relaxation time, mapping it to a distribution $p(\mathrm{Tr})$ of local Trouton ratios via the CaBER-measured $\mathrm{Tr}(\epsilon_{\mathrm{Hencky}})$ [Fig.~\ref{fig:exptl}(b)], and computing the local flow-field-weighted average $\langle \mathrm{Tr} \rangle_{t,V} =\int_{\mathrm{Tr}_0}^{\mathrm{Tr}_\infty}\mathrm{Tr} \ p(\mathrm{Tr})\ d\mathrm{Tr}$~\cite{supplementary_ref}. With the extensional contribution included, Eq.~\eqref{Eq:CaberPB}---evaluated using only fluid rheology and pore-scale flow field measurements---quantitatively recovers the measured flow thickening in both 2D arrays, as shown by the solid colored curves in Fig.~\ref{fig:2darrays}(c,d)~\cite{supplementary_ref}.

\begin{figure}
\centering
\includegraphics{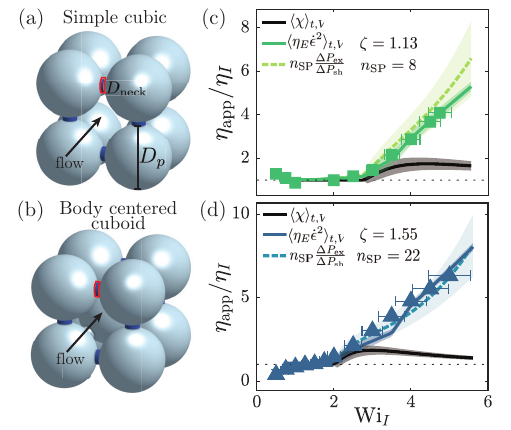}
\caption{\label{fig:3d} Experiments in 3D ordered sphere packings. (a, b) Schematics of the simple-cubic (SC) and body-centered-cuboid (BC) unit cells, where $D_p$ is the bead diameter and $D_\mathrm{neck}$ is the diameter of the cylindrical consolidation fusing adjacent grains. (c-d) Normalized apparent viscosity $\eta_{\mathrm{app}}/\eta_I$ versus Weissenberg number. Points: bulk pressure drop measurements reproduced from~\cite{chen_stagnation_2024}. Black curves: elastic instability contribution alone [second term of Eq.~\eqref{Eq:CaberPB}]. Solid colored curves: full Eq.~\eqref{Eq:CaberPB} with the extensional contribution, with $\zeta=1.13$ (SC) and $1.55$ (BC) corresponding to an extensional length scale $D_p/\zeta\approx0.7-0.9D_p$, smaller than the grain diameter and consistent with polymer extension being localized to the neck regions between consolidated grains~\cite{chen_stagnation_2024,supplementary_ref}. Dashed colored curves: Eq.~\eqref{eq:oscerPB} with $n_{\mathrm{SP}}=8$ (SC) and $22$ (BC) counted directly from the unit-cell geometry, including both the leading- and trailing-sphere stagnation points and those generated by the neck-like consolidations between grains. Shaded regions reflect error propagation from the experimental measurements.
}
\end{figure}

To test the generality of Eq.~\eqref{Eq:CaberPB}, we repeat this analysis for flow through the 3D consolidated sphere packings [Fig.~\ref{fig:3d}(a,b)]. Flow thickening again sets in coincident with the elastic instability, here at $\mathrm{Wi}_c \approx 2-3$ [points in Fig.~\ref{fig:3d}(c,d)], but the elastic instability term is again insufficient to account for the measured resistance [black curves]. Instead, just as in the 2D case, adding the extensional contribution---evaluated from the measured pore-scale velocity fields---recovers the measured resistance in both packings [solid colored curves]~\cite{supplementary_ref}. Altogether, the agreement of Eq.~\eqref{Eq:CaberPB} with our experiments across 2D and 3D geometries, spanning porosities $\phi$ from 0.28 to 0.85, establishes that polymer extension is the dominant additional contribution required to capture flow thickening in ordered porous media beyond the shear and elastic-instability terms.

Where in the pore space does this extension manifest? One might expect pore constrictions (so-called ``throats'') to be the locations of strongest polymer stretching: each throat compresses streamlines into a smaller cross section, generating an extensional velocity gradient that stretches polymer chains as they pass through~\cite{rothstein_axisymmetric_2001,james_slow_2012,zamani_effect_2015}. However, the maximum accumulated Hencky strain that throats alone can generate in each unit cell is $\epsilon_{\mathrm{Hencky}}\approx0.3-1.5$~\cite{supplementary_ref}---much smaller than the measured strains, which routinely exceed the polymer coil-stretch threshold of $\epsilon_{\mathrm{Hencky}}\approx2-3$~\cite{de_gennes_coilstretch_1974,perkins_single_1997}. Instead, prior work~\cite{haward_stagnation_2021,chen_stagnation_2024} demonstrated that stagnation points in the pore space, not throats, generate the strongest local velocity gradients and therefore most effectively stretch polymer chains~\cite{chilcott_creeping_1988,carrington_how_1996,haward_instabilities_2013,haward_microfluidic_2016}. If extension is stagnation point-dominated, then each stagnation point should contribute to flow thickening additively, and the macroscopic extensional resistance should scale linearly with the number of stagnation points per unit cell, $n_{\mathrm{SP}}$. 

We test this hypothesis by measuring the resistance from polymer extension at a \emph{single} stagnation point and then multiplying by $n_{\mathrm{SP}}$, counted \emph{a priori} from the geometry of each medium~\cite{supplementary_ref}. We use an optimized cross-slot rheometer~\cite{haward_optimized_2012,galindo-rosales_optimized_2014} as a single-stagnation point ``model pore'': in extensional flow mode, one free stagnation point sits at the device center, generating a planar extensional flow whose excess pressure drop relative to the corresponding shear-only configuration, $\Delta P_{\mathrm{ex}}/\Delta P_{\mathrm{sh}}$, isolates the per-stagnation point contribution of polymer extension~\cite{supplementary_ref}. Decomposing the Darcy pressure drop into shear and extensional parts then yields our second main result:
\begin{equation}
\label{eq:oscerPB}
\begin{aligned}
    \frac{\eta_\mathrm{app}}{\eta_I} &= \frac{k (\Delta P_{\mathrm{ex}} + \Delta P_{\mathrm{sh}})}{(Q/A)\eta_I L} \\
    & = \bigg[1+\frac{k\langle \chi \rangle_{t,V}}{(Q/A)^2 \eta_I}\bigg]\bigg[1+\frac{\Delta P_{\mathrm{ex}}}{\Delta P_{\mathrm{sh}}}\cdot \ n_{\mathrm{SP}}\bigg].
\end{aligned}
\end{equation}
As shown by the dashed colored curves in Figs.~\ref{fig:2darrays}(c,d) and~\ref{fig:3d}(c,d), Eq.~\eqref{eq:oscerPB} again recovers the measured flow resistance across all the 2D and 3D geometries tested, with all input parameters measured independently. This agreement between the theory and experiment establishes that the macroscopic flow resistance of viscoelastic polymer solutions in ordered porous media is set, additively, by polymer extension at stagnation points.

\begin{figure}[t]
\includegraphics{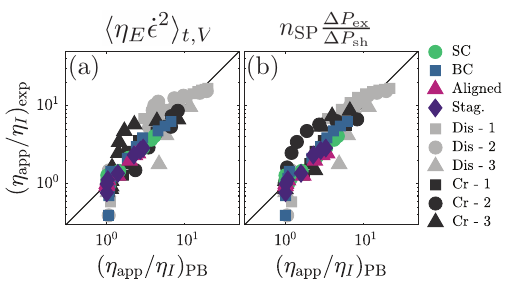}
\caption{\label{fig:collapse} Collapse of measured macroscopic flow resistance $(\eta_\mathrm{app}/\eta_I)_{\mathrm{exp}}$ against experimentally-evaluated power-balance predictions $(\eta_\mathrm{app}/\eta_I)_{\mathrm{PB}}$ across ordered (colored) and disordered (grayscale) porous media. (a) Predictions of Eq.~\eqref{Eq:CaberPB}, evaluated with $\zeta$ from the unit-cell geometry in the ordered cases and from a single fit per medium in the disordered cases; fitted disordered $\zeta$ cluster in the narrow band $D_p/\zeta\approx0.6-0.8D_p$, between the pore-throat scale ($\sim 0.16D_p$ for random bead packings~\cite{al-raoush_extraction_2005,thompson_application_2008}) and the grain diameter. (b) Predictions of Eq.~\eqref{eq:oscerPB}, with $n_{SP}$ counted directly from the unit-cell geometry in the ordered cases and fit to the data in the disordered cases. In disordered media, $n_{SP}$ becomes an effective parameter combining the stagnation point density with the per-point contribution to the resistance, and varies across replicates due to pore-to-pore heterogeneity~\cite{supplementary_ref}.
}
\end{figure}

Our experiments so far have focused on ordered porous media. We next test whether the framework extends to disordered porous media as well using experiments in two model 3D geometries: lightly sintered packings of near-monodisperse glass beads or of crushed glass, both designed to be refractive-index matched, enabling in situ flow visualization. As detailed in \cite{supplementary_ref}, in both cases, the elastic instability contribution [second term of Eq.~\eqref{Eq:CaberPB}] captures the resistance near the onset of flow thickening but progressively underpredicts the measured increase in $\eta_\mathrm{app}$ at higher $\mathrm{Wi}$---suggesting that polymer extension also contributes to flow thickening in disordered media. Our analysis confirms this idea: Eq.~\eqref{Eq:CaberPB}, evaluated using the experimentally-measured Hencky-strain distribution, captures the macroscopic resistance across the full $\mathrm{Wi}$ range in the disordered packings [Fig.~\ref{fig:collapse}]---again with $D_p/\zeta\approx0.6-0.8D_p$ characteristic of an extensional length scale set by the local pore geometry. Our framework therefore captures flow thickening across porous media of different geometries.

We have focused on shear, elastic instability-generated fluctuation, and extensional contributions to flow thickening in porous media, and in doing so have necessarily made three key simplifications. First, we assumed that shear-generated normal-stress differences contribute negligibly, as supported by an upper-bound estimate combining the bulk measurements of $N_1$ with the measured pore-scale velocity-gradient field~\cite{supplementary_ref}; incorporating this contribution explicitly~\cite{james_n1_2016} is a natural extension of our framework. Second, our continuum description of the fluid neglects polymer-wall interactions, justified here because the polymer size (radius of gyration $R_g \approx 200$~nm) is more than three orders of magnitude smaller than the pore throats  ($\gtrsim300~\upmu$m). These interactions will likely become more important in more confined media~\cite{jones_flow_1979,zitha_unsteady-state_2001}. Finally, we have not included the additional contribution $\boldsymbol\tau_p : \nabla \mathbf{u}$, where $\boldsymbol\tau_p$ is the polymer stress tensor, that accounts for the storage and dissipation of elastic energy during polymer deformation and relaxation~\cite{bird_dynamics_1987,larson_structure_1999,snoeijer_relationship_2020}, which could be quantified using birefringence~\cite{lodge_network_1956,cantow_flow_1969,sridhar_birefringence_2000,rothstein_comparison_2002,haward_optimized_2012} or by reconstructing polymer stress fields along Lagrangian fluid trajectories~\cite{corona_fingerprinting_2022,kumar_lagrangian_2023,kumar_stress_2023}, both of which remain active research frontiers. Even with these simplifications, our framework provides a quantitative pore-to-macroscale link for flow thickening across both ordered and disordered porous media---which could thus guide efforts to engineer targeted flow behavior via fluid rheology~\cite{ewoldt_designing_2022,richards_optimizing_2024} and pore geometry~\cite{stone_engineering_2004,browne_porescale_2020}.
\\
\begin{acknowledgments}
We thank H. A. Stone and R. J. Poole for insightful discussions. The authors acknowledge the use of the Imaging and Analysis Center (IAC) operated by the Princeton Materials Institute at Princeton University, which is supported in part by the Princeton Center for Complex Materials (PCCM), a National Science Foundation (NSF) Materials Research Science and Engineering Center (MRSEC; DMR-2011750). S.S.D. also acknowledges support from the Camille Dreyfus Teacher-Scholar Program of the Camille and Henry Dreyfus Foundation. S.J.H. and A.Q.S. also acknowledge financial support from the Japanese Society for the Promotion of Science (JSPS, Grant Nos. 24K07332, and 24K00810). 
\end{acknowledgments}



%

\clearpage
\onecolumngrid

\setcounter{figure}{0}
\setcounter{table}{0}
\setcounter{equation}{0}
\setcounter{section}{0}
\renewcommand{\thefigure}{S\arabic{figure}}
\renewcommand{\thetable}{S\arabic{table}}
\renewcommand{\theequation}{S\arabic{equation}}
\renewcommand{\thesection}{S\arabic{section}}

\begin{center}
    \emph{\textbf{Supplementary Material: }}\textbf{Polymer extension at stagnation points}\\ \textbf{governs flow thickening of polymer solutions in ordered porous media} \\  \vspace{0.5 cm}
    \scriptsize
    Emily Y. Chen$^1$, Simon J. Haward$^2$, Amy Q. Shen$^2$ and Sujit S. Datta$^{3, 1}$ \\ \vspace{0.5 cm}
     $^1$ Department of Chemical and Biological Engineering, Princeton University, Princeton, NJ 08544, USA \\
    $^2$ Micro/Bio/Nanofluidics Unit, Okinawa Institute of Science and Technology Graduate University, \\Onna-son, Okinawa 904-0495, Japan \\
    $^3$ Division of Chemistry and Chemical Engineering, California Institute of Technology, Pasadena, CA 91125, USA \normalfont
\end{center}

\section{Polymer solution}
To prepare the polymer solution, we gently mix a stock solution of 5000 ppm (by weight) partially hydrolyzed polyacrylamide (HPAM, 18 MDa, 30\% hydrolysis; Polysciences) in ultrapure Millipore water on a rotor mixer for 24 hours. We then dilute this stock solution to 300 ppm HPAM with the remaining solvent components (by weight): 82.6\% glycerol (ThermoFisher Scientific), 10.4\% dimethyl sulfoxide (Sigma-Aldrich), and 1\% sodium chloride (Sigma-Aldrich) and mix for another 24 hours on a stir plate at 60 rpm to avoid mechanical degradation of the polymer. The solution is then allowed to rest for 24 hours before any rheology measurements or flow experiments. All polymer solutions are used within 1 month of preparation. 

The addition of excess salt (170 mM NaCl) sufficiently screens the polyelectrolyte charges such that HPAM is expected to adopt a flexible coil configuration in solution~\cite{dobrynin_scaling_1995}. For this solvent, the overlap concentration at which polymer intermolecular interactions occur is $c^* = 600$ ppm as determined from shear rheology of a dilution series to determine the intrinsic viscosity~\cite{browne_elastic_2021}; thus, the polymer solution is in the dilute regime, $c/c^* = 0.5$. From the overlap concentration $c^* = \frac{M_W}{4/3 \pi R_g^3 N_A}$, we estimate the radius of gyration as $R_g = 220$ nm, where $M_W$ is molecular weight and $N_A$ is Avogadro's number. The fully extended length of the polymer chain is given by the contour length, $L_c = N L_{\mathrm{mon}} \approx 40 \ \upmu$m with $N$ the degree of polymerization and $L_{\mathrm{mon}}$ the monomer length.

\section{Fluid rheology}
We measure the steady shear rheology of the polymer solution using a stress-controlled Anton Paar MCR501 rheometer with a truncated cone-plate geometry (CP50-2: 50 mm diameter, 2$^\circ$, 53 $\upmu$m gap), temperature-controlled at 25$^\circ$C. We measure steady-shear flow curves of the shear stress, $\sigma(\dot\gamma)$, and first normal stress difference, $N_1(\dot\gamma)$, over shear rates of $\dot\gamma = 0.1-30 \: \mathrm{s}^{-1}$. For shear rates $\dot\gamma > 30 \ \mathrm{s}^{-1} $, we observe an elastic instability in the cone-plate geometry marked by an apparent thickening of the shear stress~\cite{shaqfeh_purely_1996,larson_instabilities_1992,howe_flow_2015,casanellas_stabilizing_2016}. The minimum resolvable $N_1$ is $\approx10$~Pa due to the normal force sensitivity of the rheometer. We perform flow curve measurements for three replicates with fresh samples for each sweep and report the standard deviation as error bars. We fit a power law model to the shear stress, $\sigma(\dot\gamma)=K \dot\gamma ^n$, and first normal stress difference, $N_1(\dot\gamma)=K_{N_1} \dot\gamma ^{n_{N_1}}$, where $K$ and $K_{N_1}$ are material prefactors and $n$ and $n_{N_1}$ are power law indices. For the dilute HPAM solution, $K = 0.323\pm0.001$ Pa$\cdot$s$^n$, $n=0.93\pm0.1$, $K_{N_1} = 0.81\pm0.05$ Pa$\cdot$s$^{n_{N_1}}$, and $n_{N_1} = 1.43\pm0.03$. In the absence of a low shear rate viscosity plateau for power law fluids, we approximate the zero-shear viscosity $\eta_0$ as $\eta|_{\dot\gamma = 0.1 \ \mathrm{s}^{-1}} \approx 340$ mPa$\cdot$s.

We characterize the transient extensional flow behavior of the polymer solution using a Haake capillary breakup extensional rheometer (CaBER). In this method, a small volume of fluid is loaded between two plates at an initial plate separation of $1~$mm, the plates are separated at $1~$m/s to a final separation of $6~$mm to create a thin filament, and the resulting thinning of the filament diameter is recorded over time~\cite{kolte_capillary_1999,anna_elasto-capillary_2001}. In the elastocapillary regime, filament thinning is governed by a balance of elastic stresses resisting break-up and the capillary forces promoting pinch-off~\cite{entov_effect_1997}. The Hencky strain over the course of the measurement is $\epsilon_{\mathrm{Hencky}} = 2\cdot\mathrm{ln}(D_0/D_{\mathrm{mid}}(t))$, where $D_0$ is the plate diameter and $D_{\mathrm{mid}}(t)$ is the filament diameter at the mid-plane. An appropriate force balance on the filament~\cite{entov_effect_1997,kolte_capillary_1999} enables determination of the Trouton ratio (Tr; ratio of transient extensional viscosity to zero-shear viscosity $\eta_E/\eta_0$) given that the surface tension and density of the fluid are known. For our polymer solution, we measure a surface tension of $61.2$ mN/m (Biolin Attension Theta; pendant drop method) and density of $1234$ kg/m$^3$ (Anton Paar DMA 35 densitometer). We measure five replicates in the CaBER device at 25$^\circ$C and find good reproducibility. We fit the measured Trouton ratio versus Hencky strain data with a logistic function to capture the low-strain Newtonian-like behavior and high-strain finite extensibility of the polymer: 
\begin{equation}
\label{eq:caberLog}
    \mathrm{Tr}(\epsilon_{\mathrm{Hencky}}) = \mathrm{Tr}_0 + \frac{ \mathrm{Tr}_\infty- \mathrm{Tr}_0}{1+\mathrm{exp}[-k(\epsilon_{\mathrm{Hencky}}-\epsilon_{H, 0.5})]},
\end{equation}
where $\mathrm{Tr}_0 = 3$ is the low-strain Newtonian limit, $\mathrm{Tr}_\infty = 1500$ is the finite extensibility limit, $k$ is the growth rate, and $\epsilon_{H, 0.5}$ is the Hencky strain at which the Trouton ratio reaches its half-maximum. Here, we determine $\mathrm{Tr}_\infty = 1500$ based on the upper limit of the elastocapillary regime. We use the resulting functional form to convert Hencky strains measured from the flow field to estimates of the ``local" apparent Trouton ratio (i.e. a pore-scale estimate of transient extensional viscosity).

\section{Porous medium fabrication}
\subsection{2D hexagonal pillar arrays}
We fabricate millifluidic 2D arrays consisting of pillars of diameter $D_p = 600 \ \upmu$m arranged in a hexagonal unit cell with center-to-center spacing of $l_p = 1.44$ mm. These dimensions correspond to a ratio of $l_p/D_p = 2.4$, which we have chosen to reproduce following the devices used by~\citeauthor{walkama_disorder_2020}~\cite{walkama_disorder_2020} and~\citeauthor{haward_stagnation_2021}~\cite{haward_stagnation_2021}. In the staggered configuration, the streamwise pillar-pillar spacing is $l_s = 2.494$ mm, while for the aligned array, $l_s = l_p$. The difference in $l_s$ for the two geometries is a result of the imposed flow direction relative to the unit cell orientation. The medium porosities are $\phi_{\mathrm{stag}} = 0.848$ and $\phi_{\mathrm{aligned}} = 0.836$. We determine the Darcy permeability of each medium, $k$, by measuring the pressure drop across the device for the flow of water applied in a stepped flow rate ramp sequence and using Darcy's law. The resulting Darcy permeabilities are $k_{\mathrm{stag}} = 895 \pm 20 \ \upmu$m$^2$ and $k_{\mathrm{aligned}} = 929 \pm 19 \ \upmu$m$^2$. The arrays have lengths of $L_{\mathrm{\mathrm{stag}}} = 33$ mm and $L_{\mathrm{aligned}} = 32.3$ mm and widths of $W = 8.4$ mm. The channel depth is 1 mm. In total, the staggered array contains 149 pillars and the aligned array contains 157 pillars.

We design the channel using CAD software (Onshape) and 3D-print the main body using a FormLabs Form 3 stereolithography 3D printer with a proprietary clear resin consisting of methacrylate oligomers and photoinitiators (FLGPCL04). We assemble the full millifluidic device by sandwiching a $\sim$1 mm layer of polydimethylsiloxane (PDMS; Dow SYLGARD 184, 8.5:1.5 by weight base to curing agent ratio) between the 3D printed base and a laser-cut transparent acrylic top sheet (McMaster-Carr) fitted with screwholes. The PDMS layer ensures a watertight seal while allowing for optical access within the channel. The device is assembled by tightly screwing the components together. We glue flexible Tygon tubing (McMaster-Carr) into the inlet and outlet of each device using a watertight 2-part epoxy (JB Marine-Weld).

\subsection{3D ordered sphere packings}
The fabrication protocol for the 3D consolidated sphere packings is described in previous work~\cite{carlson_volumetric_2022}. The bead diameter in both packings is $D_p = 780 \ \upmu$m, and the diameter of the cylindrical consolidations is $D_{\mathrm{neck}} = 170 \ \upmu$m ($D_{\mathrm{neck}}/D_p = 0.22$). Neglecting the volume of the consolidations, the simple cubic (SC) unit cell has dimensions of $D_p \times D_p \times D_p$. Similarly, the body-centered cuboid unit cell has dimensions of $D_p \times D_p \times \sqrt{2}D_p$. Note that the BC geometry is a rectangular cuboid unit cell rather than an idealized cubic body-centered unit cell. The medium porosities are $\phi_{SC} = 0.476$ and $\phi_{\mathrm{BC}} = 0.284$. The medium Darcy permeabilities are $k_{\mathrm{SC}} = 567 \pm 9 \ \upmu$m$^2$ and $k_{\mathrm{BC}} = 197 \pm 2 \ \upmu$m$^2$. The total lengths of each packing are $L_{\mathrm{SC}} = 3.84$ cm and $L_{\mathrm{BC}} = 2.74$ cm, and the cross-sectional area is $A = 0.16$ cm$^2$.

\section{Flow parameters}
For the 2D pillar arrays, we define the characteristic shear rate as $\dot\gamma = U/R_p$, where $U = Q/(HW\phi)$ is the average flow velocity and $R_p$ is the pillar radius. For the 3D sphere packings, we define the characteristic interstitial shear rate as $\dot\gamma_I = Q/(A\sqrt{k \phi})$~\cite{mitchell_viscoelastic_2016,zami-pierre_transition_2016,berg_shear_2017} and characteristic interstitial velocity as $U_I = Q/(A\phi)$.

We define the Reynolds number, the ratio of inertial to viscous stresses, as Re $=\rho U D_p/\eta_I$, where $\rho$ is the fluid mass density, $U$ is the characteristic flow speed, and $\eta_I$ is the bulk shear viscosity evaluated at the characteristic shear rate. In our experiments, Re $< 10^{-2}$ which renders the effects of inertia negligible. Thus, the onset of flow instability is assumed to arise solely from fluid elasticity. Additionally, given the weak shear-thinning nature of the polymer solution ($n = 0.93$), we do not expect shear-thinning to influence the flow in our experiments.

The Weissenberg number quantifies the ratio of elastic to viscous stresses in the fluid, which we define here as Wi $=N_1 (\dot\gamma)/[2\sigma(\dot\gamma)]$. For a given imposed flow rate $Q$, we determine a nominal imposed Wi$_I$ evaluated using the fluid shear rheology and characteristic shear rate. 

\section{Flow imaging and pressure drop measurements}
\label{sec:exptprotocol}
We prepare each experiment by first flushing the pristine media with ultrapure Millipore water to saturate the pore space and remove any air bubbles. In the 2D devices, we sequentially flush with Millipore water and pure glycerol before pre-filling the medium with the polymer solution at a constant low flow rate of $Q = 1$ mL/hr for at least 3 hours using a Harvard Apparatus PHD2000 syringe pump. In the 3D devices, we sequentially flush with Millipore water, isopropanol, Millipore water, and pure glycerol before pre-filling the medium with the polymer solution at a constant low flow rate of $Q = 1$ mL/hr for at least 3 hours. This procedure ensures that the pore space is entirely saturated with the polymer solution prior to the flow experiment.

\textbf{Flow imaging.}
For each flow imaging experiment, we apply a stepped flow rate ramp spanning a range of Wi to probe steady and unsteady flow conditions. After each step change in flow rate, we let the flow equilibrate for at least one hour before recording any data. To visualize the flow, we seed the fluid with 1 $\upmu$m carboxylate-modified, fluorescent, polystyrene tracer particles (Invitrogen; yellow-green) that have excitation at 480-510 nm and emission at 515 nm. The tracer particle seeding concentrations for the 2D and 3D flow experiments are 10 ppm and 7 ppm, respectively. We place the medium on the stage of a Nikon A1R+ laser scanning confocal fluorescence microscope, where particles are excited by a 488 nm laser and detected with a 500-550 nm sensor. We record 2 min videos per imaging location at 30 fps using the confocal resonant scanner with a raster scanning path.

In the 2D arrays, we use a 4$\times$ objective to interrogate a 3166.85 $\upmu$m x 3166.85 $\upmu$m field of view with optical section thickness $36.95 \ \upmu$m and pixel size 6.19 $\upmu$m/px (512 x 512 image). We focus the imaging plane at the channel half-height to avoid wall effects.

In the 3D packings, we use a 10$\times$ objective to interrogate a 1272.79 $\upmu$m x 1272.79 $\upmu$m field of view with optical section thickness $7.39 \ \upmu$m and pixel size 2.46 $\upmu$m/px (512 x 512 image). To capture the $z$-variations in the flow field, given the 2D flow imaging constraints of confocal microscopy, we image nine selected $z$-planes spanning the full unit cell ($\delta z = 50 \ \upmu$m) corresponding to depths ranging from $\sim 400-1200 \: \upmu$m from the bottom of the medium. We aggregate the measured flow fields across $z$ at each Wi for analysis.

In each device, we image multiple unit cells located at various positions throughout the pore space in each geometry. Sufficiently far from the inlet ($\sim$ 1-2 unit cells) and walls, we do not observe spatial variations in the flow field at lengthscales larger than the characteristic unit cell. Thus, for all subsequent analysis, we analyze flow fields corresponding to one unit cell located in the middle of each porous medium.

\textbf{Image processing.}
We use a MATLAB particle image velocimetry toolbox (PIVlab)~\cite{thielicke_pivlab_2014} to obtain time-resolved, 2D velocity fields with a temporal resolution of 33 ms. In PIVlab, we use the FFT window deformation algorithm with 4 passes of decreasing interrogation area (64-32-16-16 pixels) and 50\% overlap. The spatial PIV grid resolutions are 50.3 $\upmu$m (2D arrays) and 20.2 $\upmu$m (3D packings).

We generate pathline images and videos of the flow using ImageJ by averaging the mean pixel intensity over successive frames.

\begin{figure}
\centering
\includegraphics[scale=1.7]{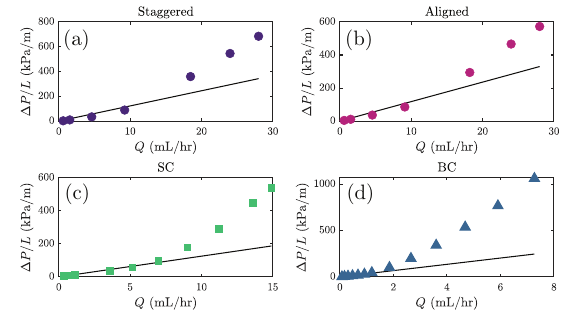}
\caption{\label{SIfig:rawpressure} Raw time-averaged pressure drop versus flow rate measurements show deviation from Darcy’s law using the measured zero-shear viscosity (solid lines) above a threshold flow rate in the (a) staggered (2D), (b) aligned (2D), (c) simple cubic (SC; 3D), and (d) body-centered cuboid (BC; 3D) porous medium geometries. Error bars corresponding to the standard deviation of time-averaged pressure drop signals are smaller than the symbol size.}
\end{figure}

\textbf{Pressure drop measurements.}
We quantify the macroscopic flow behavior using pressure drop measurements obtained by connecting a pressure transducer (Omega; 0-30 psi) in parallel to the inlet and outlet of each device. We impose a stepped sequence of increasing flow rate using a syringe pump, allowing the flow at each flow rate to equilibrate for at least one hour. We then time-average the pressure drop signal (10 Hz sampling rate) after the equilibration period and adjust for measurement noise by subtracting the pressure drop for no-flow conditions. Figure~\ref{SIfig:rawpressure} presents the raw pressure drop-flow rate measurements for the viscoelastic polymer solution in the four porous media. The solid line corresponds to Darcy's law evaluated using the zero-shear viscosity, $\eta_0$.  In all devices, the pressure drop follows the linear Darcy prediction at low flow rates; however, above a threshold flow rate, a pronounced positive deviation from Darcy's law is observed—a ``flow-thickening" of the polymer solution. We convert the pressure drop measurements into a dimensionless measure of flow resistance by calculating an apparent viscosity from Darcy's law and normalizing by the bulk shear viscosity measured from rheology, $\frac{\eta_{\mathrm{app}}}{\eta_I } = \frac{k\Delta P}{L(Q/A)\eta_I}$. A value of $\frac{\eta_{\mathrm{app}}}{\eta_I } = 1$ indicates that the apparent viscosity of the fluid in the porous media is equivalent to the bulk shear viscosity as measured in a cone-plate shear rheometer; values of $\frac{\eta_{\mathrm{app}}}{\eta_I } > 1$ correspond to a higher viscosity than predicted by bulk shear rheology alone. We do not observe any enhanced porous medium drag reduction effects ($\frac{\eta_{\mathrm{app}}}{\eta_I } < 1$) for this HPAM solution in our experiments.

\section{Derivation of power balance incorporating an apparent extensional viscosity}
To derive the dimensionless flow resistance model incorporating the local apparent extensional viscosity, we start from the mechanical energy balance obtained by dotting the Cauchy momentum equation with velocity: 
\begin{equation*}
    \frac{\partial(\frac{1}{2} \rho u^2)}{\partial t} +\nabla \cdot (\frac{1}{2} \rho u^2 \mathbf{u}) = P(\nabla \cdot \mathbf{u})-\nabla \cdot (P\mathbf{u}) +\nabla \cdot ( \boldsymbol{\tau} \cdot \mathbf{u})-\boldsymbol{\tau} : \nabla \mathbf{u}+\rho \mathbf{gu},
\end{equation*}
where $\mathbf{u}$ is velocity, $P$ is pressure, $\boldsymbol{\tau}$ is the fluid stress tensor, and $\mathbf{g}$ is gravitational acceleration. Following the derivation by~\citeauthor{browne_elastic_2021}~\cite{browne_elastic_2021}, we neglect the inertial kinetic energy terms (Re $\ll 1$), assume incompressible flow ($\nabla \cdot \mathbf{u} = 0$), neglect viscous work at the walls of the porous medium, and neglect gravitational work. Integrating the resulting energy balance and averaging over volume and time, we obtain a macroscopic energy balance,
\begin{equation*}
    \frac{\langle \Delta P \rangle_t}{L} = \frac{\langle \boldsymbol{\tau}: \nabla\mathbf{u} \rangle_{t,V}}{(Q/A)},
\end{equation*}
which balances the pressure work done by the fluid to the energy dissipation in the fluid. Employing Darcy's law, this expression can be recast in terms of the apparent viscosity in a porous medium:
\begin{equation*}
    \frac{\eta_{\mathrm{app}}}{\eta_I} = \frac{k\langle \boldsymbol{\tau}: \nabla\mathbf{u} \rangle_{t,V}}{(Q/A)^2 \eta_I},
\end{equation*}
where we have normalized by the bulk shear viscosity of the fluid to quantify a dimensionless measure of the flow resistance. 

Previously,~\citeauthor{browne_elastic_2021}~\cite{browne_elastic_2021} used a Reynolds decomposition for the viscous dissipation term, $\langle \boldsymbol{\tau}: \nabla\mathbf{u} \rangle_{t,V}$, to separate the mean and fluctuating flow components. In doing so, they derive the following power balance that accounts for resistances from the shear viscosity and unstable viscous dissipation arising from an elastic instability:
\begin{equation}
\label{eq:patchesPB}
    \frac{\eta_{\mathrm{app}}}{\eta_I} = 1 + \frac{k \langle \chi \rangle_{t,V}}{(Q/A)^2\eta_I}.
\end{equation}
Here, $\langle \chi \rangle_{t,V} \sim \langle \mathbf{s'} :  \mathbf{s'} \rangle_{t,V}$ is the unstable viscous dissipation directly measured from the pore-scale flow fields and $\mathbf{s'}$ is the fluctuating components of the rate-of-strain tensor. The growth in the unstable viscous dissipation is described by a power law: $\langle \chi \rangle_{t,V} = \alpha (\mathrm{Wi}-\mathrm{Wi}_c)^\beta$, where $\alpha$ and $\beta$ are fitting parameters and Wi$_c$ is the critical Wi for the onset of instability. This derivation assumes a \textit{generalized Newtonian fluid model}—specifically, a power law fluid—which captures the non-Newtonian shear-thinning behavior of polymer solutions, but neglects any elasticity or enhanced extensional viscosity beyond that of a Newtonian fluid ($\mathrm{Tr}_{\mathrm{Newt}}  =3$ for uniaxial extension~\cite{trouton_coefficient_1906,petrie_extensional_2006,haward_extensional_2023}). A large increase in extensional viscosity can be observed for elastic polymer solutions resulting from the alignment and stretching of flexible polymer chains under flow~\cite{macosko_rheology_1994,larson_structure_1999,mckinley_filament-stretching_2002}. The increase in extensional viscosity directly results from an increase in the polymeric tensile stresses, $\tau_{p,ii}$, which are strongly flow history-dependent. The relaxation of polymer stresses over finite timescales imparts the fluid with a ``memory" of its deformation history. Experimentally, it is challenging to directly resolve these tensile stresses $\tau_{p,ii}$ and this problem remains a subject of ongoing research~\cite{corona_fingerprinting_2022,kumar_lagrangian_2023,kumar_stress_2023}. Thus, we adopt an alternative approach to evaluating the enhanced extensional stresses through a \textit{local apparent extensional viscosity} rather than attempting to evaluate the polymer stresses directly.

Returning to the decomposition of the viscous dissipation term proposed by~\citeauthor{browne_elastic_2021}~\cite{browne_elastic_2021}:
\begin{equation*}
    \langle \boldsymbol{\tau}: \nabla\mathbf{u} \rangle_{t,V} = \underbrace{(\boldsymbol{\tau} : \nabla \mathbf{u})_0}_{\mathrm{mean}} + \underbrace{\langle \chi \rangle_{t,V}}_{\mathrm{fluctuating}},
\end{equation*}
we now further decompose the mean dissipation term to separate the shear and extensional stresses, producing,
\begin{equation}
\label{eq:stressdecomp}
\begin{aligned}
    (\boldsymbol{\tau} : \nabla \mathbf{u})_0 &= \biggr [\tau_{xx} \frac{\partial u}{\partial x}+\tau_{yy} \frac{\partial v}{\partial y}+\tau_{zz} \frac{\partial w}{\partial z} \biggr]_0 \\
    &+ \biggr[\tau_{xy} \bigg(\frac{\partial u}{\partial y}+\frac{\partial v}{\partial x}\bigg)+\tau_{yz} \bigg(\frac{\partial w}{\partial y}+\frac{\partial v}{\partial z}\bigg)+\tau_{xz} \bigg(\frac{\partial w}{\partial x}+\frac{\partial u}{\partial z}\bigg) \biggr]_0.
\end{aligned}
\end{equation}
The terms in the first bracket on the right side of Eq.~\eqref{eq:stressdecomp} contain the tensile stresses and the terms in second bracket contain the shear stresses. Here, we assume that the terms containing shear stresses are adequately captured by the shear viscosity, $\eta_I(\dot\gamma)$, from bulk rheology. The extensional stress terms can be rewritten as,
\begin{equation}
\label{eq:extdecomp}
    \biggr [\tau_{xx} \frac{\partial u}{\partial x}+\tau_{yy} \frac{\partial v}{\partial y}+\tau_{zz} \frac{\partial w}{\partial z} \biggr]_0 = \biggr[(\tau_{xx}-\tau_{yy})\frac{\partial u}{\partial x} + (\tau_{yy}-\tau_{zz})\biggr(\frac{\partial u}{\partial x}+\frac{\partial v}{\partial y}\biggr)\biggr]_0.
\end{equation}
The term $(\tau_{yy}-\tau_{zz})$ is the second normal stress difference, $N_2$, which is typically negligible for dilute polymer solutions~\cite{bird_dynamics_1987,maklad_review_2021}; hence, we neglect the second term on the right side of Eq.~\eqref{eq:extdecomp}. Finally, we introduce the extensional viscosity, $\eta_E = \frac{\tau_{xx}-\tau_{yy}}{\dot\epsilon}$, and strain rate, $\dot\epsilon = \frac{\partial u}{\partial x}$ to arrive at the modified power balance:
\begin{equation}
\label{eq:ExPB}
    \frac{\eta_{\mathrm{app}}}{\eta_I} = \underbrace{1}_{\mathrm{mean \ shear}} + \underbrace{\frac{k \langle \chi \rangle_{t,V}}{(Q/A)^2\eta_I}}_{\mathrm{unstable \ visc. \ diss.}} + \underbrace{\frac{k \langle \eta_E \ \dot\epsilon^2 \rangle_{t,V}}{(Q/A)^2\eta_I}}_{\mathrm{enhanced \ ext. \ stress}}.
\end{equation}
The last term in Eq.~\eqref{eq:ExPB} now accounts for enhanced extensional stresses arising from polymer stretching quantified via an apparent extensional viscosity.

\section{Hencky strain and determination of local Trouton ratio}
\label{SIsec:hencky}
\subsection{Measurement of Hencky strain}
Our approach to evaluating the extensional resistance term in Eq.~\eqref{eq:ExPB} first requires measuring distributions of pore-scale accumulated Hencky strains which quantifies the net extensional deformation of polymers. We determine the Hencky strain for a fluid particle by initializing the particle at a given location within a pore and integrating forward in time using the experimental velocity field over a total interval equal to the longest relaxation time of the polymer; the longest relaxation time of the 300 ppm HPAM solution is $\lambda_0 = 0.5$ s from capillary break-up rheometry. The longest relaxation time corresponds to the maximum time to which polymeric elastic stresses can persist before being dissipated by relaxation mechanisms. Mathematically, the Hencky strain is,
\begin{equation*}
    \epsilon_{\mathrm{Hencky}} = \int_{t_0}^{t_0 + \lambda_0} |\dot\epsilon| \mathrm{d}t,
\end{equation*}
which we numerically approximate as,
\begin{equation*}
    \epsilon_{\mathrm{Hencky}} \approx \sum_{n=1}^{\lambda_0/\delta t} |\dot\epsilon|_n \delta t.
\end{equation*}
Here, $|\dot\epsilon| = |\frac{\partial u}{\partial x}| +|\frac{\partial v}{\partial y}|$ is the local strain rate and $\delta t$ is the time step set by the temporal resolution of the experimental flow field. We use the absolute value of the strain rate to track the cumulative extensional deformation over stretching events. For each time step, we evaluate $|\dot\epsilon|$ at the current fluid particle position and use the corresponding fluid velocity $\mathbf{u}(\mathbf{x},t)$ to advance the particle's position to the next time step, $\delta \mathbf{x} = \mathbf{u}(\mathbf{x},t)\delta t$. In total, we repeat this Hencky strain calculation for 15 randomly selected time points in the imaging window and for 5 randomly chosen initial particle positions to generate a probability density function (p.d.f) of accumulated Hencky strains for a given imposed Wi$_I$. Note that the limited imaging field-of-view sets a spatial constraint on the maximum distance a fluid particle may travel over the relaxation time integration interval.

\begin{figure*}
\centering
\includegraphics[scale=1.2]{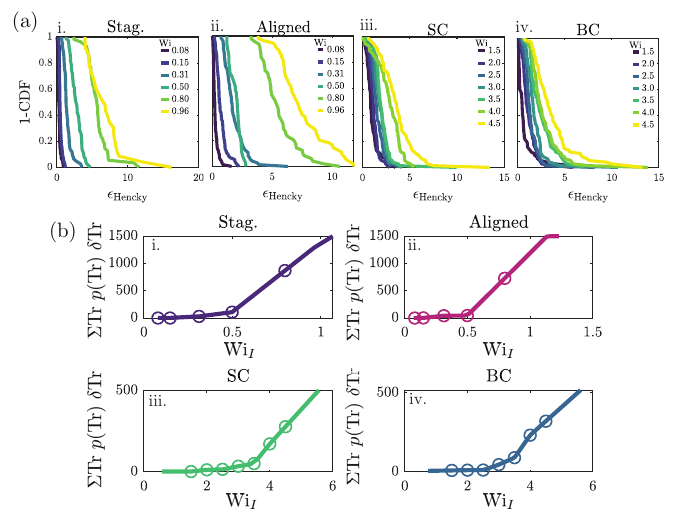}
\caption{\label{SIfig:hencky} Experimentally-measured accumulated Hencky strains. (a) Inverse cumulative distribution functions (1-CDF) of the accumulated Hencky strain over one polymer relaxation time ($\lambda_0 = 0.5$ s) as a function of Wi in each of the geometries. Each distribution comprises 75 measurements of the Hencky strain across randomly selected time points and locations within each pore. (b) Weighted apparent Trouton ratio as a function of Wi obtained by integrating over the Hencky strain probability density distributions in (a). Hollow circles are the weighted Tr ratios determined from each experimental flow rate condition and solid curves are linear interpolations/extrapolations.}
\end{figure*}

Figure~\ref{SIfig:hencky}(a) presents the inverse cumulative probability density functions (1-CDF) of Hencky strains in the four porous medium geometries. Generally, the magnitude of accumulated Hencky strain increases with increasing Wi$_I$. Importantly, we observe that the Hencky strains can exceed values of $\epsilon_{\mathrm{Hencky}} = 2-3$, which is a benchmark threshold for the polymer coil-stretch transition~\cite{de_gennes_coilstretch_1974,perkins_single_1997}. Above the coil-stretch transition, velocity gradients are sufficiently large enough to orient and stretch polymer chains from their equilibrium coil configurations, producing a concomitant increase in the fluid extensional viscosity relative to the shear viscosity~\cite{anna_interlaboratory_2001,mckinley_filament-stretching_2002}.

\subsection{Weighted Trouton ratio calculation}
\label{subsec:wtdTr}
Using the distributions of measured Hencky strains, we can further determine distributions of local apparent Trouton ratios by using the relation $\mathrm{Tr}(\epsilon_{\mathrm{Hencky}})$ in Eq.~\eqref{eq:caberLog} from capillary break-up rheometry. Note that the specific parameters in the logistic equation for $\mathrm{Tr}(\epsilon_{\mathrm{Hencky}})$ are fluid-specific. For each imposed Wi$_I$, we obtain a probability density distribution of local apparent Trouton ratios, $p(\mathrm{Tr})$, describing fluid particles that are advected and deformed by the flow.

Finally, to evaluate the extensional resistance term in Eq.~\eqref{eq:ExPB}, we first substitute for the transient extensional viscosity, $\eta_E = \eta_0 \mathrm{Tr}$, and nominal strain rate, $\dot\epsilon = \frac{Q\zeta}{A\phi D_p}$, where $\zeta$ is a dimensionless, geometry-dependent parameter (see \S\ref{subsec:zeta}). The resistance term can be rewritten as,
\begin{equation}
\label{eq:caberterm}
    \frac{k \langle \eta_E \dot\epsilon^2 \rangle_{t,V}}{(Q/A)^2\eta_I} = \frac{k \eta_0 \langle \mathrm{Tr} \rangle_{t,V} \zeta^2}{\phi^2 D_p ^2 \eta_I}.
\end{equation}
We use the probability density distributions, $p(\mathrm{Tr})$, to calculate $\langle \mathrm{Tr} \rangle_{t,V}$ by calculating the probability density-weighted expectation value using numerical integration:
\begin{equation*}
\begin{aligned}
    \langle \mathrm{Tr} \rangle_{t,V} &= \int_{\mathrm{Tr}_0}^{\mathrm{Tr}_\infty} \mathrm{Tr} \ p(\mathrm{Tr}) \ \mathrm{dTr} \\ 
    & \approx \sum_i \mathrm{Tr}_i \ p(\mathrm{Tr}_i) \ \delta \mathrm{Tr}.
\end{aligned}
\end{equation*} We determine $\langle \mathrm{Tr} \rangle_{t,V}$ for each imposed Wi$_I$ in the flow experiments and linearly interpolate or extrapolate between these discrete points to evaluate the power balance model; Fig.~\ref{SIfig:hencky}(b) shows the flow field-weighted Trouton ratios as a function of Wi$_I$ in the four porous medium geometries. Open circles are the experimental values and the solid curves are linear interpolation/extrapolations. Physically, $\langle \mathrm{Tr} \rangle_{t,V}$ can be interpreted as a medium-averaged value of the local apparent Trouton ratio weighted by the pore-scale flow field—this weighting captures variability in the extensional flow behavior within the pore space as different fluid particles experience different flow kinematics depending on their Lagrangian trajectory.

\subsection{Theoretical Hencky strain due to contractions set by geometry} \label{subsec:theoHencky}
\begin{figure}
\centering
\includegraphics{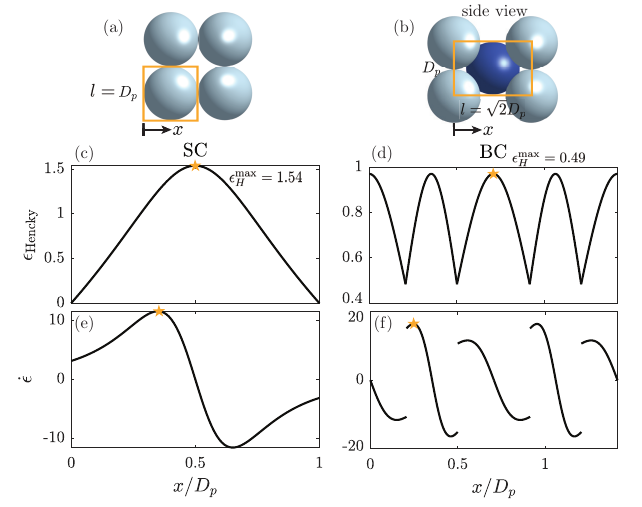}
\caption{\label{SIfig:henckygeo_3D}Analytical Hencky strains in 3D sphere packings generated by contractions in the pore geometry alone. (a) SC and (b) BC unit cells, where the yellow rectangle marks the unit cell used to determine the analytical Hencky strain. Note that the BC unit cell in this work is a rectangular cuboid rather than an ideal cubic unit cell. (c-d) Position-dependent Hencky strain, determined from $\epsilon_{\mathrm{Hencky}}(x) = \mathrm{ln}(A_\mathrm{max}/A(x))$, where $A(x)$ is the cross-sectional area of the void space. Yellow stars mark the maximum Hencky strain. (e-f) Position dependent extension rate, $\dot\epsilon = \frac{\mathrm{d}V(x)}{\mathrm{d}x}$, non-dimensionalized by $Q/D_p^3$ where $Q$ is the volumetric flow rate and $V(x) = Q/A(x)$. Yellow stars mark the maximum extension rate predicted from contractions in the pore space alone (i.e. neglecting extensional flow generated at stagnation points).}
\end{figure}

A useful feature of the ordered porous medium geometries in this work is that the unit cell void spaces can be described mathematically using geometrical considerations. As such, we can analytically derive the predicted Hencky strains for extension caused by \textit{contractions} in the pore space alone (i.e. neglecting extensional flow at stagnation points). Given the position-dependent cross-sectional area of the void space, $A(x)$, the position-dependent Hencky strain is, $\epsilon_{\mathrm{Hencky}}(x) = \mathrm{ln}\big(\frac{A_{\mathrm{max}}}{A(x)}\big)$.
 Here, $x$ denotes the streamwise position along the length of the unit cell. For the 2D arrays, the Hencky strain is equivalent to $\epsilon_{\mathrm{Hencky}} = \mathrm{ln}\big(\frac{L_{\mathrm{max}}}{L(x)}\big)$,  where $L(x)$ is the length of the void space at a given position $x$ obtained by trivially dividing $A(x)$ by the constant channel height. Equations~\eqref{eq:stagL}-~\eqref{eq:BCA} below present the mathematical expressions for the void space in each unit cell. Here, $l_p$ is the pillar center-center distance, $l$ is the streamwise length of the BC unit cell ($l=\sqrt{2}D_p$) [Fig.~\ref{SIfig:henckygeo_3D}(b)], and $D_p$ is the pillar (2D) or grain diameter (3D).
\begin{equation}
\label{eq:stagL}
L_{\mathrm{stag}}(x) = \begin{cases}
l_p-2\sqrt{(\frac{D_p}{2})^2-x^2} & 0\leq x < \frac{D_p}{2},\\
l_p  & \frac{D_p}{2}\leq x < \frac{\sqrt{3}}{2}l_p-\frac{D_p}{2},\\
l_p-2\sqrt{(\frac{D_p}{2})^2-(x-\frac{\sqrt{3}}{2}l_p)^2}  &  \frac{\sqrt{3}}{2}l_p-\frac{D_p}{2}\leq x < \frac{\sqrt{3}}{2}l_p+\frac{D_p}{2},\\
l_p  &\frac{\sqrt{3}}{2}l_p+\frac{D_p}{2}\leq x < \sqrt{3}l_p-\frac{D_p}{2} \\
l_p-2\sqrt{(\frac{D_p}{2})^2-(x-\sqrt{3}l_p)^2} & \sqrt{3}l_p-\frac{D_p}{2}\leq x \leq \sqrt{3}l_p.
\end{cases}
\end{equation}

\begin{equation}
\label{eq:alignedL}
L_{\mathrm{aligned}}(x) = \begin{cases}
\sqrt{3}l_p-2\sqrt{(\frac{D_p}{2})^2-x^2} & 0\leq x < \frac{D_p}{2},\\
\sqrt{3}l_p  & \frac{D_p}{2}\leq x < \frac{l_p}{2}-\frac{D_p}{2},\\
\sqrt{3}l_p-2\sqrt{(\frac{D_p}{2})^2-(x-\frac{l_p}{2})^2}  &   \frac{l_p}{2}-\frac{D_p}{2}\leq x <  \frac{l_p}{2}+\frac{D_p}{2}\\
\sqrt{3}l_p  &\frac{l_p}{2}+\frac{D_p}{2}\leq x < l_p-\frac{D_p}{2} \\
\sqrt{3}l_p-2\sqrt{(\frac{D_p}{2})^2-(x-l_p)^2} & l_p-\frac{D_p}{2}\leq x \leq l_p.
\end{cases}
\end{equation}

\begin{equation}
\label{eq:SCA}
A_{\mathrm{SC}}(x) = D_p^2-\pi\biggr(\frac{D_p}{2}\biggr)^2+\pi\biggr(x-\frac{D_p}{2}\biggr)^2 \: \: \: \text{for} \: 0\leq x \leq D_p
\end{equation}

\begin{equation}
\label{eq:BCA}
A_{\mathrm{BC}}(x) = \begin{cases}
D_p^2-\pi\big((\frac{D_p}{2})^2-x^2\big) & 0\leq x < \frac{l}{2}-\frac{D_p}{2},\\
D_p^2-\pi\big((\frac{D_p}{2})^2-x^2\big)-\pi\big((\frac{D_p}{2})^2-(x-\frac{l}{2})^2\big)  & \frac{l}{2}-\frac{D_p}{2}\leq x < \frac{D_p}{2},\\
D_p^2-\pi\big((\frac{D_p}{2})^2-(x-\frac{l}{2})^2\big)  & \frac{D_p}{2} \leq x < \frac{l}{2},\\
D_p^2-\pi\big((\frac{D_p}{2})^2-(x-\frac{l}{2})^2\big)  & \frac{l}{2} \leq x < l-\frac{D_p}{2},\\
D_p^2-\pi\big((\frac{D_p}{2})^2-(x-\frac{l}{2})^2\big)-\pi\big((\frac{D_p}{2})^2-(x-l)^2\big)  &  l-\frac{D_p}{2} \leq x < \frac{l}{2}+\frac{D_p}{2},\\
D_p^2-\pi\big((\frac{D_p}{2})^2-(x-l)^2\big) & \frac{l}{2}+\frac{D_p}{2} \leq x \leq l.
\end{cases}
\end{equation}

\begin{figure}
\centering
\includegraphics{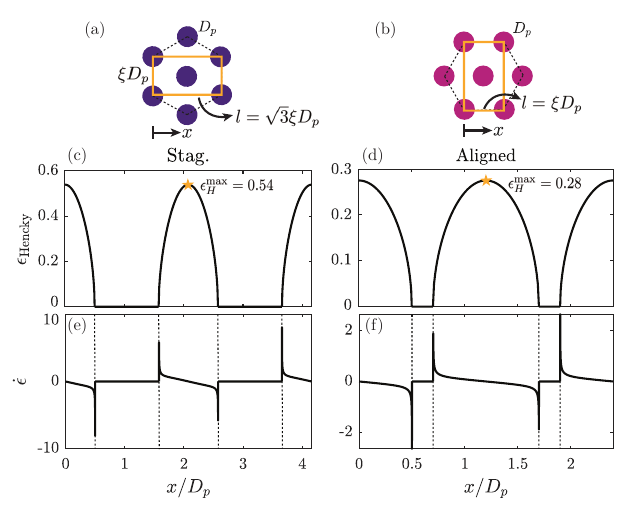}
\caption{\label{SIfig:henckygeo_2D}Analytical Hencky strains in 2D hexagonal arrays generated by contractions in the pore geometry. (a) Staggered and (b) aligned hexagonal unit cells, where the yellow rectangle marks the region used to determine the analytical Hencky strain and $\xi$ is the ratio of the pillar center-to-center spacing $l_p$ to the pillar diameter $D_p$. (c-d) Position-dependent Hencky strain in each unit cell, determined from $\epsilon_{\mathrm{Hencky}}(x) = \mathrm{ln}(L_\mathrm{max}/L(x))$, where $L(x)$ is the length of the void space. Yellow stars mark the maximum Hencky strain. (e-f) Position-dependent extension rate, $\dot\epsilon = \frac{\mathrm{d}V(x)}{\mathrm{d}x}$, non-dimensionalized by $q/D_p^2$ where $q$ is the 2D flux and $V(x) = q/L(x)$. Dotted lines indicate locations where the analytical strain rate diverges due to the circular nature of pillar geometry.}
\end{figure}

Figures~\ref{SIfig:henckygeo_2D}(a, b) show schematics of the 2D pillar array unit cell geometries and characteristic length scales. Note that here we use a truncated rectangular unit cell rather than the full hexagonal unit cell for ease of calculation. Figures~\ref{SIfig:henckygeo_2D}(c, d) present the local Hencky strain profiles for the staggered and aligned unit cells. Yellow stars mark the maximum values of Hencky strain: $\epsilon_{\mathrm{Hencky}}^{\mathrm{max}}(\mathrm{stag.}) = 0.54$ and $\epsilon_{\mathrm{Hencky}}^{\mathrm{max}}(\mathrm{aligned}) = 0.28$. This is a key result: the measured Hencky strains from the experimental flow fields [Fig.~\ref{SIfig:hencky}(a)] are much larger than the maximum Hencky strains associated with contraction-generated extensional flow [Figs.~\ref{SIfig:henckygeo_2D}(c, d)]; thus, we hypothesize that extensional flow in the 2D arrays is dominated by stagnation points rather than contractions in the pore space geometry—ultimately motivating the resistance-per-stagnation point model in \S\ref{sec:oscerDer}.

Additionally, we use the analytical expressions of the void space cross-sectional area to estimate a position-dependent strain rate in each unit cell following~\citeauthor{haward_viscosity_2003}~\cite{haward_viscosity_2003,odell_viscosity_2006}. For constant imposed 2D volumetric flux $q$, conservation of mass requires $q = \bar V \bar L = V(x) L(x)$, where $V(x)$ is the position-dependent flow speed and an overbar denotes a spatially-averaged quantity. Rearranging, we find $V(x) = q/L(x)$ and differentiation produces the local strain rate, $\dot\epsilon (x) = \frac{\mathrm{d}V(x)}{\mathrm{d}x}$. The resulting strain rate profiles (non-dimensionalized by $q/D_p^2$) are shown in Figs.~\ref{SIfig:henckygeo_2D}(e, f). Because of the circular nature of the pillars in this geometry, there is a numerical artifact where the analytical expressions for $\dot\epsilon (x)$ diverge, denoted by the dotted lines on the plots.

We perform a similar analysis for the 3D sphere packing unit cells [Figs.~\ref{SIfig:henckygeo_3D}(a, b)], now using the full void space cross-sectional area, $A(x)$. Figures~\ref{SIfig:henckygeo_3D}(c, d) present the local Hencky strain profiles with maximum values of $\epsilon_{\mathrm{Hencky}}^{\mathrm{max}}(\mathrm{SC}) = 1.54$ and $\epsilon_{\mathrm{Hencky}}^{\mathrm{max}}(\mathrm{BC}) = 0.49$. Again, these maximum values are smaller than the Hencky strain values we measure from the flow fields in Fig.~\ref{SIfig:hencky}(a), and we conclude that extensional flow in the 3D packings is dominated by stagnation points rather than contractions in the pore space geometry. Figures~\ref{SIfig:henckygeo_3D}(e, f) present the local strain rate profiles in each unit cell non-dimensionalized by $Q/D_p^3$, determined from $\dot\epsilon = \frac{\mathrm{d}V(x)}{\mathrm{d}x}$ and $V(x) = Q/A(x)$ with $Q$ a constant volumetric imposed flow rate. Yellow stars mark the locations of maximum strain rate. In this analysis, we neglect the consolidations to simplify the geometric expressions for the void space; however, we do not expect our conclusions regarding the contraction-generated versus stagnation point-generated extensional strains to change substantially with the inclusion of the consolidations given their small contribution to the total solid volume.

\subsection{\label{subsec:zeta}Physical interpretation of effective strain rate}
As mentioned in \S\ref{subsec:wtdTr}, we define the nominal strain rate in the porous medium as,
\begin{equation}
\label{eq:nomstrainrate}
    \dot\epsilon = \frac{Q\zeta}{\phi A D_p},
\end{equation}
where we use $\zeta$ in Eq.~\eqref{eq:caberterm} as a fitting parameter in the power balance model [Eq.~\eqref{eq:ExPB}]. Similar approaches have been used to define an effective interstitial shear rate~\cite{mitchell_viscoelastic_2016,zami-pierre_transition_2016,berg_shear_2017} and nominal extensional strain rates~\cite{haas_viscoelastic_1982,farinato_polymer_1987,dyakonova_macromolecular_1996,nguyen_rheology_1999} in prior literature, where parameters analogous to $\zeta$ are used as geometry-dependent fitting parameters. However, previous analyses, to our knowledge, do not provide a physical interpretation of the corresponding fitting parameters. Here, we provide a physical interpretation and rationalization of the identified $\zeta$ values in the extensional viscosity-modified power balance model for the four ordered porous medium geometries. 

Table~\ref{SItab:zetas} lists the experimental $\zeta$ values from the power balance models in each of the four porous medium geometries. A nominal strain rate can be approximated as a streamwise velocity difference over a characteristic length scale of extension: $\frac{\partial u}{\partial x} \approx \frac{\Delta u}{\Delta l}$. Our definition implies a velocity difference of $\Delta u \approx (U_I-0)= \frac{Q}{\phi A}$ over a length scale $\Delta l \approx D_p/\zeta$, where $U_I$ is the characteristic interstitial velocity. Table~\ref{SItab:zetas} summarizes the resulting extensional length scales $\Delta l \approx D_p/\zeta$ proposed by our analysis. 

\begin{table}
\small
\centering
\caption{\label{SItab:zetas} Comparison of theoretical and experimental $\zeta$ values. The parameter $\alpha$ describes the weight on $D_p$ in a linear combination of geometric length scales in the 2D arrays: $D_p$ and $l_s$ (2D).}
\begin{tabular}{ccccc}
\hline \hline
Geometry&Theory&Expt.&$1/\zeta$ & $\alpha$\\
\hline
Stag. (2D)& 0.443 & 0.53 & 1.89 &0.72\\
Aligned (2D)& 0.275 & 0.47& 2.13&0.18\\
SC (3D)& 0-5.52 & 1.13& 0.89&---\\
BC (3D)& 0-4.25 & 1.55& 0.65&---\\
\hline \hline
\end{tabular}
\end{table}

In the 2D arrays, the extensional length scale in both configurations is larger than the pillar diameter $D_p$ but smaller than the streamwise pillar spacing $l_s$. Previous work studying viscoelastic flows in cylindrical arrays has suggested that the relevant length scale of extension depends on the density of pillars (or solidity): small diameter pillars ($\frac{D_p}{l_s} \ll 1$) act as ``point sources" of stagnation points producing a dominant extensional length scale equal to the distance between point sources $l_s$, while extension in arrays of larger pillars with small spacing ($\frac{D_p}{l_s} \lessapprox 1$) is controlled by $D_p$~\cite{chmielewski_effect_1992,chmielewski_elastic_1993,nguyen_rheology_1999,james_slow_2012}. Our results align with these prior studies, suggesting that the appropriate extensional length scale for the hexagonal arrays used in this work fall somewhere between these two extremes. Indeed, we may consider $D_p/\zeta$ as a linear combination of the two length scales $D_p$ and $l_s$ given by, $\frac{D_p}{\zeta} = \alpha D_p +(1-\alpha) l_s$,
with $\alpha$ a weighting parameter ranging from $0-1$. We find $\alpha_{\mathrm{stag}} = 0.72$ and $\alpha_{\mathrm{aligned}} = 0.18$. A lower value of $\alpha$ (smaller weight on $D_p$ compared to $l_s$) in the aligned array compared to the staggered array can be rationalized by the increased screening of stagnation points with increasing Wi, as proposed by~\citeauthor{haward_stagnation_2021}~\cite{haward_stagnation_2021}. 

In the 3D packings, the values of $D_p/\zeta$ suggest that the relevant extensional length scale is smaller than the grain diameter $D_p$. Intuitively, this aligns with our previous work where we find that elastic instabilities in these consolidated sphere packings are generated from stagnation points arising at \textit{grain consolidations} rather than stagnation points on the grain surfaces themselves~\cite{chen_stagnation_2024}. As a result, we expect that the effective length scale of extension in the 3D ordered sphere packings lies between the diameter of the consolidations $D_{\mathrm{neck}}$ and the diameter of the grains $D_p$. For the specific geometries in this work, $D_{\mathrm{neck}}/D_p \approx 0.22$, satisfying our hypothesis that $D_{\mathrm{neck}}/D_p \leq 1/ \zeta \leq 1$.


\subsection{Theoretical prediction of $\zeta$}
The local strain rate analysis in \S\ref{subsec:theoHencky} also enables theoretical prediction of $\zeta$ using purely geometrical considerations. Using the analytical expressions of the local strain rate $\dot\epsilon(x)$, we obtain expressions for the maximum (3D) or characteristic (2D) strain rate in a given unit cell and equate this with our definition of the nominal strain rate [Eq.~\eqref{eq:nomstrainrate}]. Solving for $\zeta$ produces an expression that can be evaluated using known numerical constants and the medium porosity. Because the expressions for the local strain rate in the 2D arrays are unbounded [Figs.~\ref{SIfig:henckygeo_2D}(e, f)], we instead define the characteristic strain rate as $\dot\epsilon = \frac{V_{\mathrm{max}}-V_{\mathrm{min}}}{L_{\mathrm{min}}-L_{\mathrm{max}}}$. For the 3D arrays, we determine expressions for the maximum local strain rate (marked in Figs.~\ref{SIfig:henckygeo_3D}(e, f)) which produces an upper bound estimate for $\zeta$. Table~\ref{SItab:zetas} summarizes the theoretical $\zeta$ predictions. We find reasonable agreement with the experimentally-determined values for all four geometries. In summary, while $\zeta$ is a fitting parameter in our extensional resistance-modified power balance model, we show that the values of $\zeta$ for the ordered geometries have a physical basis by rationalizing the effective length scales for extensional strain in each unit cell geometry and comparing with theoretical predictions.

\section{\label{sec:oscerDer}Derivation of modified Darcy's law incorporating an extensional resistance-per-stagnation point in ordered geometries}
The analysis in \S\ref{subsec:theoHencky} suggests that the extensional flow in the 2D pillar arrays and 3D sphere packings is predominantly controlled by stagnation points rather than contractions in the pore space. Accordingly, we seek to develop an alternative resistance model that captures the enhanced extensional flow resistance on a resistance-per-stagnation point basis. 

\begin{figure}
\centering
\includegraphics[scale=1.7]{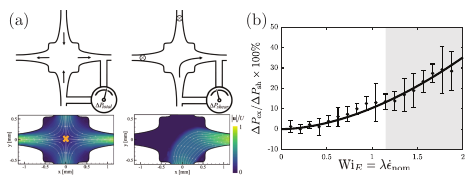}
\caption{\label{SIfig:oscer}Optimized cross-slot device and excess extensional pressure drop. (a) Schematic of OSCER device operating in extensional flow (left) and shear flow (right) modes with example flow fields. One free stagnation point (yellow $x$) is located at the center of the device when operating in planar extensional flow mode. (b) Ratio of extensional excess pressure drop to shear pressure drop as a function of the nominal imposed extensional $\mathrm{Wi}_E = \lambda_0 \dot\epsilon$. The solid curve is a power law fit: $\Delta P_\mathrm{ex}/\Delta P_\mathrm{sh} \times 100 \% = 10.4 \mathrm{Wi}_E ^{1.8}$. Error bars represent one standard deviation in the time-averaged signal of the pressure drop. The gray shaded region indicates the onset of a mobile stagnation point associated with elastic instability-induced flow asymmetry~\cite{haward_optimized_2012,galindo-rosales_optimized_2014}.}
\end{figure}

We treat the optimized cross slot extensional rheometer (OSCER)~\cite{haward_optimized_2012,galindo-rosales_optimized_2014} as a model pore containing one free stagnation point (yellow $x$ marked in Fig.~\ref{SIfig:oscer}(a)) at the center of the device. In planar extensional flow mode, flow is driven through two opposing inlets and exits through two perpendicular opposing outlets, thereby generating a strong extensional velocity gradient in the outflow channels. We measure the total pressure drop across the device, $\Delta P_{\mathrm{tot}}$, while applying a constant flow rate ramp. The device can also operate in a shear flow mode: one inlet and one outlet are closed off such that the fluid passes through a predominantly shearing corner flow. We measure the pressure drop associated with shearing in this mode as $\Delta P_{\mathrm{sh}}$. Assuming the excess pressure drop across the device results from purely extensional effects, we then obtain the extensional pressure drop in the device as $\Delta P_{\mathrm{ex}}=\Delta P_{\mathrm{tot}}-\Delta P_{\mathrm{sh}}$. Figure~\ref{SIfig:oscer}(b) plots the ratio of the extensional to shear pressure drop in the OSCER device as a function of the imposed extensional Weissenberg number, $\mathrm{Wi}_E$, defined as the product of the longest polymer relaxation time ($\lambda$) and a nominal strain rate in the device ($\dot\epsilon_{\mathrm{nom}}$). Error bars represent one standard deviation from time-averaging the pressure drop signal. The shaded gray region indicates the onset of an elastic instability characterized by asymmetry in the inflow/outflow velocities and a mobile stagnation point~\cite{arratia_elastic_2006,sousa_purely-elastic_2018,davoodi_control_2019,yokokoji_rheological_2023}. As $\mathrm{Wi}_E$ increases, the extensional contribution to the pressure drop increases greatly, resulting from the onset of polymer stretching by the extensional velocity gradient originating at the stagnation point. We fit a power law to the pressure ratio measurements, $\frac{\Delta P_\mathrm{ex}}{\Delta P_\mathrm{sh}} \times 100 \% = 10.4 \mathrm{Wi}_E ^{1.8}$, which we take to model the extensional flow resistance-per-stagnation point. Note that the specific power law fitting parameters are fluid-dependent.

We now return to Darcy's law and explicitly separate the pressure drop into extensional and shear components,
\begin{equation*}
    \frac{\Delta P}{L} = \frac{\Delta P_\mathrm{ex}+ \Delta P_\mathrm{sh}}{L} = \frac{\eta_{\mathrm{app}}}{k} \frac{Q}{A}.
\end{equation*} Factoring out the shear pressure drop, rearranging for the apparent viscosity, and normalizing by the bulk shear viscosity produces,
\begin{equation*}
    \frac{\eta_{\mathrm{app}}}{\eta_I} = \biggr[\frac{k\Delta P_\mathrm{sh}}{(Q/A)L\eta_I} \biggr]\biggr[1+\frac{\Delta P_\mathrm{ex}}{\Delta P_\mathrm{sh}}\biggr].
\end{equation*} The first bracketed term is equivalent to the power balance model accounting for the unstable viscous dissipation only [Eq.~\eqref{eq:patchesPB}], and the right bracketed term is a factor accounting for the enhanced extensional resistance from \textit{one} stagnation point. To correctly account for multiple stagnation points, we multiply the pressure ratio by the number of stagnation points per unit cell, $n_{\mathrm{SP}}$, which is known \textit{a priori} for the ordered geometries. The resulting modified Darcy's law accounting for extensional flow resistance is,
\begin{equation}
\label{SIEq:OscerPB}
    \frac{\eta_{\mathrm{app}}}{\eta_I} = \biggr[1 + \frac{k \langle \chi \rangle_{t,V}}{(Q/A)^2\eta_I} \biggr]\biggr[1+\frac{\Delta P_\mathrm{ex}}{\Delta P_\mathrm{sh}} n_{\mathrm{SP}}\biggr].
\end{equation} This model can be evaluated directly using the local flow fields ($\langle \chi \rangle_{t,V}$), fluid rheology ($\eta_I$), and measurement of the pressure ratio in the OSCER device ($\frac{\Delta P_\mathrm{ex}}{\Delta P_\mathrm{sh}}$) without requiring any additional fitting parameters given that $n_{\mathrm{SP}}$ is known. In the 2D hexagonal arrays, $n_{\mathrm{SP}} = 6$. Table~\ref{SItab:nSP} provides a breakdown of the number of stagnation points per unit cell in the 3D packings, accounting for upstream and downstream stagnation points located at both the consolidations and the grain surfaces.

\begin{table}
\small
\centering
\caption{\label{SItab:nSP} Enumeration of stagnation points for the 3D consolidated sphere packing unit cells, accounting for both upstream and downstream stagnation points.}
\begin{tabular}{cccc}
\hline \hline
Geometry&Consolidations&Grains&Total\\
\hline
SC (3D)& 8 & 0 & 8\\
BC (3D)& 12 & 10 & 22\\
\hline \hline
\end{tabular}
\end{table}

\section{Disordered packings: power balances}
\subsection{Disordered sphere packings}
\label{SIsec:disPB}
\begin{figure}
\centering
\includegraphics[scale=1.3]{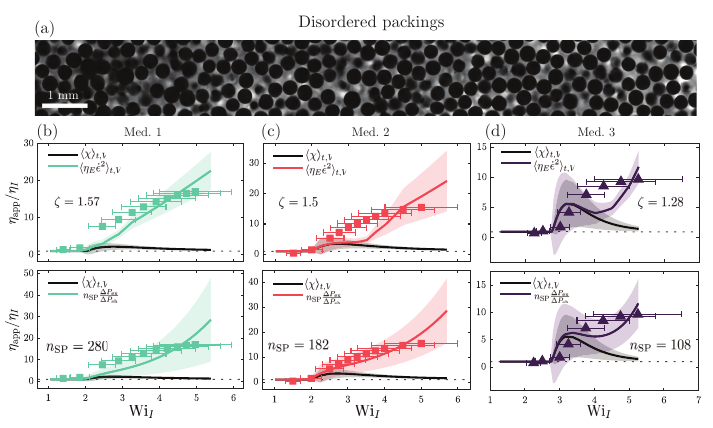}
\caption{\label{SIfig:disordered}Extensional resistance-modified power balance models for disordered bead packings ($D_p = 300-355~\upmu$m) exhibit greater variability in accuracy. (a) Example confocal micrograph of pore space in disordered bead packings. Black regions correspond to the solid grain matrix. (b-d) Power balance models presented in order of decreasing accuracy for experiments in 3 different porous media. Top row: local apparent extensional viscosity power balance with fitted $\zeta$ values. Bottom row: resistance-per-stagnation point model with fitted $n_\mathrm{SP}$. Here, we simply use $n_\mathrm{SP}$ as a fitting parameter to modulate the extensional contribution to the flow resistance. Shaded regions and error bars are error propagation from fluid rheology and flow measurements.}
\end{figure}

As many porous media in nature and in industrial applications are inherently disordered, we now investigate the applicability of the extensional resistance-modified power balance models towards predicting the flow resistance in two types of disordered porous media: lightly-sintered disordered bead packings [Fig.~\ref{SIfig:disordered}(a)] and lightly-sintered crushed glass packings [Fig.~\ref{SIfig:crushed}(a)].  

\begin{table}
\small
\centering
\caption{\label{SItab:disordered} Disordered bead packings: medium parameters.}
\begin{tabular}{ccccc}
\hline \hline
Medium&$\phi$&$k$ ($\upmu$m$^2$)&$L$ (cm)&$A$ (mm$^2$)\\
\hline
1 & 0.321 & 82.9$\pm$0.8 & 2.62 &  9\\
2 & 0.339 & 93.6$\pm$0.8 & 2.4 &  9\\
3 & 0.332 & 83.3$\pm$0.7 & 2.02 &  9\\
\hline \hline
\end{tabular}
\end{table}

Following our previous work~\cite{browne_elastic_2021,browne_homogenizing_2023,browne_harnessing_2024}, we fabricate disordered bead packings by lightly sintering randomly-packed borosilicate glass beads ($D_p = 300-355$ $\upmu$m; Mo-Sci) in a 3$\times$3 mm rectangular quartz capillary. Figure~\ref{SIfig:disordered}(a) presents a sample confocal micrograph of a $z$-slice within one packing showing the disordered nature of the pore space geometry. Table~\ref{SItab:disordered} summarizes the relevant medium parameters. We repeat the same experimental flow protocol described in \S\ref{sec:exptprotocol}, with one difference: due to the disordered nature of the pore space—where there are no longer characteristic unit cells—we randomly select and image 15 different pores within the porous medium in order to more accurately capture the pore-scale flow behavior. 

Figures~\ref{SIfig:disordered}(b-d) show the extensional resistance-modified power balances [Eq.~\eqref{eq:ExPB}] for 3 different experimental replicates in the disordered sphere packings, presented in order of decreasing model accuracy. The top row of plots corresponds to the local apparent extensional viscosity power balance model using a locally-weighted Trouton ratio [Eq.~\eqref{eq:ExPB}]. The power balance captures the macroscopic flow resistance in medium 1; in media 2 and 3, the model–data discrepancy is larger. The kink in the model curves for media 2 and 3 reflects mis-prediction of the onset of the extensional-resistance term: the abrupt uptick at $\mathrm{Wi}\approx3.5$ (medium 2) and $4.5$ (medium 3) signals the crossover from a regime dominated by unstable viscous dissipation to one dominated by the extensional-resistance term. The relative weights of the two resistance terms are set by the critical Weissenberg number (Wi$_c$) for the onset of elastic instability and by the flow-field-weighted Trouton ratio ($\sum \mathrm{Tr} \ p(\mathrm{Tr}) \ \delta \mathrm{Tr}$). Because (Wi$_c$) can be determined reliably from flow-field fluctuations, we attribute the model variability primarily to uncertainty in the weighted apparent Trouton ratio. 

Consistent with prior measurements by ~\citeauthor{browne_elastic_2021}~\cite{browne_elastic_2021}, we find that the unstable viscous dissipation power balance (black curves) captures the macroscopic flow resistance near the onset of flow thickening but progressively underpredicts flow thickening at higher Wi---motivating the inclusion of an extensional-resistance term, as in the ordered geometries. Notably, the extent of underprediction varies across the three media (e.g., the unstable-dissipation balance tracks the data more closely near onset for medium 3, more similar to the results of ~\citeauthor{browne_elastic_2021}~\cite{browne_elastic_2021}, than for medium 1, which differs more substantially). Two additional features of our data point to medium-to-medium heterogeneity as the origin of this variability: (i) the large-Wi plateau values of $\eta_{\mathrm{app}}/\eta_I$ differ across replicates by more than experimental noise alone, and (ii) the Wi at which flow-thickening onsets shifts modestly between media prepared with the same polymer solution. We attribute this variability to the ``porous individualism'' of disordered packings~\cite{browne_elastic_2021}: each finite-volume realization samples a different local distribution of stagnation points, Hencky strains, and chaotic-flow regions. By contrast, ordered geometries yield reproducible flow behavior across unit-cell locations sufficiently far from the inlet and outlet, because of the periodic unit-cell geometry. We therefore expect that a statistically robust evaluation of the power-balance models in disordered media will require larger pore-scale samples ($n \gg 15$ pores) than used here.

The fitted $\zeta$ values are similar across the 3 disordered packings ($\zeta = 1.28-1.57$), corresponding to a characteristic extensional length scale $\sim 0.6-0.8 \ D_p$. This length scale falls between the pore throat diameter ($0.16 D_p$~\cite{al-raoush_extraction_2005,thompson_application_2008}) and grain diameter ($D_p$), which correspond to extensional flow generated in pore throat constrictions and stagnation points arising at grain surfaces, respectively.

The bottom row in Figs.~\ref{SIfig:disordered}(b-d) presents the resistance-per-stagnation point power balance model [Eq.~\eqref{SIEq:OscerPB}] for the 3 disordered packings. Here, we simply use $n_{\mathrm{SP}}$ as a fitting parameter to estimate extensional effects as it is challenging to (1) define a representative unit cell volume in the disordered packings and (2) accurately identify the locations of stagnation points within the 3D pore space. The models appear to suitably capture the correct order of magnitude of the flow resistance; however, the development of this resistance model was specifically founded on the premise of primarily \textit{stagnation point-dominated extensional flow} in the ordered geometries, which is in general not applicable to disordered packings with highly varying flow topology~\cite{nguyen_rheology_1999,de_viscoelastic_2017,de_viscoelastic_2017-1}. The fitted values of $n_{\mathrm{SP}}$ (number of stagnation points per characteristic unit cell) also vary greatly across different experiments and have seemingly large values compared to the ordered porous medium geometries. 
\begin{table}
\small
\centering
\caption{\label{SItab:crushed} Crushed packings: medium parameters.}
\begin{tabular}{ccccc}
\hline \hline
Medium&$\phi$&$k$ ($\upmu$m$^2$)&$L$ (cm)&$A$ (mm$^2$)\\
\hline
1 & 0.297 & 365$\pm$4 & 2.91 &  9\\
2 & 0.321 & 218$\pm$2 & 3.11&  9\\
3 & 0.292 & 233$\pm$2 & 2.94 &  9\\
\hline \hline
\end{tabular}
\end{table}

\subsection{Crushed glass packings: power balances}
\begin{figure}
\centering
\includegraphics[scale=1.3]{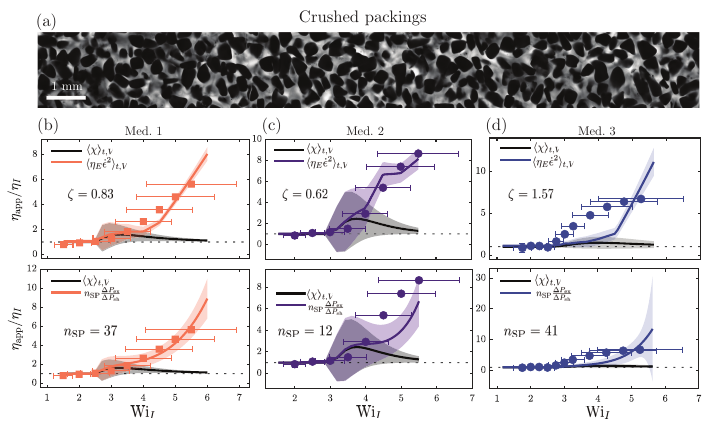}
\caption{\label{SIfig:crushed}Power balance models in crushed glass packings (average grain size: 300-400 $\upmu$m) exhibit greater variability in accuracy. (a) Example confocal micrograph of pore space in the crushed glass packings. Black regions correspond to the solid grain matrix. (b-d) Power balance models presented in order of decreasing accuracy for experiments in 3 different porous media. Top row: local apparent extensional viscosity power balance with fitted $\zeta$ values. Bottom row: resistance-per-stagnation point model with fitted $n_\mathrm{SP}$, using $n_\mathrm{SP}$ as a fitting parameter to modulate the extensional contribution to the flow resistance. Shaded regions and error bars are error propagation from fluid rheology and flow measurements.}
\end{figure}

We next examine the polymer solution flow resistance in packings of crushed glass beads, which produce pore space morphologies more representative of geological porous media encountered in subsurface flow applications. Briefly, we manually crush 1 mm borosilicate beads, sieve the particulates to achieve a target size range of 300-400 $\upmu$m, and lightly sinter packings of the glass fragments in a 3$\times$3 rectangular quartz capillary. Figure~\ref{SIfig:crushed}(a) shows a sample confocal micrograph of a $z$-slice within one of the crushed glass packings—note the enhanced pore space heterogeneity compared to the disordered bead packings [Fig.~\ref{SIfig:disordered}(a)]. Table~\ref{SItab:crushed} summarizes the relevant medium parameters for three different experimental replicates.
Figures~\ref{SIfig:crushed}(b-d) present the resulting extensional-modified power balance models for the crushed glass packings, where the top row shows the local apparent extensional viscosity model [Eq.~\eqref{eq:ExPB}] and the bottom row shows the resistance-per-stagnation point model [Eq.~\eqref{SIEq:OscerPB}], using $n_{\mathrm{SP}}$ as a fitting parameter. Experiments in panels (b-d) for different media are presented in order of decreasing accuracy. For media 1 and 2, the local apparent extensional viscosity model recapitulates the flow resistance in the crushed packings. The kink in the model curve for medium 2 results from the weighted Trouton ratio calculation, where the data for interpolation are not entirely smooth due to experimental variability. In contrast to media 1 and 2, the flow resistance of medium 3 is not captured well by the local extensional viscosity model, as there seems to be a mismatch in the onset of the extensional resistance term predicted by the model (see discussion in \S\ref{SIsec:disPB}). The values of the fitting parameter $\zeta$ in the crushed packings are more varied compared to the disordered bead packings, which may reflect the enhanced heterogeneity of the pore space and subsequently a broadening in the distribution of length scales for extensional flow.

As discussed in \S\ref{SIsec:disPB}, applying the resistance-per-stagnation point model to disordered packing geometries is no longer appropriate as the flow topology in the crushed packings likely consists of a complex mix of shear and extensional flow—even from pore-to-pore at the same imposed flow rate. We present the resistance-per-stagnation point models here for thoroughness in the bottom row of Figs.~\ref{SIfig:crushed}(b-d), using $n_{\mathrm{SP}}$ as a fitting parameter rather than as a known geometrical parameter. The resulting models again show large variability in accuracy, and the fitted values of $n_{\mathrm{SP}}$ also vary greatly across different experiments. 

In conclusion, our analyses in the disordered bead packings and crushed glass packings highlight the increased complexity in translating pore-scale flow behavior to the macroscopic flow resistance due to (1) increased spatial heterogeneity at the pore-scale requiring adequate flow sampling and (2) the mixed flow topology, where the flow is not dominated by extensional flow as we have found for the ordered porous medium geometries. Nevertheless, we expect that multiple mechanisms of flow resistance still contribute to the flow-thickening phenomenon in disordered porous media and expanding on the proposed resistance models to improve our prediction of these contributions will be a subject of future work.

\subsection{Counting stagnation points in disordered packings}
The fitted $n_{SP}$ values for the disordered packings are an order of magnitude or more larger than the counted $n_{SP}$ for the ordered geometries: $n_{SP}=280, 182, 108$ across the three disordered bead packings and $n_{SP}=37,12,41$ across the three crushed-glass packings. Three observations help interpret these values. First, in the absence of a well-defined unit cell, $n_{SP}$ no longer represents a counted geometric integer but an \emph{effective} parameter that combines the stagnation-point density per unit volume with the per-point contribution to the macroscopic resistance. Second, disordered packings contain many more stagnation points per unit volume than the ordered geometries studied here---grain-grain contacts occur throughout the medium, and the tortuous, irregular flow paths generate stagnation regions wherever streamlines bend sharply---making the order-of-magnitude increase in fitted $n_{SP}$ relative to the ordered case geometrically plausible. Third, the fitted $n_{SP}$ is smaller for the crushed-glass packings than for the bead packings, despite the sharper edges and more irregular geometry of the former; the contrast likely reflects differences in pore-scale geometry, permeability, and porosity between the two media, though a quantitative decomposition of these effects into a stagnation-point density and a per-point efficacy is beyond the scope of this work. The replicate-to-replicate variability within each medium type (a factor of approximately 3 in the bead packings, larger in the crushed-glass packings) reflects the ``porous individualism'' of disordered packings: each finite-volume realization samples a different local distribution of stagnation points and Hencky strains. Equation~2 from the main text thus applies in disordered media only with $n_{SP}$ as a fit parameter and with appropriate caution, in contrast to its parameter-free transferability across ordered geometries. Flows in disordered porous media are a complex mixture of shear and extension~\cite{de_viscoelastic_2017-1}, which violates the assumptions underlying Eq.~(2) of the main text.

\section{Resistance from shear-generated normal stresses}
\subsection{Distinguishing between shear and extensionally-generated polymeric normal stresses} 
\begin{figure}
\centering
\includegraphics[scale=1.3]{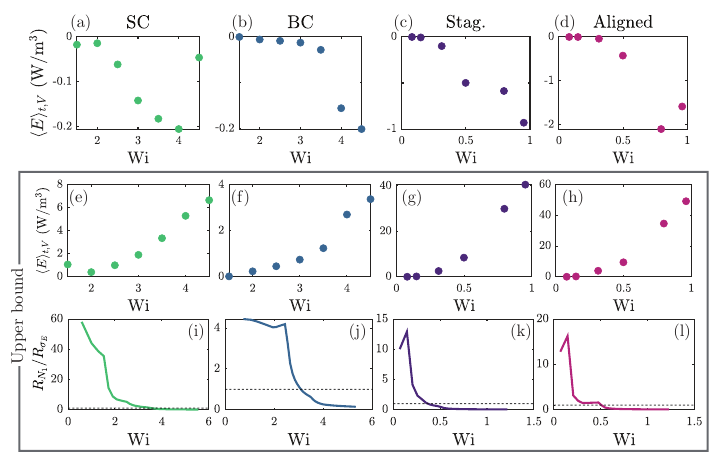}
\caption{\label{SIfig:N1resistance} Estimating energy dissipation by shear-generated polymeric normal stresses. (a-d) Energy dissipated, $\langle E \rangle_{t,V} = \langle N_1 (\dot{\gamma}) \frac{\partial u }{ \partial x} \rangle_{t,V}$, by the first normal stress difference $N_1$. Negative values indicate net energy dissipation, while positive values indicate net energy storage. The magnitude of dissipation is much smaller compared to the dissipation by the elastic instability, suggesting that dissipation by $N_1$ stresses is likely to be negligible compared to dissipation from the elastic instability and extensional stresses, $\sigma_E$. $N_1$ is evaluated using the instantaneous shear rate field and steady-state power law model from bulk rheology. (e-h) Maximum upper bound estimate of shear-induced normal stress dissipation, obtained by taking the absolute value of $\frac{\partial u}{\partial x}$. (i-l) Upper bound estimate of the ratio of the flow resistance contributed by the shear-induced normal stresses $R_{N_1}$ and extension-generated tensile stresses $R_{\sigma_E}$, evaluated using the upper bound values in (e-h). The extension-generated contribution to the flow resistance rapidly begins to dominate over the shear contribution with increasing $\mathrm{Wi}$ in all ordered porous medium geometries. Note that the true resistance ratio should fall below these curves.}
\end{figure}

In Eq.~\eqref{eq:extdecomp}, the term $(\tau_{xx}-\tau_{yy}) \frac{\partial u}{\partial x}$ contains the tensile stress difference $\tau_{xx}-\tau_{yy}$. In our derivation, we substitute this stress difference with the apparent extensional viscosity arising from \textit{extensional} flow. However, these tensile stresses may also arise from \textit{shearing} flow, described by the first normal stress difference, $N_1$. This distinction was delineated by~\citeauthor{james_slow_2012}~\cite{james_slow_2012,james_n1_2016}. Accordingly, we can introduce resistance contributions from both $\eta_E = (\tau_{xx}-\tau_{yy})_{\mathrm{ext}}/\dot{\epsilon}$ and $N_1 = (\tau_{xx}-\tau_{yy})_{\mathrm{shear}}$ by separating out the stresses generated by shear and extensional flow. The revised resistance model follows as,
\begin{equation}
\label{eq:PBwN1}
    \frac{\eta_{\mathrm{app}}}{\eta_I} = \underbrace{1}_{\mathrm{mean \ shear}} + \underbrace{\frac{k \langle \chi \rangle_{t,V}}{(Q/A)^2\eta_I}}_{\mathrm{unstable \ visc. \ diss.}} + \underbrace{\frac{k \langle \eta_E \ \dot\epsilon^2 \rangle_{t,V}}{(Q/A)^2\eta_I}}_{\mathrm{tensile \ stress \ from \ ext. \ flow}} + \underbrace{\frac{k \langle N_1 \ \partial u/\partial x \rangle_{t,V}}{(Q/A)^2\eta_I}}_{\mathrm{tensile \ stress \ from \ shear \ flow}}.
\end{equation}
A first approximation to evaluating the normal stress difference term in Eq.~\eqref{eq:PBwN1} is simply to evaluate the steady-state $N_1$ power law model obtained in bulk rheology using the instantaneous local pore-scale shear rate fields. Additionally, the values of $\partial u/\partial x$ can take on both negative and positive signs: negative indicates compression and positive indicates extension. For positive $N_1$, the sign of the product $N_1\partial u/\partial x$ then determines whether the shear-generated tensile stress dissipates or absorbs work from the flow. Figs.~\ref{SIfig:N1resistance}(a-d) show the average $E = N_1 \frac{\partial u}{\partial x}$ values as a function of $\mathrm{Wi}$ measured from the flow fields in the different ordered geometries, where negative values indicate net energy dissipation, while positive values indicate net energy accumulation due to storage of elastic stresses in the polymer chains. For all medium geometries, the net dissipation increases with $\mathrm{Wi}$; however, the magnitude of this dissipation is quite small.

Here, we estimate the maximum possible upper bound to the dissipation from $N_1$ by taking the absolute value of the measured gradient $\frac{\partial u}{\partial x}$. Figs.~\ref{SIfig:N1resistance}(e-h) present the resulting  $E = N_1 |\frac{\partial u}{\partial x}|$ curves as a function of $\mathrm{Wi}$. We use these upper bound estimates to evaluate the full resistance model in Eq.~\eqref{eq:PBwN1}. Figs.~\ref{SIfig:N1resistance}(i-l) show the ratio of the resistances contributed by the normal stress difference (fourth term on the right in Eq.~\eqref{eq:PBwN1}) and extensional viscosity (third term on the right in Eq.~\eqref{eq:PBwN1}). At low $\mathrm{Wi}$, the $N_1$ contribution dominates. As $\mathrm{Wi}$ increases, this ratio rapidly becomes dominated by the extensional viscosity resistance across all of the ordered geometries. Note that the data shown correspond to an upper bound estimate of $N_1$; hence, the true contribution from $N_1$ is likely lower. We thus conclude that, in the ordered porous medium geometries, the tensile stresses arising from \textit{extensional} flow govern the flow-thickening behavior compared to shear-generated tensile stresses.

\subsection{Assessment of transient versus steady estimates of $N_1$} 
\begin{figure}
\centering
\includegraphics[scale=0.9]{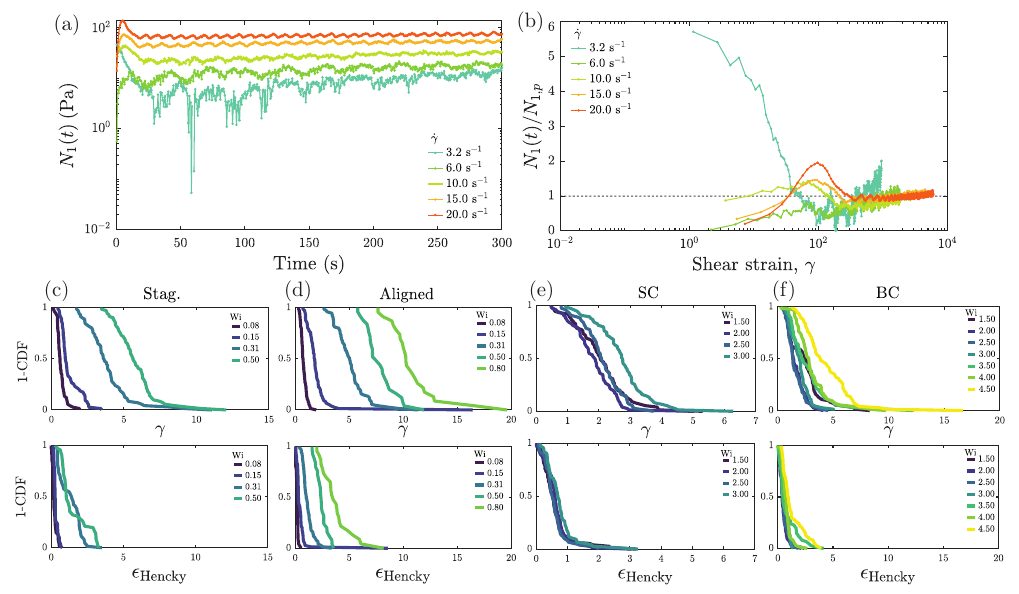}
\caption{\label{SIfig:shearstrain} Accounting for the transient development of polymeric tensile stresses from shearing flow. (a-b) Transient step-shear rheology measurements of $N_1$ at different imposed shear rates as a function of (a) time and (b) shear strain with rescaling by the steady-state plateau value $N_{1,p}$. The collapse of curves in (b) suggests that steady-state values of $N_1$ are reached after 300 strain units for all shear rates. Note, there is an upper limit of 25 $\mathrm{s^{-1}}$ to these measurements due to the onset of an elastic instability at high shear rates in the cone-plate geometry. (c-f) Inverse cumulative distributions of the accumulated shear strain and extensional Hencky strain in the ordered porous medium geometries. In all cases, using the transient rheology in (a-b) as a mapping from shear strain to transient $N_1$, the corresponding $N_1$ values estimated for the pore-scale flow fields are all below the steady-state plateau values.}
\end{figure}
The first normal stress difference is fundamentally a transient property that depends on the history of shearing. Thus, evaluating the steady-state $N_1$ power law model from bulk rheology using the local instantaneous pore-scale shear rate fields may be inaccurate. To capture the transient nature of $N_1$ in evaluation of the resistance model, we first perform transient bulk shear rheology measurements by imposing a step shear strain and recording the stress response over time at various shear rates (Fig.~\ref{SIfig:shearstrain}(a)). The range of resolvable shear rates is limited on the lower end by the instrument resolution and on the higher end by the onset of an elastic instability in the cone-plate geometry. Rescaling the $N_1$ data by the steady-state plateau value $N_{1,p}$ and plotting against the accumulated shear strain $\gamma = \dot{\gamma}t$, we observe a collapse in the transient $N_1$ response in Fig.~\ref{SIfig:shearstrain}(b). This collapse can be used as a mapping from measured shear strain to transient $N_1$, where the shear strains can be measured from the spatiotemporal flow fields as $\gamma = \int_{t=0}^{t=\lambda}\dot\gamma \ \mathrm{d}t$.

Figs.~\ref{SIfig:shearstrain}(c-f) present distributions of the measured shear strains, $\gamma$, and extensional Hencky strains, $\epsilon_{\mathrm{Hencky}}$, in each of the four ordered porous medium geometries. Generally, both types of strains increase with $\mathrm{Wi}$. Importantly, comparing the maximum values of $\gamma$ with the transient rheology in Fig.~\ref{SIfig:shearstrain}(b), we find that the accumulated shear strains in the pore-scale flow fields ($\gamma \approx 10-20$) are smaller than the nominal shear strain ($\gamma \approx 300$) for $N_1$ to reach a steady state. This result suggests that the calculations in Fig.~\ref{SIfig:N1resistance} are indeed an upper bound estimate for the resistance contributed by shear-generated normal stresses, neglecting the stress overshoot behavior at around 100 strain units. Further, while the accumulated strain distributions indicate that accumulated shear strains are larger in magnitude than the extensional strains, the full power balance calculation presented in Figs.~\ref{SIfig:N1resistance}(i-l) indicate that energy dissipation by the associated transient extensional stresses is larger than the shear stresses, particularly with increasing $\mathrm{Wi}$. We thus conclude that—in the ordered porous medium geometries—the polymeric tensile stresses $\tau_{\mathrm{xx}}-\tau_{\mathrm{yy}}$ primarily arise from extensional flow rather than shearing flow. We expect that this may not be the case in other types of porous medium geometries, as altogether our results indicate that the flow topology is intricately linked to the pore space geometry and thus the macroscopic flow resistance.


\begin{thebibliography}{144}%
\makeatletter
\providecommand \@ifxundefined [1]{%
 \@ifx{#1\undefined}
}%
\providecommand \@ifnum [1]{%
 \ifnum #1\expandafter \@firstoftwo
 \else \expandafter \@secondoftwo
 \fi
}%
\providecommand \@ifx [1]{%
 \ifx #1\expandafter \@firstoftwo
 \else \expandafter \@secondoftwo
 \fi
}%
\providecommand \natexlab [1]{#1}%
\providecommand \enquote  [1]{``#1''}%
\providecommand \bibnamefont  [1]{#1}%
\providecommand \bibfnamefont [1]{#1}%
\providecommand \citenamefont [1]{#1}%
\providecommand \href@noop [0]{\@secondoftwo}%
\providecommand \href [0]{\begingroup \@sanitize@url \@href}%
\providecommand \@href[1]{\@@startlink{#1}\@@href}%
\providecommand \@@href[1]{\endgroup#1\@@endlink}%
\providecommand \@sanitize@url [0]{\catcode `\\12\catcode `\$12\catcode `\&12\catcode `\#12\catcode `\^12\catcode `\_12\catcode `\%12\relax}%
\providecommand \@@startlink[1]{}%
\providecommand \@@endlink[0]{}%
\providecommand \url  [0]{\begingroup\@sanitize@url \@url }%
\providecommand \@url [1]{\endgroup\@href {#1}{\urlprefix }}%
\providecommand \urlprefix  [0]{URL }%
\providecommand \Eprint [0]{\href }%
\providecommand \doibase [0]{https://doi.org/}%
\providecommand \selectlanguage [0]{\@gobble}%
\providecommand \bibinfo  [0]{\@secondoftwo}%
\providecommand \bibfield  [0]{\@secondoftwo}%
\providecommand \translation [1]{[#1]}%
\providecommand \BibitemOpen [0]{}%
\providecommand \bibitemStop [0]{}%
\providecommand \bibitemNoStop [0]{.\EOS\space}%
\providecommand \EOS [0]{\spacefactor3000\relax}%
\providecommand \BibitemShut  [1]{\csname bibitem#1\endcsname}%
\let\auto@bib@innerbib\@empty
\bibitem [{\citenamefont {Petrie}\ and\ \citenamefont {Denn}(1976)}]{petrie_instabilities_1976}%
  \BibitemOpen
  \bibfield  {author} {\bibinfo {author} {\bibfnamefont {C.~J.~S.}\ \bibnamefont {Petrie}}\ and\ \bibinfo {author} {\bibfnamefont {M.~M.}\ \bibnamefont {Denn}},\ }\bibfield  {title} {\bibinfo {title} {Instabilities in polymer processing},\ }\href@noop {} {\bibfield  {journal} {\bibinfo  {journal} {AIChE J}\ }\textbf {\bibinfo {volume} {22}},\ \bibinfo {pages} {209} (\bibinfo {year} {1976})}\BibitemShut {NoStop}%
\bibitem [{\citenamefont {Denn}(2008)}]{denn_polymer_2008}%
  \BibitemOpen
  \bibfield  {author} {\bibinfo {author} {\bibfnamefont {M.~M.}\ \bibnamefont {Denn}},\ }\href@noop {} {\emph {\bibinfo {title} {Polymer {Melt} {Processing}: {Foundations} in {Fluid} {Mechanics} and {Heat} {Transfer}}}}\ (\bibinfo  {publisher} {Cambridge University Press},\ \bibinfo {year} {2008})\BibitemShut {NoStop}%
\bibitem [{\citenamefont {Turner}\ \emph {et~al.}(2014)\citenamefont {Turner}, \citenamefont {Strong},\ and\ \citenamefont {Gold}}]{turner_review_2014}%
  \BibitemOpen
  \bibfield  {author} {\bibinfo {author} {\bibfnamefont {B.~N.}\ \bibnamefont {Turner}}, \bibinfo {author} {\bibfnamefont {R.}~\bibnamefont {Strong}},\ and\ \bibinfo {author} {\bibfnamefont {S.~A.}\ \bibnamefont {Gold}},\ }\bibfield  {title} {\bibinfo {title} {A review of melt extrusion additive manufacturing processes: {I}. {Process} design and modeling},\ }\href@noop {} {\bibfield  {journal} {\bibinfo  {journal} {Rapid Protyp. J.}\ }\textbf {\bibinfo {volume} {20}},\ \bibinfo {pages} {192} (\bibinfo {year} {2014})}\BibitemShut {NoStop}%
\bibitem [{\citenamefont {Elbadawi}(2018)}]{elbadawi_polymeric_2018}%
  \BibitemOpen
  \bibfield  {author} {\bibinfo {author} {\bibfnamefont {M.}~\bibnamefont {Elbadawi}},\ }\bibfield  {title} {\bibinfo {title} {Polymeric {Additive} {Manufacturing}: {The} {Necessity} and {Utility} of {Rheology}},\ }in\ \href@noop {} {\emph {\bibinfo {booktitle} {Polymer {Rheology}}}}\ (\bibinfo  {publisher} {InTechOpen},\ \bibinfo {year} {2018})\BibitemShut {NoStop}%
\bibitem [{\citenamefont {Agassant}\ \emph {et~al.}(2006)\citenamefont {Agassant}, \citenamefont {Arda}, \citenamefont {Combeaud}, \citenamefont {Merten}, \citenamefont {Münstedt}, \citenamefont {Mackley}, \citenamefont {Robert},\ and\ \citenamefont {Vergnes}}]{agassant_polymer_2006}%
  \BibitemOpen
  \bibfield  {author} {\bibinfo {author} {\bibfnamefont {J.-F.}\ \bibnamefont {Agassant}}, \bibinfo {author} {\bibfnamefont {D.~R.}\ \bibnamefont {Arda}}, \bibinfo {author} {\bibfnamefont {C.}~\bibnamefont {Combeaud}}, \bibinfo {author} {\bibfnamefont {A.}~\bibnamefont {Merten}}, \bibinfo {author} {\bibfnamefont {H.}~\bibnamefont {Münstedt}}, \bibinfo {author} {\bibfnamefont {M.~R.}\ \bibnamefont {Mackley}}, \bibinfo {author} {\bibfnamefont {L.}~\bibnamefont {Robert}},\ and\ \bibinfo {author} {\bibfnamefont {B.}~\bibnamefont {Vergnes}},\ }\bibfield  {title} {\bibinfo {title} {Polymer {Processing} {Extrusion} {Instabilities} and {Methods} for their {Elimination} or {Minimisation}},\ }\href@noop {} {\bibfield  {journal} {\bibinfo  {journal} {Int. Polym. Proc.}\ }\textbf {\bibinfo {volume} {21}},\ \bibinfo {pages} {239} (\bibinfo {year} {2006})}\BibitemShut {NoStop}%
\bibitem [{\citenamefont {Das}\ \emph {et~al.}(2021)\citenamefont {Das}, \citenamefont {Gilmer}, \citenamefont {Biria},\ and\ \citenamefont {Bortner}}]{das_importance_2021}%
  \BibitemOpen
  \bibfield  {author} {\bibinfo {author} {\bibfnamefont {A.}~\bibnamefont {Das}}, \bibinfo {author} {\bibfnamefont {E.~L.}\ \bibnamefont {Gilmer}}, \bibinfo {author} {\bibfnamefont {S.}~\bibnamefont {Biria}},\ and\ \bibinfo {author} {\bibfnamefont {M.~J.}\ \bibnamefont {Bortner}},\ }\bibfield  {title} {\bibinfo {title} {Importance of {Polymer} {Rheology} on {Material} {Extrusion} {Additive} {Manufacturing}: {Correlating} {Process} {Physics} to {Print} {Properties}},\ }\href@noop {} {\bibfield  {journal} {\bibinfo  {journal} {ACS Appl. Polym. Mater.}\ }\textbf {\bibinfo {volume} {3}},\ \bibinfo {pages} {1218} (\bibinfo {year} {2021})}\BibitemShut {NoStop}%
\bibitem [{\citenamefont {Kozicki}\ and\ \citenamefont {Slegr}(1994)}]{kozicki_filtration_1994}%
  \BibitemOpen
  \bibfield  {author} {\bibinfo {author} {\bibfnamefont {W.}~\bibnamefont {Kozicki}}\ and\ \bibinfo {author} {\bibfnamefont {H.}~\bibnamefont {Slegr}},\ }\bibfield  {title} {\bibinfo {title} {Filtration in viscoelastic continua},\ }\href@noop {} {\bibfield  {journal} {\bibinfo  {journal} {J. Non-Newton. Fluid Mech.}\ }\textbf {\bibinfo {volume} {53}},\ \bibinfo {pages} {129} (\bibinfo {year} {1994})}\BibitemShut {NoStop}%
\bibitem [{\citenamefont {Luo}\ and\ \citenamefont {Teraoka}(1996)}]{luo_high_1996}%
  \BibitemOpen
  \bibfield  {author} {\bibinfo {author} {\bibfnamefont {M.}~\bibnamefont {Luo}}\ and\ \bibinfo {author} {\bibfnamefont {I.}~\bibnamefont {Teraoka}},\ }\bibfield  {title} {\bibinfo {title} {High {Osmotic} {Pressure} {Chromatography} for {Large}-{Scale} {Fractionation} of {Polymers}},\ }\href@noop {} {\bibfield  {journal} {\bibinfo  {journal} {Macromolecules}\ }\textbf {\bibinfo {volume} {29}},\ \bibinfo {pages} {4226} (\bibinfo {year} {1996})}\BibitemShut {NoStop}%
\bibitem [{\citenamefont {Bourgeat}\ \emph {et~al.}(2003)\citenamefont {Bourgeat}, \citenamefont {Gipouloux},\ and\ \citenamefont {Marusic-Paloka}}]{bourgeat_filtration_2003}%
  \BibitemOpen
  \bibfield  {author} {\bibinfo {author} {\bibfnamefont {A.}~\bibnamefont {Bourgeat}}, \bibinfo {author} {\bibfnamefont {O.}~\bibnamefont {Gipouloux}},\ and\ \bibinfo {author} {\bibfnamefont {E.}~\bibnamefont {Marusic-Paloka}},\ }\bibfield  {title} {\bibinfo {title} {Filtration {Law} for {Polymer} {Flow} {Through} {Porous} {Media}},\ }\href@noop {} {\bibfield  {journal} {\bibinfo  {journal} {Multiscale Model. Simul.}\ }\textbf {\bibinfo {volume} {1}},\ \bibinfo {pages} {432} (\bibinfo {year} {2003})}\BibitemShut {NoStop}%
\bibitem [{\citenamefont {Gritti}\ \emph {et~al.}(2024)\citenamefont {Gritti}, \citenamefont {Chen},\ and\ \citenamefont {Datta}}]{gritti_harnessing_2024}%
  \BibitemOpen
  \bibfield  {author} {\bibinfo {author} {\bibfnamefont {F.}~\bibnamefont {Gritti}}, \bibinfo {author} {\bibfnamefont {E.~Y.}\ \bibnamefont {Chen}},\ and\ \bibinfo {author} {\bibfnamefont {S.~S.}\ \bibnamefont {Datta}},\ }\bibfield  {title} {\bibinfo {title} {Harnessing an elastic flow instability to improve the kinetic performance of chromatographic columns},\ }\href@noop {} {\bibfield  {journal} {\bibinfo  {journal} {J. Chromatogr. A}\ }\textbf {\bibinfo {volume} {1735}},\ \bibinfo {pages} {465326} (\bibinfo {year} {2024})}\BibitemShut {NoStop}%
\bibitem [{\citenamefont {Durst}\ \emph {et~al.}(1981)\citenamefont {Durst}, \citenamefont {Haas},\ and\ \citenamefont {Kaczmar}}]{durst_flows_1981}%
  \BibitemOpen
  \bibfield  {author} {\bibinfo {author} {\bibfnamefont {F.}~\bibnamefont {Durst}}, \bibinfo {author} {\bibfnamefont {R.}~\bibnamefont {Haas}},\ and\ \bibinfo {author} {\bibfnamefont {B.~U.}\ \bibnamefont {Kaczmar}},\ }\bibfield  {title} {\bibinfo {title} {Flows of dilute hydrolyzed polyacrylamide solutions in porous media under various solvent conditions},\ }\href@noop {} {\bibfield  {journal} {\bibinfo  {journal} {J. Appl. Polym. Sci.}\ }\textbf {\bibinfo {volume} {26}},\ \bibinfo {pages} {3125} (\bibinfo {year} {1981})}\BibitemShut {NoStop}%
\bibitem [{\citenamefont {Okandan}(1984)}]{okandan_heavy_1984}%
  \BibitemOpen
  \bibinfo {editor} {\bibfnamefont {E.}~\bibnamefont {Okandan}},\ ed.,\ \href@noop {} {\emph {\bibinfo {title} {Heavy {Crude} {Oil} {Recovery}}}}\ (\bibinfo  {publisher} {Springer Netherlands},\ \bibinfo {address} {Dordrecht},\ \bibinfo {year} {1984})\BibitemShut {NoStop}%
\bibitem [{\citenamefont {Seright}\ \emph {et~al.}(2010)\citenamefont {Seright}, \citenamefont {Fan}, \citenamefont {Wavrik},\ and\ \citenamefont {de~Carvalho~Balaban}}]{seright_new_2010}%
  \BibitemOpen
  \bibfield  {author} {\bibinfo {author} {\bibfnamefont {R.~S.}\ \bibnamefont {Seright}}, \bibinfo {author} {\bibfnamefont {T.}~\bibnamefont {Fan}}, \bibinfo {author} {\bibfnamefont {K.}~\bibnamefont {Wavrik}},\ and\ \bibinfo {author} {\bibfnamefont {R.}~\bibnamefont {de~Carvalho~Balaban}},\ }\bibfield  {title} {\bibinfo {title} {New {Insights} into {Polymer} {Rheology} in {Porous} {Media}},\ }\href@noop {} {\bibfield  {journal} {\bibinfo  {journal} {SPE J.}\ }\textbf {\bibinfo {volume} {16}},\ \bibinfo {pages} {35} (\bibinfo {year} {2010})}\BibitemShut {NoStop}%
\bibitem [{\citenamefont {Sorbie}(2013)}]{sorbie_polymer-improved_2013}%
  \BibitemOpen
  \bibfield  {author} {\bibinfo {author} {\bibfnamefont {K.~S.}\ \bibnamefont {Sorbie}},\ }\href@noop {} {\emph {\bibinfo {title} {Polymer-{Improved} {Oil} {Recovery}}}},\ \bibinfo {edition} {1st}\ ed.\ (\bibinfo  {publisher} {Springer Dordrecht},\ \bibinfo {year} {2013})\BibitemShut {NoStop}%
\bibitem [{\citenamefont {Clarke}\ \emph {et~al.}(2016)\citenamefont {Clarke}, \citenamefont {Howe}, \citenamefont {Mitchell}, \citenamefont {Staniland},\ and\ \citenamefont {Hawkes}}]{clarke_how_2016}%
  \BibitemOpen
  \bibfield  {author} {\bibinfo {author} {\bibfnamefont {A.}~\bibnamefont {Clarke}}, \bibinfo {author} {\bibfnamefont {A.~M.}\ \bibnamefont {Howe}}, \bibinfo {author} {\bibfnamefont {J.}~\bibnamefont {Mitchell}}, \bibinfo {author} {\bibfnamefont {J.}~\bibnamefont {Staniland}},\ and\ \bibinfo {author} {\bibfnamefont {L.~A.}\ \bibnamefont {Hawkes}},\ }\bibfield  {title} {\bibinfo {title} {How {Viscoelastic}-{Polymer} {Flooding} {Enhances} {Displacement} {Efficiency}},\ }\href@noop {} {\bibfield  {journal} {\bibinfo  {journal} {SPE J.}\ }\textbf {\bibinfo {volume} {21}},\ \bibinfo {pages} {675} (\bibinfo {year} {2016})}\BibitemShut {NoStop}%
\bibitem [{\citenamefont {Pogaku}\ \emph {et~al.}(2018)\citenamefont {Pogaku}, \citenamefont {Mohd~Fuat}, \citenamefont {Sakar}, \citenamefont {Cha}, \citenamefont {Musa}, \citenamefont {Awang~Tajudin},\ and\ \citenamefont {Morris}}]{pogaku_polymer_2018}%
  \BibitemOpen
  \bibfield  {author} {\bibinfo {author} {\bibfnamefont {R.}~\bibnamefont {Pogaku}}, \bibinfo {author} {\bibfnamefont {N.~H.}\ \bibnamefont {Mohd~Fuat}}, \bibinfo {author} {\bibfnamefont {S.}~\bibnamefont {Sakar}}, \bibinfo {author} {\bibfnamefont {Z.~W.}\ \bibnamefont {Cha}}, \bibinfo {author} {\bibfnamefont {N.}~\bibnamefont {Musa}}, \bibinfo {author} {\bibfnamefont {D.~N.~A.}\ \bibnamefont {Awang~Tajudin}},\ and\ \bibinfo {author} {\bibfnamefont {L.~O.}\ \bibnamefont {Morris}},\ }\bibfield  {title} {\bibinfo {title} {Polymer flooding and its combinations with other chemical injection methods in enhanced oil recovery},\ }\href@noop {} {\bibfield  {journal} {\bibinfo  {journal} {Polym. Bull.}\ }\textbf {\bibinfo {volume} {75}},\ \bibinfo {pages} {1753} (\bibinfo {year} {2018})}\BibitemShut {NoStop}%
\bibitem [{\citenamefont {Mirzaie~Yegane}\ \emph {et~al.}(2022)\citenamefont {Mirzaie~Yegane}, \citenamefont {Boukany},\ and\ \citenamefont {Zitha}}]{mirzaie_yegane_fundamentals_2022}%
  \BibitemOpen
  \bibfield  {author} {\bibinfo {author} {\bibfnamefont {M.}~\bibnamefont {Mirzaie~Yegane}}, \bibinfo {author} {\bibfnamefont {P.~E.}\ \bibnamefont {Boukany}},\ and\ \bibinfo {author} {\bibfnamefont {P.}~\bibnamefont {Zitha}},\ }\bibfield  {title} {\bibinfo {title} {Fundamentals and {Recent} {Progress} in the {Flow} of {Water}-{Soluble} {Polymers} in {Porous} {Media} for {Enhanced} {Oil} {Recovery}},\ }\href@noop {} {\bibfield  {journal} {\bibinfo  {journal} {Energies}\ }\textbf {\bibinfo {volume} {15}},\ \bibinfo {pages} {8575} (\bibinfo {year} {2022})}\BibitemShut {NoStop}%
\bibitem [{\citenamefont {Di~Dato}\ \emph {et~al.}(2022)\citenamefont {Di~Dato}, \citenamefont {D’Angelo}, \citenamefont {Casasso},\ and\ \citenamefont {Zarlenga}}]{di_dato_impact_2022}%
  \BibitemOpen
  \bibfield  {author} {\bibinfo {author} {\bibfnamefont {M.}~\bibnamefont {Di~Dato}}, \bibinfo {author} {\bibfnamefont {C.}~\bibnamefont {D’Angelo}}, \bibinfo {author} {\bibfnamefont {A.}~\bibnamefont {Casasso}},\ and\ \bibinfo {author} {\bibfnamefont {A.}~\bibnamefont {Zarlenga}},\ }\bibfield  {title} {\bibinfo {title} {The impact of porous medium heterogeneity on the thermal feedback of open-loop shallow geothermal systems},\ }\href@noop {} {\bibfield  {journal} {\bibinfo  {journal} {J. Hydrol.}\ }\textbf {\bibinfo {volume} {604}},\ \bibinfo {pages} {127205} (\bibinfo {year} {2022})}\BibitemShut {NoStop}%
\bibitem [{\citenamefont {Roote}(1998)}]{roote_technology_1998}%
  \BibitemOpen
  \bibfield  {author} {\bibinfo {author} {\bibfnamefont {D.~S.}\ \bibnamefont {Roote}},\ }\href@noop {} {\emph {\bibinfo {title} {Technology {Status} {Report}: {In} {Situ} {Flushing}}}},\ \bibinfo {type} {Tech. Rep.}\ (\bibinfo  {institution} {Ground-Water Remediation Technologies Analysis Center},\ \bibinfo {year} {1998})\BibitemShut {NoStop}%
\bibitem [{\citenamefont {Smith}\ \emph {et~al.}(2008)\citenamefont {Smith}, \citenamefont {Silva}, \citenamefont {Munakata-Marr},\ and\ \citenamefont {McCray}}]{smith_compatibility_2008}%
  \BibitemOpen
  \bibfield  {author} {\bibinfo {author} {\bibfnamefont {M.~M.}\ \bibnamefont {Smith}}, \bibinfo {author} {\bibfnamefont {J.~A.~K.}\ \bibnamefont {Silva}}, \bibinfo {author} {\bibfnamefont {J.}~\bibnamefont {Munakata-Marr}},\ and\ \bibinfo {author} {\bibfnamefont {J.~E.}\ \bibnamefont {McCray}},\ }\bibfield  {title} {\bibinfo {title} {Compatibility of {Polymers} and {Chemical} {Oxidants} for {Enhanced} {Groundwater} {Remediation}},\ }\href@noop {} {\bibfield  {journal} {\bibinfo  {journal} {Environ. Sci. Technol.}\ }\textbf {\bibinfo {volume} {42}},\ \bibinfo {pages} {9296} (\bibinfo {year} {2008})}\BibitemShut {NoStop}%
\bibitem [{\citenamefont {Huo}\ \emph {et~al.}(2020)\citenamefont {Huo}, \citenamefont {Liu}, \citenamefont {Yang}, \citenamefont {Ahmad},\ and\ \citenamefont {Zhong}}]{huo_surfactant-enhanced_2020}%
  \BibitemOpen
  \bibfield  {author} {\bibinfo {author} {\bibfnamefont {L.}~\bibnamefont {Huo}}, \bibinfo {author} {\bibfnamefont {G.}~\bibnamefont {Liu}}, \bibinfo {author} {\bibfnamefont {X.}~\bibnamefont {Yang}}, \bibinfo {author} {\bibfnamefont {Z.}~\bibnamefont {Ahmad}},\ and\ \bibinfo {author} {\bibfnamefont {H.}~\bibnamefont {Zhong}},\ }\bibfield  {title} {\bibinfo {title} {Surfactant-enhanced aquifer remediation: {Mechanisms}, influences, limitations and the countermeasures},\ }\href@noop {} {\bibfield  {journal} {\bibinfo  {journal} {Chemosphere}\ }\textbf {\bibinfo {volume} {252}},\ \bibinfo {pages} {126620} (\bibinfo {year} {2020})}\BibitemShut {NoStop}%
\bibitem [{\citenamefont {Hartmann}\ \emph {et~al.}(2021)\citenamefont {Hartmann}, \citenamefont {Jasechko}, \citenamefont {Gleeson}, \citenamefont {Wada}, \citenamefont {Andreo}, \citenamefont {Barberá}, \citenamefont {Brielmann}, \citenamefont {Bouchaou}, \citenamefont {Charlier}, \citenamefont {Darling}, \citenamefont {Filippini}, \citenamefont {Garvelmann}, \citenamefont {Goldscheider}, \citenamefont {Kralik}, \citenamefont {Kunstmann}, \citenamefont {Ladouche}, \citenamefont {Lange}, \citenamefont {Lucianetti}, \citenamefont {Martín}, \citenamefont {Mudarra}, \citenamefont {Sánchez}, \citenamefont {Stumpp}, \citenamefont {Zagana},\ and\ \citenamefont {Wagener}}]{hartmann_risk_2021}%
  \BibitemOpen
  \bibfield  {author} {\bibinfo {author} {\bibfnamefont {A.}~\bibnamefont {Hartmann}}, \bibinfo {author} {\bibfnamefont {S.}~\bibnamefont {Jasechko}}, \bibinfo {author} {\bibfnamefont {T.}~\bibnamefont {Gleeson}}, \bibinfo {author} {\bibfnamefont {Y.}~\bibnamefont {Wada}}, \bibinfo {author} {\bibfnamefont {B.}~\bibnamefont {Andreo}}, \bibinfo {author} {\bibfnamefont {J.~A.}\ \bibnamefont {Barberá}}, \bibinfo {author} {\bibfnamefont {H.}~\bibnamefont {Brielmann}}, \bibinfo {author} {\bibfnamefont {L.}~\bibnamefont {Bouchaou}}, \bibinfo {author} {\bibfnamefont {J.-B.}\ \bibnamefont {Charlier}}, \bibinfo {author} {\bibfnamefont {W.~G.}\ \bibnamefont {Darling}}, \bibinfo {author} {\bibfnamefont {M.}~\bibnamefont {Filippini}}, \bibinfo {author} {\bibfnamefont {J.}~\bibnamefont {Garvelmann}}, \bibinfo {author} {\bibfnamefont {N.}~\bibnamefont {Goldscheider}}, \bibinfo {author} {\bibfnamefont {M.}~\bibnamefont {Kralik}}, \bibinfo {author} {\bibfnamefont {H.}~\bibnamefont {Kunstmann}}, \bibinfo {author}
  {\bibfnamefont {B.}~\bibnamefont {Ladouche}}, \bibinfo {author} {\bibfnamefont {J.}~\bibnamefont {Lange}}, \bibinfo {author} {\bibfnamefont {G.}~\bibnamefont {Lucianetti}}, \bibinfo {author} {\bibfnamefont {J.~F.}\ \bibnamefont {Martín}}, \bibinfo {author} {\bibfnamefont {M.}~\bibnamefont {Mudarra}}, \bibinfo {author} {\bibfnamefont {D.}~\bibnamefont {Sánchez}}, \bibinfo {author} {\bibfnamefont {C.}~\bibnamefont {Stumpp}}, \bibinfo {author} {\bibfnamefont {E.}~\bibnamefont {Zagana}},\ and\ \bibinfo {author} {\bibfnamefont {T.}~\bibnamefont {Wagener}},\ }\bibfield  {title} {\bibinfo {title} {Risk of groundwater contamination widely underestimated because of fast flow into aquifers},\ }\href@noop {} {\bibfield  {journal} {\bibinfo  {journal} {Proc. Natl. Acad. Sci. U.S.A.}\ }\textbf {\bibinfo {volume} {118}},\ \bibinfo {pages} {e2024492118} (\bibinfo {year} {2021})}\BibitemShut {NoStop}%
\bibitem [{\citenamefont {Whitaker}(1986)}]{whitaker_flow_1986}%
  \BibitemOpen
  \bibfield  {author} {\bibinfo {author} {\bibfnamefont {S.}~\bibnamefont {Whitaker}},\ }\bibfield  {title} {\bibinfo {title} {Flow in porous media {I}: {A} theoretical derivation of {Darcy}'s law},\ }\href@noop {} {\bibfield  {journal} {\bibinfo  {journal} {Transp Porous Med}\ }\textbf {\bibinfo {volume} {1}},\ \bibinfo {pages} {3} (\bibinfo {year} {1986})}\BibitemShut {NoStop}%
\bibitem [{\citenamefont {Pye}(1964)}]{pye_improved_1964}%
  \BibitemOpen
  \bibfield  {author} {\bibinfo {author} {\bibfnamefont {D.~J.}\ \bibnamefont {Pye}},\ }\bibfield  {title} {\bibinfo {title} {Improved {Secondary} {Recovery} by {Control} of {Water} {Mobility}},\ }\href@noop {} {\bibfield  {journal} {\bibinfo  {journal} {J. Pet. Technol.}\ }\textbf {\bibinfo {volume} {16}},\ \bibinfo {pages} {911} (\bibinfo {year} {1964})}\BibitemShut {NoStop}%
\bibitem [{\citenamefont {Sandiford}(1964)}]{sandiford_laboratory_1964}%
  \BibitemOpen
  \bibfield  {author} {\bibinfo {author} {\bibfnamefont {B.}~\bibnamefont {Sandiford}},\ }\bibfield  {title} {\bibinfo {title} {Laboratory and {Field} {Studies} of {Water} {Floods} {Using} {Polymer} {Solutions} to {Increase} {Oil} {Recoveries}},\ }\href@noop {} {\bibfield  {journal} {\bibinfo  {journal} {J. Pet. Technol.}\ }\textbf {\bibinfo {volume} {16}},\ \bibinfo {pages} {917} (\bibinfo {year} {1964})}\BibitemShut {NoStop}%
\bibitem [{\citenamefont {Jones}\ and\ \citenamefont {Maddock}(1966)}]{jones_flow_1966}%
  \BibitemOpen
  \bibfield  {author} {\bibinfo {author} {\bibfnamefont {W.~M.}\ \bibnamefont {Jones}}\ and\ \bibinfo {author} {\bibfnamefont {J.~L.}\ \bibnamefont {Maddock}},\ }\bibfield  {title} {\bibinfo {title} {Flow of viscoelastic liquids: Comparison of departures from laminar flow in porous beds and in tubes},\ }in\ \href@noop {} {\emph {\bibinfo {booktitle} {SPE Symposium on Mechanics of Rheologically Complex Fluids}}}\ (\bibinfo {year} {1966})\ pp.\ \bibinfo {pages} {SPE--1686--MS}\BibitemShut {NoStop}%
\bibitem [{\citenamefont {Dauben}\ and\ \citenamefont {Menzie}(1967)}]{dauben_flow_1967}%
  \BibitemOpen
  \bibfield  {author} {\bibinfo {author} {\bibfnamefont {D.~L.}\ \bibnamefont {Dauben}}\ and\ \bibinfo {author} {\bibfnamefont {D.~E.}\ \bibnamefont {Menzie}},\ }\bibfield  {title} {\bibinfo {title} {Flow of {Polymer} {Solutions} {Through} {Porous} {Media}},\ }\href@noop {} {\bibfield  {journal} {\bibinfo  {journal} {J. Pet. Technol.}\ }\textbf {\bibinfo {volume} {19}},\ \bibinfo {pages} {1065} (\bibinfo {year} {1967})}\BibitemShut {NoStop}%
\bibitem [{\citenamefont {Marshall}\ and\ \citenamefont {Metzner}(1967)}]{marshall_flow_1967}%
  \BibitemOpen
  \bibfield  {author} {\bibinfo {author} {\bibfnamefont {R.~J.}\ \bibnamefont {Marshall}}\ and\ \bibinfo {author} {\bibfnamefont {A.~B.}\ \bibnamefont {Metzner}},\ }\bibfield  {title} {\bibinfo {title} {Flow of {Viscoelastic} {Fluids} through {Porous} {Media}},\ }\href@noop {} {\bibfield  {journal} {\bibinfo  {journal} {Ind. Eng. Chem. Fund.}\ }\textbf {\bibinfo {volume} {6}} (\bibinfo {year} {1967})}\BibitemShut {NoStop}%
\bibitem [{\citenamefont {Savins}(1968)}]{savins_shear_1968}%
  \BibitemOpen
  \bibfield  {author} {\bibinfo {author} {\bibfnamefont {J.~G.}\ \bibnamefont {Savins}},\ }\bibfield  {title} {\bibinfo {title} {Shear thickening phenomena in poly(vinyl)alcohol-borate complexes},\ }\href@noop {} {\bibfield  {journal} {\bibinfo  {journal} {Rheol Acta}\ }\textbf {\bibinfo {volume} {7}},\ \bibinfo {pages} {87} (\bibinfo {year} {1968})}\BibitemShut {NoStop}%
\bibitem [{\citenamefont {Savins}(1969)}]{savins_non-newtonian_1969}%
  \BibitemOpen
  \bibfield  {author} {\bibinfo {author} {\bibfnamefont {J.~G.}\ \bibnamefont {Savins}},\ }\bibfield  {title} {\bibinfo {title} {Non-{Newtonian} {Flow} {Through} {Porous} {Media}},\ }\href@noop {} {\bibfield  {journal} {\bibinfo  {journal} {Ind. Eng. Chem.}\ }\textbf {\bibinfo {volume} {61}},\ \bibinfo {pages} {18} (\bibinfo {year} {1969})}\BibitemShut {NoStop}%
\bibitem [{\citenamefont {Gogarty}\ \emph {et~al.}(1972)\citenamefont {Gogarty}, \citenamefont {Levy},\ and\ \citenamefont {Fox}}]{gogarty_viscoelastic_1972}%
  \BibitemOpen
  \bibfield  {author} {\bibinfo {author} {\bibfnamefont {W.~B.}\ \bibnamefont {Gogarty}}, \bibinfo {author} {\bibfnamefont {G.~L.}\ \bibnamefont {Levy}},\ and\ \bibinfo {author} {\bibfnamefont {V.~G.}\ \bibnamefont {Fox}},\ }\bibfield  {title} {\bibinfo {title} {Viscoelastic effects in polymer flow through porous media},\ }in\ \href@noop {} {\emph {\bibinfo {booktitle} {Fall Meeting of the Society of Petroleum Engineers of AIME}}},\ \bibinfo {series and number} {SPE Annual Technical Conference and Exhibition}\ (\bibinfo {year} {1972})\ pp.\ \bibinfo {pages} {SPE--4025--MS}\BibitemShut {NoStop}%
\bibitem [{\citenamefont {Kulicke}\ and\ \citenamefont {Haas}(1984)}]{kulicke_flow_1984}%
  \BibitemOpen
  \bibfield  {author} {\bibinfo {author} {\bibfnamefont {W.~M.}\ \bibnamefont {Kulicke}}\ and\ \bibinfo {author} {\bibfnamefont {R.}~\bibnamefont {Haas}},\ }\bibfield  {title} {\bibinfo {title} {Flow behavior of dilute polyacrylamide solutions through porous media. 1. {Influence} of chain length, concentration, and thermodynamic quality of the solvent},\ }\href@noop {} {\bibfield  {journal} {\bibinfo  {journal} {Ind. Eng. Chem. Fund.}\ }\textbf {\bibinfo {volume} {23}},\ \bibinfo {pages} {308} (\bibinfo {year} {1984})}\BibitemShut {NoStop}%
\bibitem [{\citenamefont {Burcik}(1965)}]{burcik_note_1965}%
  \BibitemOpen
  \bibfield  {author} {\bibinfo {author} {\bibfnamefont {E.~J.}\ \bibnamefont {Burcik}},\ }\bibfield  {title} {\bibinfo {title} {A note on the flow behavior of polyacrylamide solutions in porous media},\ }\href@noop {} {\bibfield  {journal} {\bibinfo  {journal} {Prod. Mon.}\ }\textbf {\bibinfo {volume} {29}},\ \bibinfo {pages} {14} (\bibinfo {year} {1965})}\BibitemShut {NoStop}%
\bibitem [{\citenamefont {Sadowski}\ and\ \citenamefont {Bird}(1965)}]{sadowski_non-newtonian_1965-1}%
  \BibitemOpen
  \bibfield  {author} {\bibinfo {author} {\bibfnamefont {T.~J.}\ \bibnamefont {Sadowski}}\ and\ \bibinfo {author} {\bibfnamefont {R.~B.}\ \bibnamefont {Bird}},\ }\bibfield  {title} {\bibinfo {title} {Non-{Newtonian} {Flow} through {Porous} {Media}. {I}. {Theoretical}},\ }\href@noop {} {\bibfield  {journal} {\bibinfo  {journal} {T. Soc. Rheol.}\ }\textbf {\bibinfo {volume} {9}},\ \bibinfo {pages} {243} (\bibinfo {year} {1965})}\BibitemShut {NoStop}%
\bibitem [{\citenamefont {Mungan}\ \emph {et~al.}(1966)\citenamefont {Mungan}, \citenamefont {Smith},\ and\ \citenamefont {Thompson}}]{mungan_aspects_1966}%
  \BibitemOpen
  \bibfield  {author} {\bibinfo {author} {\bibfnamefont {N.}~\bibnamefont {Mungan}}, \bibinfo {author} {\bibfnamefont {F.}~\bibnamefont {Smith}},\ and\ \bibinfo {author} {\bibfnamefont {J.}~\bibnamefont {Thompson}},\ }\bibfield  {title} {\bibinfo {title} {Some {Aspects} of {Polymer} {Floods}},\ }\href@noop {} {\bibfield  {journal} {\bibinfo  {journal} {J. Pet. Technol.}\ }\textbf {\bibinfo {volume} {18}},\ \bibinfo {pages} {1143} (\bibinfo {year} {1966})}\BibitemShut {NoStop}%
\bibitem [{\citenamefont {Hirasaki}\ and\ \citenamefont {Pope}(1974)}]{hirasaki_analysis_1974}%
  \BibitemOpen
  \bibfield  {author} {\bibinfo {author} {\bibfnamefont {G.}~\bibnamefont {Hirasaki}}\ and\ \bibinfo {author} {\bibfnamefont {G.}~\bibnamefont {Pope}},\ }\bibfield  {title} {\bibinfo {title} {Analysis of {Factors} {Influencing} {Mobility} and {Adsorption} in the {Flow} of {Polymer} {Solution} {Through} {Porous} {Media}},\ }\href@noop {} {\bibfield  {journal} {\bibinfo  {journal} {SPE J.}\ }\textbf {\bibinfo {volume} {14}},\ \bibinfo {pages} {337} (\bibinfo {year} {1974})}\BibitemShut {NoStop}%
\bibitem [{\citenamefont {Dominguez}\ and\ \citenamefont {Willhite}(1977)}]{dominguez_retention_1977}%
  \BibitemOpen
  \bibfield  {author} {\bibinfo {author} {\bibfnamefont {J.}~\bibnamefont {Dominguez}}\ and\ \bibinfo {author} {\bibfnamefont {G.}~\bibnamefont {Willhite}},\ }\bibfield  {title} {\bibinfo {title} {Retention and {Flow} {Characteristics} of {Polymer} {Solutions} in {Porous} {Media}},\ }\href@noop {} {\bibfield  {journal} {\bibinfo  {journal} {SPE J.}\ }\textbf {\bibinfo {volume} {17}},\ \bibinfo {pages} {111} (\bibinfo {year} {1977})}\BibitemShut {NoStop}%
\bibitem [{\citenamefont {Jones}\ and\ \citenamefont {Ho}(1979)}]{jones_flow_1979}%
  \BibitemOpen
  \bibfield  {author} {\bibinfo {author} {\bibfnamefont {W.~M.}\ \bibnamefont {Jones}}\ and\ \bibinfo {author} {\bibfnamefont {S.~P.}\ \bibnamefont {Ho}},\ }\bibfield  {title} {\bibinfo {title} {The flow of dilute aqueous solutions of macromolecules in various geometries: {VII}. {Mechanisms} of resistance in porous media},\ }\href@noop {} {\bibfield  {journal} {\bibinfo  {journal} {J. Phys. D: Appl. Phys.}\ }\textbf {\bibinfo {volume} {12}},\ \bibinfo {pages} {383} (\bibinfo {year} {1979})}\BibitemShut {NoStop}%
\bibitem [{\citenamefont {Bagassi}\ \emph {et~al.}(1989)\citenamefont {Bagassi}, \citenamefont {Chauveteau}, \citenamefont {Lecourtier}, \citenamefont {Englert},\ and\ \citenamefont {Tirrell}}]{bagassi_behavior_1989}%
  \BibitemOpen
  \bibfield  {author} {\bibinfo {author} {\bibfnamefont {M.}~\bibnamefont {Bagassi}}, \bibinfo {author} {\bibfnamefont {G.}~\bibnamefont {Chauveteau}}, \bibinfo {author} {\bibfnamefont {J.}~\bibnamefont {Lecourtier}}, \bibinfo {author} {\bibfnamefont {J.}~\bibnamefont {Englert}},\ and\ \bibinfo {author} {\bibfnamefont {M.}~\bibnamefont {Tirrell}},\ }\bibfield  {title} {\bibinfo {title} {Behavior of adsorbed polymer layers in shear and elongational flows},\ }\href@noop {} {\bibfield  {journal} {\bibinfo  {journal} {Macromolecules}\ }\textbf {\bibinfo {volume} {22}},\ \bibinfo {pages} {262} (\bibinfo {year} {1989})}\BibitemShut {NoStop}%
\bibitem [{\citenamefont {Tam}\ \emph {et~al.}(1991)\citenamefont {Tam}, \citenamefont {Tiu},\ and\ \citenamefont {Fang}}]{tam_remarks_1991}%
  \BibitemOpen
  \bibfield  {author} {\bibinfo {author} {\bibfnamefont {K.~C.}\ \bibnamefont {Tam}}, \bibinfo {author} {\bibfnamefont {C.}~\bibnamefont {Tiu}},\ and\ \bibinfo {author} {\bibfnamefont {T.~N.}\ \bibnamefont {Fang}},\ }\bibfield  {title} {\bibinfo {title} {Remarks on the {Shear}-{Thickening} {Behavior} of {Dilute} {Polymer} {Solutions}},\ }\href@noop {} {\bibfield  {journal} {\bibinfo  {journal} {Polym-Plast Technol}\ }\textbf {\bibinfo {volume} {30}},\ \bibinfo {pages} {145} (\bibinfo {year} {1991})}\BibitemShut {NoStop}%
\bibitem [{\citenamefont {Zitha}\ \emph {et~al.}(2001)\citenamefont {Zitha}, \citenamefont {Chauveteau},\ and\ \citenamefont {Léger}}]{zitha_unsteady-state_2001}%
  \BibitemOpen
  \bibfield  {author} {\bibinfo {author} {\bibfnamefont {P.~L.}\ \bibnamefont {Zitha}}, \bibinfo {author} {\bibfnamefont {G.}~\bibnamefont {Chauveteau}},\ and\ \bibinfo {author} {\bibfnamefont {L.}~\bibnamefont {Léger}},\ }\bibfield  {title} {\bibinfo {title} {Unsteady-{State} {Flow} of {Flexible} {Polymers} in {Porous} {Media}},\ }\href@noop {} {\bibfield  {journal} {\bibinfo  {journal} {J. Colloid Interf. Sci.}\ }\textbf {\bibinfo {volume} {234}},\ \bibinfo {pages} {269} (\bibinfo {year} {2001})}\BibitemShut {NoStop}%
\bibitem [{\citenamefont {Sadowski}(1965)}]{sadowski_non-newtonian_1965}%
  \BibitemOpen
  \bibfield  {author} {\bibinfo {author} {\bibfnamefont {T.~J.}\ \bibnamefont {Sadowski}},\ }\bibfield  {title} {\bibinfo {title} {Non-{Newtonian} {Flow} through {Porous} {Media}. {II}. {Experimental}},\ }\href@noop {} {\bibfield  {journal} {\bibinfo  {journal} {T. Soc. Rheol.}\ }\textbf {\bibinfo {volume} {9}},\ \bibinfo {pages} {251} (\bibinfo {year} {1965})}\BibitemShut {NoStop}%
\bibitem [{\citenamefont {James}\ and\ \citenamefont {McLaren}(1975)}]{james_laminar_1975}%
  \BibitemOpen
  \bibfield  {author} {\bibinfo {author} {\bibfnamefont {D.~F.}\ \bibnamefont {James}}\ and\ \bibinfo {author} {\bibfnamefont {D.~R.}\ \bibnamefont {McLaren}},\ }\bibfield  {title} {\bibinfo {title} {The laminar flow of dilute polymer solutions through porous media},\ }\href@noop {} {\bibfield  {journal} {\bibinfo  {journal} {J. Fluid Mech.}\ }\textbf {\bibinfo {volume} {70}},\ \bibinfo {pages} {733} (\bibinfo {year} {1975})}\BibitemShut {NoStop}%
\bibitem [{\citenamefont {Moan}\ \emph {et~al.}(1979)\citenamefont {Moan}, \citenamefont {Chauveteau},\ and\ \citenamefont {Ghoniem}}]{moan_entrance_1979}%
  \BibitemOpen
  \bibfield  {author} {\bibinfo {author} {\bibfnamefont {M.}~\bibnamefont {Moan}}, \bibinfo {author} {\bibfnamefont {G.}~\bibnamefont {Chauveteau}},\ and\ \bibinfo {author} {\bibfnamefont {S.}~\bibnamefont {Ghoniem}},\ }\bibfield  {title} {\bibinfo {title} {Entrance effect in capillary flow of dilute and semi-dilute polymer solutions},\ }\href@noop {} {\bibfield  {journal} {\bibinfo  {journal} {J. Non-Newton. Fluid Mech.}\ } (\bibinfo {year} {1979})}\BibitemShut {NoStop}%
\bibitem [{\citenamefont {Chauveteau}\ and\ \citenamefont {Moan}(1981)}]{chauveteau_onset_1981}%
  \BibitemOpen
  \bibfield  {author} {\bibinfo {author} {\bibfnamefont {G.}~\bibnamefont {Chauveteau}}\ and\ \bibinfo {author} {\bibfnamefont {M.}~\bibnamefont {Moan}},\ }\bibfield  {title} {\bibinfo {title} {The onset of dilatant behaviour in non-inertial flow of dilute polymer solutions through channels with varying cross-sections},\ }\href@noop {} {\bibfield  {journal} {\bibinfo  {journal} {J. Phys. Lett.-Paris}\ }\textbf {\bibinfo {volume} {42}},\ \bibinfo {pages} {201} (\bibinfo {year} {1981})}\BibitemShut {NoStop}%
\bibitem [{\citenamefont {Haas}\ and\ \citenamefont {Durst}(1982)}]{haas_viscoelastic_1982}%
  \BibitemOpen
  \bibfield  {author} {\bibinfo {author} {\bibfnamefont {R.}~\bibnamefont {Haas}}\ and\ \bibinfo {author} {\bibfnamefont {F.}~\bibnamefont {Durst}},\ }\bibfield  {title} {\bibinfo {title} {Viscoelastic flow of dilute polymer solutions in regularly packed beds},\ }\href@noop {} {\bibfield  {journal} {\bibinfo  {journal} {Rheol Acta}\ }\textbf {\bibinfo {volume} {21}},\ \bibinfo {pages} {566} (\bibinfo {year} {1982})}\BibitemShut {NoStop}%
\bibitem [{\citenamefont {Chauveteau}\ \emph {et~al.}(1984)\citenamefont {Chauveteau}, \citenamefont {Moan},\ and\ \citenamefont {Magueur}}]{chauveteau_thickening_1984}%
  \BibitemOpen
  \bibfield  {author} {\bibinfo {author} {\bibfnamefont {G.}~\bibnamefont {Chauveteau}}, \bibinfo {author} {\bibfnamefont {M.}~\bibnamefont {Moan}},\ and\ \bibinfo {author} {\bibfnamefont {A.}~\bibnamefont {Magueur}},\ }\bibfield  {title} {\bibinfo {title} {Thickening behaviour of dilute polymer solutions in non-inertial elongational flows},\ }\href@noop {} {\bibfield  {journal} {\bibinfo  {journal} {J. Non-Newton. Fluid Mech.}\ }\textbf {\bibinfo {volume} {16}},\ \bibinfo {pages} {315} (\bibinfo {year} {1984})}\BibitemShut {NoStop}%
\bibitem [{\citenamefont {Gupta}\ and\ \citenamefont {Sridhar}(1985)}]{gupta_viscoelastic_1985}%
  \BibitemOpen
  \bibfield  {author} {\bibinfo {author} {\bibfnamefont {R.~K.}\ \bibnamefont {Gupta}}\ and\ \bibinfo {author} {\bibfnamefont {T.}~\bibnamefont {Sridhar}},\ }\bibfield  {title} {\bibinfo {title} {Viscoelastic effects in non-{Newtonian} flows through porous media},\ }\href@noop {} {\bibfield  {journal} {\bibinfo  {journal} {Rheol Acta}\ }\textbf {\bibinfo {volume} {24}},\ \bibinfo {pages} {148} (\bibinfo {year} {1985})}\BibitemShut {NoStop}%
\bibitem [{\citenamefont {Magueur}\ \emph {et~al.}(1985)\citenamefont {Magueur}, \citenamefont {Moan},\ and\ \citenamefont {Chauveteau}}]{magueur_effect_1985}%
  \BibitemOpen
  \bibfield  {author} {\bibinfo {author} {\bibfnamefont {A.}~\bibnamefont {Magueur}}, \bibinfo {author} {\bibfnamefont {M.}~\bibnamefont {Moan}},\ and\ \bibinfo {author} {\bibfnamefont {G.}~\bibnamefont {Chauveteau}},\ }\bibfield  {title} {\bibinfo {title} {Effect of successive contractions and expansions on the apparent viscosity of dilute polymer solutions},\ }\href@noop {} {\bibfield  {journal} {\bibinfo  {journal} {Chem. Eng. Commun.}\ }\textbf {\bibinfo {volume} {36}},\ \bibinfo {pages} {351} (\bibinfo {year} {1985})}\BibitemShut {NoStop}%
\bibitem [{\citenamefont {Ghoniem}(1985)}]{ghoniem_extensional_1985}%
  \BibitemOpen
  \bibfield  {author} {\bibinfo {author} {\bibfnamefont {S.~A.-A.}\ \bibnamefont {Ghoniem}},\ }\bibfield  {title} {\bibinfo {title} {Extensional flow of polymer solutions through porous media},\ }\href@noop {} {\bibfield  {journal} {\bibinfo  {journal} {Rheol Acta}\ }\textbf {\bibinfo {volume} {24}},\ \bibinfo {pages} {588} (\bibinfo {year} {1985})}\BibitemShut {NoStop}%
\bibitem [{\citenamefont {Durst}\ \emph {et~al.}(1987)\citenamefont {Durst}, \citenamefont {Haas},\ and\ \citenamefont {Interthal}}]{durst_nature_1987}%
  \BibitemOpen
  \bibfield  {author} {\bibinfo {author} {\bibfnamefont {F.}~\bibnamefont {Durst}}, \bibinfo {author} {\bibfnamefont {R.}~\bibnamefont {Haas}},\ and\ \bibinfo {author} {\bibfnamefont {W.}~\bibnamefont {Interthal}},\ }\bibfield  {title} {\bibinfo {title} {The nature of flows through porous media},\ }\href@noop {} {\bibfield  {journal} {\bibinfo  {journal} {J. Non-Newton. Fluid Mech.}\ }\textbf {\bibinfo {volume} {22}},\ \bibinfo {pages} {169} (\bibinfo {year} {1987})}\BibitemShut {NoStop}%
\bibitem [{\citenamefont {Chmielewski}\ and\ \citenamefont {Jayaraman}(1992)}]{chmielewski_effect_1992}%
  \BibitemOpen
  \bibfield  {author} {\bibinfo {author} {\bibfnamefont {C.}~\bibnamefont {Chmielewski}}\ and\ \bibinfo {author} {\bibfnamefont {K.}~\bibnamefont {Jayaraman}},\ }\bibfield  {title} {\bibinfo {title} {The effect of polymer extensibility on crossflow of polymer solutions through cylinder arrays},\ }\href@noop {} {\bibfield  {journal} {\bibinfo  {journal} {J. Rheol.}\ }\textbf {\bibinfo {volume} {36}},\ \bibinfo {pages} {1105} (\bibinfo {year} {1992})}\BibitemShut {NoStop}%
\bibitem [{\citenamefont {Skartsis}\ \emph {et~al.}(1992)\citenamefont {Skartsis}, \citenamefont {Khomami},\ and\ \citenamefont {Kardos}}]{skartsis_polymeric_1992}%
  \BibitemOpen
  \bibfield  {author} {\bibinfo {author} {\bibfnamefont {L.}~\bibnamefont {Skartsis}}, \bibinfo {author} {\bibfnamefont {B.}~\bibnamefont {Khomami}},\ and\ \bibinfo {author} {\bibfnamefont {J.~L.}\ \bibnamefont {Kardos}},\ }\bibfield  {title} {\bibinfo {title} {Polymeric flow through fibrous media},\ }\href@noop {} {\bibfield  {journal} {\bibinfo  {journal} {Journal of Rheology}\ }\textbf {\bibinfo {volume} {36}},\ \bibinfo {pages} {589} (\bibinfo {year} {1992})}\BibitemShut {NoStop}%
\bibitem [{\citenamefont {Kozicki}(2001)}]{kozicki_viscoelastic_2001}%
  \BibitemOpen
  \bibfield  {author} {\bibinfo {author} {\bibfnamefont {W.}~\bibnamefont {Kozicki}},\ }\bibfield  {title} {\bibinfo {title} {Viscoelastic flow in packed beds or porous media},\ }\href@noop {} {\bibfield  {journal} {\bibinfo  {journal} {Can J Chem Eng}\ }\textbf {\bibinfo {volume} {79}},\ \bibinfo {pages} {124} (\bibinfo {year} {2001})}\BibitemShut {NoStop}%
\bibitem [{\citenamefont {Rothstein}\ and\ \citenamefont {McKinley}(2001)}]{rothstein_axisymmetric_2001}%
  \BibitemOpen
  \bibfield  {author} {\bibinfo {author} {\bibfnamefont {J.~P.}\ \bibnamefont {Rothstein}}\ and\ \bibinfo {author} {\bibfnamefont {G.~H.}\ \bibnamefont {McKinley}},\ }\bibfield  {title} {\bibinfo {title} {The axisymmetric contraction–expansion: the role of extensional rheology on vortex growth dynamics and the enhanced pressure drop},\ }\href@noop {} {\bibfield  {journal} {\bibinfo  {journal} {J. Non-Newton. Fluid Mech.}\ }\textbf {\bibinfo {volume} {98}},\ \bibinfo {pages} {33} (\bibinfo {year} {2001})}\BibitemShut {NoStop}%
\bibitem [{\citenamefont {Kozicki}(2002)}]{kozicki_flow_2002}%
  \BibitemOpen
  \bibfield  {author} {\bibinfo {author} {\bibfnamefont {W.}~\bibnamefont {Kozicki}},\ }\bibfield  {title} {\bibinfo {title} {Flow of a {FENE} {Fluid} in {Packed} {Beds} or {Porous} {Media}},\ }\href@noop {} {\bibfield  {journal} {\bibinfo  {journal} {Can J Chem Eng}\ }\textbf {\bibinfo {volume} {80}},\ \bibinfo {pages} {818} (\bibinfo {year} {2002})}\BibitemShut {NoStop}%
\bibitem [{\citenamefont {Haward}\ and\ \citenamefont {Odell}(2003)}]{haward_viscosity_2003}%
  \BibitemOpen
  \bibfield  {author} {\bibinfo {author} {\bibfnamefont {S.~J.}\ \bibnamefont {Haward}}\ and\ \bibinfo {author} {\bibfnamefont {J.~A.}\ \bibnamefont {Odell}},\ }\bibfield  {title} {\bibinfo {title} {Viscosity enhancement in non-{Newtonian} flow of dilute polymer solutions through crystallographic porous media},\ }\href@noop {} {\bibfield  {journal} {\bibinfo  {journal} {Rheol. Acta}\ }\textbf {\bibinfo {volume} {42}},\ \bibinfo {pages} {516} (\bibinfo {year} {2003})}\BibitemShut {NoStop}%
\bibitem [{\citenamefont {Odell}\ and\ \citenamefont {Haward}(2006)}]{odell_viscosity_2006}%
  \BibitemOpen
  \bibfield  {author} {\bibinfo {author} {\bibfnamefont {J.~A.}\ \bibnamefont {Odell}}\ and\ \bibinfo {author} {\bibfnamefont {S.~J.}\ \bibnamefont {Haward}},\ }\bibfield  {title} {\bibinfo {title} {Viscosity enhancement in non-{Newtonian} flow of dilute aqueous polymer solutions through crystallographic and random porous media},\ }\href@noop {} {\bibfield  {journal} {\bibinfo  {journal} {Rheol. Acta}\ }\textbf {\bibinfo {volume} {45}},\ \bibinfo {pages} {853} (\bibinfo {year} {2006})}\BibitemShut {NoStop}%
\bibitem [{\citenamefont {Walters}\ \emph {et~al.}(2009)\citenamefont {Walters}, \citenamefont {Tamaddon-Jahromi}, \citenamefont {Webster}, \citenamefont {Tomé},\ and\ \citenamefont {McKee}}]{walters_competing_2009}%
  \BibitemOpen
  \bibfield  {author} {\bibinfo {author} {\bibfnamefont {K.}~\bibnamefont {Walters}}, \bibinfo {author} {\bibfnamefont {H.~R.}\ \bibnamefont {Tamaddon-Jahromi}}, \bibinfo {author} {\bibfnamefont {M.~F.}\ \bibnamefont {Webster}}, \bibinfo {author} {\bibfnamefont {M.~F.}\ \bibnamefont {Tomé}},\ and\ \bibinfo {author} {\bibfnamefont {S.}~\bibnamefont {McKee}},\ }\bibfield  {title} {\bibinfo {title} {The competing roles of extensional viscosity and normal stress differences in complex flows of elastic liquids},\ }\href@noop {} {\bibfield  {journal} {\bibinfo  {journal} {Korea Aus. Rheol. J.}\ }\textbf {\bibinfo {volume} {21}} (\bibinfo {year} {2009})}\BibitemShut {NoStop}%
\bibitem [{\citenamefont {Zamani}\ \emph {et~al.}(2015)\citenamefont {Zamani}, \citenamefont {Bondino}, \citenamefont {Kaufmann},\ and\ \citenamefont {Skauge}}]{zamani_effect_2015}%
  \BibitemOpen
  \bibfield  {author} {\bibinfo {author} {\bibfnamefont {N.}~\bibnamefont {Zamani}}, \bibinfo {author} {\bibfnamefont {I.}~\bibnamefont {Bondino}}, \bibinfo {author} {\bibfnamefont {R.}~\bibnamefont {Kaufmann}},\ and\ \bibinfo {author} {\bibfnamefont {A.}~\bibnamefont {Skauge}},\ }\bibfield  {title} {\bibinfo {title} {Effect of porous media properties on the onset of polymer extensional viscosity},\ }\href@noop {} {\bibfield  {journal} {\bibinfo  {journal} {J. Pet. Sci. Eng.}\ }\textbf {\bibinfo {volume} {133}},\ \bibinfo {pages} {483} (\bibinfo {year} {2015})}\BibitemShut {NoStop}%
\bibitem [{\citenamefont {Skauge}\ \emph {et~al.}(2018)\citenamefont {Skauge}, \citenamefont {Zamani}, \citenamefont {Gausdal~Jacobsen}, \citenamefont {Shaker~Shiran}, \citenamefont {Al-Shakry},\ and\ \citenamefont {Skauge}}]{skauge_polymer_2018}%
  \BibitemOpen
  \bibfield  {author} {\bibinfo {author} {\bibfnamefont {A.}~\bibnamefont {Skauge}}, \bibinfo {author} {\bibfnamefont {N.}~\bibnamefont {Zamani}}, \bibinfo {author} {\bibfnamefont {J.}~\bibnamefont {Gausdal~Jacobsen}}, \bibinfo {author} {\bibfnamefont {B.}~\bibnamefont {Shaker~Shiran}}, \bibinfo {author} {\bibfnamefont {B.}~\bibnamefont {Al-Shakry}},\ and\ \bibinfo {author} {\bibfnamefont {T.}~\bibnamefont {Skauge}},\ }\bibfield  {title} {\bibinfo {title} {Polymer {Flow} in {Porous} {Media}: {Relevance} to {Enhanced} {Oil} {Recovery}},\ }\href@noop {} {\bibfield  {journal} {\bibinfo  {journal} {Colloids and Interfaces}\ }\textbf {\bibinfo {volume} {2}},\ \bibinfo {pages} {27} (\bibinfo {year} {2018})}\BibitemShut {NoStop}%
\bibitem [{\citenamefont {Ibezim}\ \emph {et~al.}(2021)\citenamefont {Ibezim}, \citenamefont {Poole},\ and\ \citenamefont {Dennis}}]{ibezim_viscoelastic_2021}%
  \BibitemOpen
  \bibfield  {author} {\bibinfo {author} {\bibfnamefont {V.~C.}\ \bibnamefont {Ibezim}}, \bibinfo {author} {\bibfnamefont {R.~J.}\ \bibnamefont {Poole}},\ and\ \bibinfo {author} {\bibfnamefont {D.~J.}\ \bibnamefont {Dennis}},\ }\bibfield  {title} {\bibinfo {title} {Viscoelastic fluid flow in microporous media},\ }\href@noop {} {\bibfield  {journal} {\bibinfo  {journal} {J. Non-Newton. Fluid Mech.}\ }\textbf {\bibinfo {volume} {296}},\ \bibinfo {pages} {104638} (\bibinfo {year} {2021})}\BibitemShut {NoStop}%
\bibitem [{\citenamefont {Mokhtari}\ \emph {et~al.}(2022{\natexlab{a}})\citenamefont {Mokhtari}, \citenamefont {Latché}, \citenamefont {Quintard},\ and\ \citenamefont {Davit}}]{mokhtari_modified_2022}%
  \BibitemOpen
  \bibfield  {author} {\bibinfo {author} {\bibfnamefont {O.}~\bibnamefont {Mokhtari}}, \bibinfo {author} {\bibfnamefont {J.-C.}\ \bibnamefont {Latché}}, \bibinfo {author} {\bibfnamefont {M.}~\bibnamefont {Quintard}},\ and\ \bibinfo {author} {\bibfnamefont {Y.}~\bibnamefont {Davit}},\ }\bibfield  {title} {\bibinfo {title} {A modified {Darcy}’s law for viscoelastic flows of highly dilute polymer solutions through porous media},\ }\href@noop {} {\bibfield  {journal} {\bibinfo  {journal} {J. Non-Newton. Fluid Mech.}\ }\textbf {\bibinfo {volume} {309}},\ \bibinfo {pages} {104919} (\bibinfo {year} {2022}{\natexlab{a}})}\BibitemShut {NoStop}%
\bibitem [{\citenamefont {Deiber}\ and\ \citenamefont {Schowalter}(1981)}]{deiber_modeling_1981}%
  \BibitemOpen
  \bibfield  {author} {\bibinfo {author} {\bibfnamefont {J.~A.}\ \bibnamefont {Deiber}}\ and\ \bibinfo {author} {\bibfnamefont {W.~R.}\ \bibnamefont {Schowalter}},\ }\bibfield  {title} {\bibinfo {title} {Modeling the flow of viscoelastic fluids through porous media},\ }\href@noop {} {\bibfield  {journal} {\bibinfo  {journal} {AIChE J.}\ }\textbf {\bibinfo {volume} {27}},\ \bibinfo {pages} {912} (\bibinfo {year} {1981})}\BibitemShut {NoStop}%
\bibitem [{\citenamefont {Vorwerk}(1994)}]{vorwerk_shearing_1994}%
  \BibitemOpen
  \bibfield  {author} {\bibinfo {author} {\bibfnamefont {J.}~\bibnamefont {Vorwerk}},\ }\bibfield  {title} {\bibinfo {title} {Shearing effects for the flow of surfactant and polymer solutions through a packed bed of spheres},\ }\href@noop {} {\bibfield  {journal} {\bibinfo  {journal} {J. Non-Newton. Fluid Mech.}\ }\textbf {\bibinfo {volume} {51}},\ \bibinfo {pages} {17} (\bibinfo {year} {1994})}\BibitemShut {NoStop}%
\bibitem [{\citenamefont {Helmreich}\ \emph {et~al.}(1995)\citenamefont {Helmreich}, \citenamefont {Vorwerk}, \citenamefont {Steger}, \citenamefont {Müller},\ and\ \citenamefont {Brunn}}]{helmreich_non-viscous_1995}%
  \BibitemOpen
  \bibfield  {author} {\bibinfo {author} {\bibfnamefont {A.}~\bibnamefont {Helmreich}}, \bibinfo {author} {\bibfnamefont {J.}~\bibnamefont {Vorwerk}}, \bibinfo {author} {\bibfnamefont {R.}~\bibnamefont {Steger}}, \bibinfo {author} {\bibfnamefont {M.}~\bibnamefont {Müller}},\ and\ \bibinfo {author} {\bibfnamefont {P.}~\bibnamefont {Brunn}},\ }\bibfield  {title} {\bibinfo {title} {Non-viscous effects in the flow of xanthan gum solutions through a packed bed of spheres},\ }\href@noop {} {\bibfield  {journal} {\bibinfo  {journal} {J. Chem. Eng.}\ }\textbf {\bibinfo {volume} {59}},\ \bibinfo {pages} {111} (\bibinfo {year} {1995})}\BibitemShut {NoStop}%
\bibitem [{\citenamefont {James}\ \emph {et~al.}(2012)\citenamefont {James}, \citenamefont {Yip},\ and\ \citenamefont {Currie}}]{james_slow_2012}%
  \BibitemOpen
  \bibfield  {author} {\bibinfo {author} {\bibfnamefont {D.~F.}\ \bibnamefont {James}}, \bibinfo {author} {\bibfnamefont {R.}~\bibnamefont {Yip}},\ and\ \bibinfo {author} {\bibfnamefont {I.~G.}\ \bibnamefont {Currie}},\ }\bibfield  {title} {\bibinfo {title} {Slow flow of {Boger} fluids through model fibrous porous media},\ }\href@noop {} {\bibfield  {journal} {\bibinfo  {journal} {J. Rheol.}\ }\textbf {\bibinfo {volume} {56}},\ \bibinfo {pages} {1249} (\bibinfo {year} {2012})}\BibitemShut {NoStop}%
\bibitem [{\citenamefont {James}(2016)}]{james_n1_2016}%
  \BibitemOpen
  \bibfield  {author} {\bibinfo {author} {\bibfnamefont {D.~F.}\ \bibnamefont {James}},\ }\bibfield  {title} {\bibinfo {title} {N1 stresses in extensional flows},\ }\href@noop {} {\bibfield  {journal} {\bibinfo  {journal} {J. Non-Newton. Fluid Mech.}\ }\textbf {\bibinfo {volume} {232}},\ \bibinfo {pages} {33} (\bibinfo {year} {2016})}\BibitemShut {NoStop}%
\bibitem [{\citenamefont {De}\ \emph {et~al.}(2017{\natexlab{a}})\citenamefont {De}, \citenamefont {Kuipers}, \citenamefont {Peters},\ and\ \citenamefont {Padding}}]{de_viscoelastic_2017}%
  \BibitemOpen
  \bibfield  {author} {\bibinfo {author} {\bibfnamefont {S.}~\bibnamefont {De}}, \bibinfo {author} {\bibfnamefont {J.~A.~M.}\ \bibnamefont {Kuipers}}, \bibinfo {author} {\bibfnamefont {E.~A. J.~F.}\ \bibnamefont {Peters}},\ and\ \bibinfo {author} {\bibfnamefont {J.~T.}\ \bibnamefont {Padding}},\ }\bibfield  {title} {\bibinfo {title} {Viscoelastic flow simulations in random porous media},\ }\href@noop {} {\bibfield  {journal} {\bibinfo  {journal} {J. Non-Newton. Fluid Mech.}\ }\textbf {\bibinfo {volume} {248}},\ \bibinfo {pages} {50} (\bibinfo {year} {2017}{\natexlab{a}})}\BibitemShut {NoStop}%
\bibitem [{\citenamefont {De}\ \emph {et~al.}(2017{\natexlab{b}})\citenamefont {De}, \citenamefont {Kuipers}, \citenamefont {Peters},\ and\ \citenamefont {Padding}}]{de_viscoelastic_2017-1}%
  \BibitemOpen
  \bibfield  {author} {\bibinfo {author} {\bibfnamefont {S.}~\bibnamefont {De}}, \bibinfo {author} {\bibfnamefont {J.~A.~M.}\ \bibnamefont {Kuipers}}, \bibinfo {author} {\bibfnamefont {E.~A. J.~F.}\ \bibnamefont {Peters}},\ and\ \bibinfo {author} {\bibfnamefont {J.~T.}\ \bibnamefont {Padding}},\ }\bibfield  {title} {\bibinfo {title} {Viscoelastic flow simulations in model porous media},\ }\href@noop {} {\bibfield  {journal} {\bibinfo  {journal} {Phys. Rev. Fluids}\ }\textbf {\bibinfo {volume} {2}},\ \bibinfo {pages} {053303} (\bibinfo {year} {2017}{\natexlab{b}})}\BibitemShut {NoStop}%
\bibitem [{\citenamefont {Liu}\ \emph {et~al.}(2017)\citenamefont {Liu}, \citenamefont {Wang},\ and\ \citenamefont {Hwang}}]{liu_flow_2017}%
  \BibitemOpen
  \bibfield  {author} {\bibinfo {author} {\bibfnamefont {H.~L.}\ \bibnamefont {Liu}}, \bibinfo {author} {\bibfnamefont {J.}~\bibnamefont {Wang}},\ and\ \bibinfo {author} {\bibfnamefont {W.~R.}\ \bibnamefont {Hwang}},\ }\bibfield  {title} {\bibinfo {title} {Flow resistance of viscoelastic flows in fibrous porous media},\ }\href@noop {} {\bibfield  {journal} {\bibinfo  {journal} {J. Non-Newton. Fluid Mech.}\ }\textbf {\bibinfo {volume} {246}},\ \bibinfo {pages} {21} (\bibinfo {year} {2017})}\BibitemShut {NoStop}%
\bibitem [{\citenamefont {Odell}\ \emph {et~al.}(1988)\citenamefont {Odell}, \citenamefont {Müller},\ and\ \citenamefont {Keller}}]{odell_non-newtonian_1988}%
  \BibitemOpen
  \bibfield  {author} {\bibinfo {author} {\bibfnamefont {J.}~\bibnamefont {Odell}}, \bibinfo {author} {\bibfnamefont {A.}~\bibnamefont {Müller}},\ and\ \bibinfo {author} {\bibfnamefont {A.}~\bibnamefont {Keller}},\ }\bibfield  {title} {\bibinfo {title} {Non-{Newtonian} behaviour of hydrolysed polyacrylamide in strong elongational flows: a transient network approach},\ }\href@noop {} {\bibfield  {journal} {\bibinfo  {journal} {Polymer}\ }\textbf {\bibinfo {volume} {29}},\ \bibinfo {pages} {1179} (\bibinfo {year} {1988})}\BibitemShut {NoStop}%
\bibitem [{\citenamefont {Rodriguez}\ \emph {et~al.}(1993)\citenamefont {Rodriguez}, \citenamefont {Romero}, \citenamefont {Sargenti}, \citenamefont {Müller}, \citenamefont {Sáez},\ and\ \citenamefont {Odell}}]{rodriguez_flow_1992}%
  \BibitemOpen
  \bibfield  {author} {\bibinfo {author} {\bibfnamefont {S.}~\bibnamefont {Rodriguez}}, \bibinfo {author} {\bibfnamefont {C.}~\bibnamefont {Romero}}, \bibinfo {author} {\bibfnamefont {M.~L.}\ \bibnamefont {Sargenti}}, \bibinfo {author} {\bibfnamefont {A.~J.}\ \bibnamefont {Müller}}, \bibinfo {author} {\bibfnamefont {A.~E.}\ \bibnamefont {Sáez}},\ and\ \bibinfo {author} {\bibfnamefont {J.~A.}\ \bibnamefont {Odell}},\ }\bibfield  {title} {\bibinfo {title} {Flow of polymer solutions through porous media},\ }\href@noop {} {\bibfield  {journal} {\bibinfo  {journal} {J. Non-Newton. Fluid Mech.}\ }\textbf {\bibinfo {volume} {49}},\ \bibinfo {pages} {63} (\bibinfo {year} {1993})}\BibitemShut {NoStop}%
\bibitem [{\citenamefont {Sáez}\ \emph {et~al.}(1994)\citenamefont {Sáez}, \citenamefont {Müller},\ and\ \citenamefont {Odell}}]{saez_flow_1994}%
  \BibitemOpen
  \bibfield  {author} {\bibinfo {author} {\bibfnamefont {A.~E.}\ \bibnamefont {Sáez}}, \bibinfo {author} {\bibfnamefont {A.~J.}\ \bibnamefont {Müller}},\ and\ \bibinfo {author} {\bibfnamefont {J.~A.}\ \bibnamefont {Odell}},\ }\bibfield  {title} {\bibinfo {title} {Flow of monodisperse polystyrene solutions through porous media},\ }\href@noop {} {\bibfield  {journal} {\bibinfo  {journal} {Colloid Polym Sci}\ }\textbf {\bibinfo {volume} {272}},\ \bibinfo {pages} {1224} (\bibinfo {year} {1994})}\BibitemShut {NoStop}%
\bibitem [{\citenamefont {Kauser}\ \emph {et~al.}(1999)\citenamefont {Kauser}, \citenamefont {Dos~Santos}, \citenamefont {Delgado}, \citenamefont {Müller},\ and\ \citenamefont {Sáez}}]{kauser_flow_1999}%
  \BibitemOpen
  \bibfield  {author} {\bibinfo {author} {\bibfnamefont {N.}~\bibnamefont {Kauser}}, \bibinfo {author} {\bibfnamefont {L.}~\bibnamefont {Dos~Santos}}, \bibinfo {author} {\bibfnamefont {M.}~\bibnamefont {Delgado}}, \bibinfo {author} {\bibfnamefont {A.~J.}\ \bibnamefont {Müller}},\ and\ \bibinfo {author} {\bibfnamefont {A.~E.}\ \bibnamefont {Sáez}},\ }\bibfield  {title} {\bibinfo {title} {Flow of mixtures of poly(ethylene oxide) and hydrolyzed polyacrylamide solutions through porous media},\ }\href@noop {} {\bibfield  {journal} {\bibinfo  {journal} {J. Appl. Polym. Sci.}\ }\textbf {\bibinfo {volume} {72}},\ \bibinfo {pages} {783} (\bibinfo {year} {1999})}\BibitemShut {NoStop}%
\bibitem [{\citenamefont {Chmielewski}\ and\ \citenamefont {Jayaraman}(1993)}]{chmielewski_elastic_1993}%
  \BibitemOpen
  \bibfield  {author} {\bibinfo {author} {\bibfnamefont {C.}~\bibnamefont {Chmielewski}}\ and\ \bibinfo {author} {\bibfnamefont {K.}~\bibnamefont {Jayaraman}},\ }\bibfield  {title} {\bibinfo {title} {Elastic instability in crossflow of polymer solutions through periodic arrays of cylinders},\ }\href@noop {} {\bibfield  {journal} {\bibinfo  {journal} {J. Non-Newton. Fluid Mech.}\ }\textbf {\bibinfo {volume} {48}},\ \bibinfo {pages} {285} (\bibinfo {year} {1993})}\BibitemShut {NoStop}%
\bibitem [{\citenamefont {Galindo-Rosales}\ \emph {et~al.}(2012)\citenamefont {Galindo-Rosales}, \citenamefont {Campo-Deaño}, \citenamefont {Pinho}, \citenamefont {Van~Bokhorst}, \citenamefont {Hamersma}, \citenamefont {Oliveira},\ and\ \citenamefont {Alves}}]{galindo-rosales_microfluidic_2012}%
  \BibitemOpen
  \bibfield  {author} {\bibinfo {author} {\bibfnamefont {F.~J.}\ \bibnamefont {Galindo-Rosales}}, \bibinfo {author} {\bibfnamefont {L.}~\bibnamefont {Campo-Deaño}}, \bibinfo {author} {\bibfnamefont {F.~T.}\ \bibnamefont {Pinho}}, \bibinfo {author} {\bibfnamefont {E.}~\bibnamefont {Van~Bokhorst}}, \bibinfo {author} {\bibfnamefont {P.~J.}\ \bibnamefont {Hamersma}}, \bibinfo {author} {\bibfnamefont {M.~S.~N.}\ \bibnamefont {Oliveira}},\ and\ \bibinfo {author} {\bibfnamefont {M.~A.}\ \bibnamefont {Alves}},\ }\bibfield  {title} {\bibinfo {title} {Microfluidic systems for the analysis of viscoelastic fluid flow phenomena in porous media},\ }\href@noop {} {\bibfield  {journal} {\bibinfo  {journal} {Microfluid Nanofluid}\ }\textbf {\bibinfo {volume} {12}},\ \bibinfo {pages} {485} (\bibinfo {year} {2012})}\BibitemShut {NoStop}%
\bibitem [{\citenamefont {Clarke}\ \emph {et~al.}(2015)\citenamefont {Clarke}, \citenamefont {Howe}, \citenamefont {Mitchell}, \citenamefont {Staniland}, \citenamefont {Hawkes},\ and\ \citenamefont {Leeper}}]{clarke_mechanism_2015}%
  \BibitemOpen
  \bibfield  {author} {\bibinfo {author} {\bibfnamefont {A.}~\bibnamefont {Clarke}}, \bibinfo {author} {\bibfnamefont {A.~M.}\ \bibnamefont {Howe}}, \bibinfo {author} {\bibfnamefont {J.}~\bibnamefont {Mitchell}}, \bibinfo {author} {\bibfnamefont {J.}~\bibnamefont {Staniland}}, \bibinfo {author} {\bibfnamefont {L.}~\bibnamefont {Hawkes}},\ and\ \bibinfo {author} {\bibfnamefont {K.}~\bibnamefont {Leeper}},\ }\bibfield  {title} {\bibinfo {title} {Mechanism of anomalously increased oil displacement with aqueous viscoelastic polymer solutions},\ }\href@noop {} {\bibfield  {journal} {\bibinfo  {journal} {Soft Matter}\ }\textbf {\bibinfo {volume} {11}},\ \bibinfo {pages} {3536} (\bibinfo {year} {2015})}\BibitemShut {NoStop}%
\bibitem [{\citenamefont {Machado}\ \emph {et~al.}(2016)\citenamefont {Machado}, \citenamefont {Bodiguel}, \citenamefont {Beaumont}, \citenamefont {Clisson},\ and\ \citenamefont {Colin}}]{machado_extra_2016}%
  \BibitemOpen
  \bibfield  {author} {\bibinfo {author} {\bibfnamefont {A.}~\bibnamefont {Machado}}, \bibinfo {author} {\bibfnamefont {H.}~\bibnamefont {Bodiguel}}, \bibinfo {author} {\bibfnamefont {J.}~\bibnamefont {Beaumont}}, \bibinfo {author} {\bibfnamefont {G.}~\bibnamefont {Clisson}},\ and\ \bibinfo {author} {\bibfnamefont {A.}~\bibnamefont {Colin}},\ }\bibfield  {title} {\bibinfo {title} {Extra dissipation and flow uniformization due to elastic instabilities of shear-thinning polymer solutions in model porous media},\ }\href@noop {} {\bibfield  {journal} {\bibinfo  {journal} {Biomicrofluidics}\ }\textbf {\bibinfo {volume} {10}} (\bibinfo {year} {2016})}\BibitemShut {NoStop}%
\bibitem [{\citenamefont {Mitchell}\ \emph {et~al.}(2016)\citenamefont {Mitchell}, \citenamefont {Lyons}, \citenamefont {Howe},\ and\ \citenamefont {Clarke}}]{mitchell_viscoelastic_2016}%
  \BibitemOpen
  \bibfield  {author} {\bibinfo {author} {\bibfnamefont {J.}~\bibnamefont {Mitchell}}, \bibinfo {author} {\bibfnamefont {K.}~\bibnamefont {Lyons}}, \bibinfo {author} {\bibfnamefont {A.~M.}\ \bibnamefont {Howe}},\ and\ \bibinfo {author} {\bibfnamefont {A.}~\bibnamefont {Clarke}},\ }\bibfield  {title} {\bibinfo {title} {Viscoelastic polymer flows and elastic turbulence in three-dimensional porous structures},\ }\href@noop {} {\bibfield  {journal} {\bibinfo  {journal} {Soft Matter}\ }\textbf {\bibinfo {volume} {12}},\ \bibinfo {pages} {460} (\bibinfo {year} {2016})}\BibitemShut {NoStop}%
\bibitem [{\citenamefont {Kawale}\ \emph {et~al.}(2017)\citenamefont {Kawale}, \citenamefont {Marques}, \citenamefont {Zitha}, \citenamefont {Kreutzer}, \citenamefont {Rossen},\ and\ \citenamefont {Boukany}}]{kawale_elastic_2017}%
  \BibitemOpen
  \bibfield  {author} {\bibinfo {author} {\bibfnamefont {D.}~\bibnamefont {Kawale}}, \bibinfo {author} {\bibfnamefont {E.}~\bibnamefont {Marques}}, \bibinfo {author} {\bibfnamefont {P.~L.~J.}\ \bibnamefont {Zitha}}, \bibinfo {author} {\bibfnamefont {M.~T.}\ \bibnamefont {Kreutzer}}, \bibinfo {author} {\bibfnamefont {W.~R.}\ \bibnamefont {Rossen}},\ and\ \bibinfo {author} {\bibfnamefont {P.~E.}\ \bibnamefont {Boukany}},\ }\bibfield  {title} {\bibinfo {title} {Elastic instabilities during the flow of hydrolyzed polyacrylamide solution in porous media: effect of pore-shape and salt},\ }\href@noop {} {\bibfield  {journal} {\bibinfo  {journal} {Soft Matter}\ }\textbf {\bibinfo {volume} {13}},\ \bibinfo {pages} {765} (\bibinfo {year} {2017})}\BibitemShut {NoStop}%
\bibitem [{\citenamefont {Hemingway}\ \emph {et~al.}(2018)\citenamefont {Hemingway}, \citenamefont {Clarke}, \citenamefont {Pearson},\ and\ \citenamefont {Fielding}}]{hemingway_thickening_2018}%
  \BibitemOpen
  \bibfield  {author} {\bibinfo {author} {\bibfnamefont {E.~J.}\ \bibnamefont {Hemingway}}, \bibinfo {author} {\bibfnamefont {A.}~\bibnamefont {Clarke}}, \bibinfo {author} {\bibfnamefont {J.~R.~A.}\ \bibnamefont {Pearson}},\ and\ \bibinfo {author} {\bibfnamefont {S.~M.}\ \bibnamefont {Fielding}},\ }\bibfield  {title} {\bibinfo {title} {Thickening of viscoelastic flow in a model porous medium},\ }\href@noop {} {\bibfield  {journal} {\bibinfo  {journal} {J. Non-Newton. Fluid Mech.}\ }\textbf {\bibinfo {volume} {251}},\ \bibinfo {pages} {56} (\bibinfo {year} {2018})}\BibitemShut {NoStop}%
\bibitem [{\citenamefont {Qin}\ \emph {et~al.}(2019{\natexlab{a}})\citenamefont {Qin}, \citenamefont {Salipante}, \citenamefont {Hudson},\ and\ \citenamefont {Arratia}}]{qin_flow_2019}%
  \BibitemOpen
  \bibfield  {author} {\bibinfo {author} {\bibfnamefont {B.}~\bibnamefont {Qin}}, \bibinfo {author} {\bibfnamefont {P.~F.}\ \bibnamefont {Salipante}}, \bibinfo {author} {\bibfnamefont {S.~D.}\ \bibnamefont {Hudson}},\ and\ \bibinfo {author} {\bibfnamefont {P.~E.}\ \bibnamefont {Arratia}},\ }\bibfield  {title} {\bibinfo {title} {Flow {Resistance} and {Structures} in {Viscoelastic} {Channel} {Flows} at {Low} {Re}},\ }\href@noop {} {\bibfield  {journal} {\bibinfo  {journal} {Phys. Rev. Lett.}\ }\textbf {\bibinfo {volume} {123}},\ \bibinfo {pages} {6} (\bibinfo {year} {2019}{\natexlab{a}})}\BibitemShut {NoStop}%
\bibitem [{\citenamefont {Ekanem}\ \emph {et~al.}(2020)\citenamefont {Ekanem}, \citenamefont {Berg}, \citenamefont {De}, \citenamefont {Fadili}, \citenamefont {Bultreys}, \citenamefont {Rücker}, \citenamefont {Southwick}, \citenamefont {Crawshaw},\ and\ \citenamefont {Luckham}}]{ekanem_signature_2020}%
  \BibitemOpen
  \bibfield  {author} {\bibinfo {author} {\bibfnamefont {E.~M.}\ \bibnamefont {Ekanem}}, \bibinfo {author} {\bibfnamefont {S.}~\bibnamefont {Berg}}, \bibinfo {author} {\bibfnamefont {S.}~\bibnamefont {De}}, \bibinfo {author} {\bibfnamefont {A.}~\bibnamefont {Fadili}}, \bibinfo {author} {\bibfnamefont {T.}~\bibnamefont {Bultreys}}, \bibinfo {author} {\bibfnamefont {M.}~\bibnamefont {Rücker}}, \bibinfo {author} {\bibfnamefont {J.}~\bibnamefont {Southwick}}, \bibinfo {author} {\bibfnamefont {J.}~\bibnamefont {Crawshaw}},\ and\ \bibinfo {author} {\bibfnamefont {P.~F.}\ \bibnamefont {Luckham}},\ }\bibfield  {title} {\bibinfo {title} {Signature of elastic turbulence of viscoelastic fluid flow in a single pore throat},\ }\href@noop {} {\bibfield  {journal} {\bibinfo  {journal} {Phys. Rev. E}\ }\textbf {\bibinfo {volume} {101}},\ \bibinfo {pages} {042605} (\bibinfo {year} {2020})}\BibitemShut {NoStop}%
\bibitem [{\citenamefont {Browne}\ and\ \citenamefont {Datta}(2021)}]{browne_elastic_2021}%
  \BibitemOpen
  \bibfield  {author} {\bibinfo {author} {\bibfnamefont {C.~A.}\ \bibnamefont {Browne}}\ and\ \bibinfo {author} {\bibfnamefont {S.~S.}\ \bibnamefont {Datta}},\ }\bibfield  {title} {\bibinfo {title} {Elastic turbulence generates anomalous flow resistance in porous media},\ }\href@noop {} {\bibfield  {journal} {\bibinfo  {journal} {Sci. Adv.}\ }\textbf {\bibinfo {volume} {7}},\ \bibinfo {pages} {11} (\bibinfo {year} {2021})}\BibitemShut {NoStop}%
\bibitem [{\citenamefont {Browne}\ \emph {et~al.}(2023)\citenamefont {Browne}, \citenamefont {Huang}, \citenamefont {Zheng},\ and\ \citenamefont {Datta}}]{browne_homogenizing_2023}%
  \BibitemOpen
  \bibfield  {author} {\bibinfo {author} {\bibfnamefont {C.~A.}\ \bibnamefont {Browne}}, \bibinfo {author} {\bibfnamefont {R.~B.}\ \bibnamefont {Huang}}, \bibinfo {author} {\bibfnamefont {C.~W.}\ \bibnamefont {Zheng}},\ and\ \bibinfo {author} {\bibfnamefont {S.~S.}\ \bibnamefont {Datta}},\ }\bibfield  {title} {\bibinfo {title} {Homogenizing fluid transport in stratified porous media using an elastic flow instability},\ }\href@noop {} {\bibfield  {journal} {\bibinfo  {journal} {J. Fluid Mech.}\ }\textbf {\bibinfo {volume} {963}},\ \bibinfo {pages} {A30} (\bibinfo {year} {2023})}\BibitemShut {NoStop}%
\bibitem [{\citenamefont {Chilcott}\ and\ \citenamefont {Rallison}(1988)}]{chilcott_creeping_1988}%
  \BibitemOpen
  \bibfield  {author} {\bibinfo {author} {\bibfnamefont {M.}~\bibnamefont {Chilcott}}\ and\ \bibinfo {author} {\bibfnamefont {J.}~\bibnamefont {Rallison}},\ }\bibfield  {title} {\bibinfo {title} {Creeping flow of dilute polymer solutions past cylinders and spheres},\ }\href@noop {} {\bibfield  {journal} {\bibinfo  {journal} {J. Non-Newton. Fluid Mech.}\ }\textbf {\bibinfo {volume} {29}},\ \bibinfo {pages} {381} (\bibinfo {year} {1988})}\BibitemShut {NoStop}%
\bibitem [{\citenamefont {Talwar}\ and\ \citenamefont {Khomami}(1995)}]{talwar_flow_1995}%
  \BibitemOpen
  \bibfield  {author} {\bibinfo {author} {\bibfnamefont {K.~K.}\ \bibnamefont {Talwar}}\ and\ \bibinfo {author} {\bibfnamefont {B.}~\bibnamefont {Khomami}},\ }\bibfield  {title} {\bibinfo {title} {Flow of viscoelastic fluids past periodic square arrays of cylinders: inertial and shear thinning viscosity and elasticity effects},\ }\href@noop {} {\bibfield  {journal} {\bibinfo  {journal} {J. Rheol.}\ }\textbf {\bibinfo {volume} {57}},\ \bibinfo {pages} {177} (\bibinfo {year} {1995})}\BibitemShut {NoStop}%
\bibitem [{\citenamefont {Khomami}\ and\ \citenamefont {Moreno}(1997)}]{khomami_stability_1997}%
  \BibitemOpen
  \bibfield  {author} {\bibinfo {author} {\bibfnamefont {B.}~\bibnamefont {Khomami}}\ and\ \bibinfo {author} {\bibfnamefont {L.~D.}\ \bibnamefont {Moreno}},\ }\bibfield  {title} {\bibinfo {title} {Stability of viscoelastic flow around periodic arrays of cylinders},\ }\href@noop {} {\bibfield  {journal} {\bibinfo  {journal} {Rheol. Acta}\ }\textbf {\bibinfo {volume} {36}},\ \bibinfo {pages} {367} (\bibinfo {year} {1997})}\BibitemShut {NoStop}%
\bibitem [{\citenamefont {Mokhtari}\ \emph {et~al.}(2022{\natexlab{b}})\citenamefont {Mokhtari}, \citenamefont {Latché}, \citenamefont {Quintard},\ and\ \citenamefont {Davit}}]{mokhtari_birefringent_2022}%
  \BibitemOpen
  \bibfield  {author} {\bibinfo {author} {\bibfnamefont {O.}~\bibnamefont {Mokhtari}}, \bibinfo {author} {\bibfnamefont {J.-C.}\ \bibnamefont {Latché}}, \bibinfo {author} {\bibfnamefont {M.}~\bibnamefont {Quintard}},\ and\ \bibinfo {author} {\bibfnamefont {Y.}~\bibnamefont {Davit}},\ }\bibfield  {title} {\bibinfo {title} {Birefringent strands drive the flow of viscoelastic fluids past obstacles},\ }\href@noop {} {\bibfield  {journal} {\bibinfo  {journal} {J. Fluid Mech.}\ }\textbf {\bibinfo {volume} {948}},\ \bibinfo {pages} {A2} (\bibinfo {year} {2022}{\natexlab{b}})}\BibitemShut {NoStop}%
\bibitem [{\citenamefont {Wissler}(1971)}]{wissler_viscoelastic_1971}%
  \BibitemOpen
  \bibfield  {author} {\bibinfo {author} {\bibfnamefont {E.~H.}\ \bibnamefont {Wissler}},\ }\bibfield  {title} {\bibinfo {title} {Viscoelastic {Effects} in the {Flow} of {Non}-{Newtonian} {Fluids} through a {Porous} {Medium}},\ }\href@noop {} {\bibfield  {journal} {\bibinfo  {journal} {Ind. Eng. Chem. Fund.}\ }\textbf {\bibinfo {volume} {10}},\ \bibinfo {pages} {411} (\bibinfo {year} {1971})}\BibitemShut {NoStop}%
\bibitem [{\citenamefont {Vossoughi}\ and\ \citenamefont {Seyer}(1974)}]{vossoughi_pressure_1974}%
  \BibitemOpen
  \bibfield  {author} {\bibinfo {author} {\bibfnamefont {S.}~\bibnamefont {Vossoughi}}\ and\ \bibinfo {author} {\bibfnamefont {F.~A.}\ \bibnamefont {Seyer}},\ }\bibfield  {title} {\bibinfo {title} {Pressure drop for flow of polymer solution in a model porous medium},\ }\href@noop {} {\bibfield  {journal} {\bibinfo  {journal} {Can. J. Chem. Eng.}\ }\textbf {\bibinfo {volume} {52}},\ \bibinfo {pages} {666} (\bibinfo {year} {1974})}\BibitemShut {NoStop}%
\bibitem [{\citenamefont {Dharmadhikari}\ and\ \citenamefont {Kale}(1985)}]{dharmadhikari_flow_1985}%
  \BibitemOpen
  \bibfield  {author} {\bibinfo {author} {\bibfnamefont {R.~V.}\ \bibnamefont {Dharmadhikari}}\ and\ \bibinfo {author} {\bibfnamefont {D.~D.}\ \bibnamefont {Kale}},\ }\bibfield  {title} {\bibinfo {title} {Flow of non-{Newtonian} fluids through porous media},\ }\href@noop {} {\bibfield  {journal} {\bibinfo  {journal} {Chem. Eng. Sci.}\ }\textbf {\bibinfo {volume} {40}},\ \bibinfo {pages} {527} (\bibinfo {year} {1985})}\BibitemShut {NoStop}%
\bibitem [{\citenamefont {Bendová}\ \emph {et~al.}(2009)\citenamefont {Bendová}, \citenamefont {Šiška},\ and\ \citenamefont {Machač}}]{bendova_pressure_2009}%
  \BibitemOpen
  \bibfield  {author} {\bibinfo {author} {\bibfnamefont {H.}~\bibnamefont {Bendová}}, \bibinfo {author} {\bibfnamefont {B.}~\bibnamefont {Šiška}},\ and\ \bibinfo {author} {\bibfnamefont {I.}~\bibnamefont {Machač}},\ }\bibfield  {title} {\bibinfo {title} {Pressure drop excess in the flow of viscoelastic liquids through fixed beds of particles},\ }\href@noop {} {\bibfield  {journal} {\bibinfo  {journal} {Chem. Eng. Process. Process Intensif.}\ }\textbf {\bibinfo {volume} {48}},\ \bibinfo {pages} {29} (\bibinfo {year} {2009})}\BibitemShut {NoStop}%
\bibitem [{\citenamefont {Sobti}\ and\ \citenamefont {Wanchoo}(2014)}]{sobti_creeping_2014}%
  \BibitemOpen
  \bibfield  {author} {\bibinfo {author} {\bibfnamefont {A.}~\bibnamefont {Sobti}}\ and\ \bibinfo {author} {\bibfnamefont {R.~K.}\ \bibnamefont {Wanchoo}},\ }\bibfield  {title} {\bibinfo {title} {Creeping {Flow} of {Viscoelastic} {Fluid} through a {Packed} {Bed}},\ }\href@noop {} {\bibfield  {journal} {\bibinfo  {journal} {Ind. Eng. Chem. Res.}\ }\textbf {\bibinfo {volume} {53}},\ \bibinfo {pages} {14508} (\bibinfo {year} {2014})}\BibitemShut {NoStop}%
\bibitem [{\citenamefont {Browne}\ and\ \citenamefont {Datta}(2024)}]{browne_harnessing_2024}%
  \BibitemOpen
  \bibfield  {author} {\bibinfo {author} {\bibfnamefont {C.~A.}\ \bibnamefont {Browne}}\ and\ \bibinfo {author} {\bibfnamefont {S.~S.}\ \bibnamefont {Datta}},\ }\bibfield  {title} {\bibinfo {title} {Harnessing elastic instabilities for enhanced mixing and reaction kinetics in porous media},\ }\href@noop {} {\bibfield  {journal} {\bibinfo  {journal} {Proc. Natl. Acad. Sci. U.S.A.}\ }\textbf {\bibinfo {volume} {121}},\ \bibinfo {pages} {e2320962121} (\bibinfo {year} {2024})}\BibitemShut {NoStop}%
\bibitem [{\citenamefont {Datta}\ \emph {et~al.}(2022)\citenamefont {Datta}, \citenamefont {Ardekani}, \citenamefont {Arratia}, \citenamefont {Beris}, \citenamefont {Bischofberger}, \citenamefont {McKinley}, \citenamefont {Eggers}, \citenamefont {L{\'o}pez-Aguilar}, \citenamefont {Fielding}, \citenamefont {Frishman} \emph {et~al.}}]{datta2022perspectives}%
  \BibitemOpen
  \bibfield  {author} {\bibinfo {author} {\bibfnamefont {S.~S.}\ \bibnamefont {Datta}}, \bibinfo {author} {\bibfnamefont {A.~M.}\ \bibnamefont {Ardekani}}, \bibinfo {author} {\bibfnamefont {P.~E.}\ \bibnamefont {Arratia}}, \bibinfo {author} {\bibfnamefont {A.~N.}\ \bibnamefont {Beris}}, \bibinfo {author} {\bibfnamefont {I.}~\bibnamefont {Bischofberger}}, \bibinfo {author} {\bibfnamefont {G.~H.}\ \bibnamefont {McKinley}}, \bibinfo {author} {\bibfnamefont {J.~G.}\ \bibnamefont {Eggers}}, \bibinfo {author} {\bibfnamefont {J.~E.}\ \bibnamefont {L{\'o}pez-Aguilar}}, \bibinfo {author} {\bibfnamefont {S.~M.}\ \bibnamefont {Fielding}}, \bibinfo {author} {\bibfnamefont {A.}~\bibnamefont {Frishman}}, \emph {et~al.},\ }\bibfield  {title} {\bibinfo {title} {Perspectives on viscoelastic flow instabilities and elastic turbulence},\ }\href@noop {} {\bibfield  {journal} {\bibinfo  {journal} {Phys. Rev. Fluids}\ }\textbf {\bibinfo {volume} {7}},\ \bibinfo {pages} {080701} (\bibinfo {year} {2022})}\BibitemShut {NoStop}%
\bibitem [{\citenamefont {Groisman}\ and\ \citenamefont {Steinberg}(2000)}]{groisman_elastic_2000}%
  \BibitemOpen
  \bibfield  {author} {\bibinfo {author} {\bibfnamefont {A.}~\bibnamefont {Groisman}}\ and\ \bibinfo {author} {\bibfnamefont {V.}~\bibnamefont {Steinberg}},\ }\bibfield  {title} {\bibinfo {title} {Elastic turbulence in a polymer solution flow},\ }\href@noop {} {\bibfield  {journal} {\bibinfo  {journal} {Nature}\ }\textbf {\bibinfo {volume} {405}},\ \bibinfo {pages} {53} (\bibinfo {year} {2000})}\BibitemShut {NoStop}%
\bibitem [{\citenamefont {Groisman}\ and\ \citenamefont {Steinberg}(2004)}]{groisman_elastic_2004}%
  \BibitemOpen
  \bibfield  {author} {\bibinfo {author} {\bibfnamefont {A.}~\bibnamefont {Groisman}}\ and\ \bibinfo {author} {\bibfnamefont {V.}~\bibnamefont {Steinberg}},\ }\bibfield  {title} {\bibinfo {title} {Elastic turbulence in curvilinear flows of polymer solutions},\ }\href {https://iopscience.iop.org/article/10.1088/1367-2630/6/1/029} {\bibfield  {journal} {\bibinfo  {journal} {New J. Phys.}\ }\textbf {\bibinfo {volume} {6}},\ \bibinfo {pages} {29} (\bibinfo {year} {2004})}\BibitemShut {NoStop}%
\bibitem [{\citenamefont {Zilz}\ \emph {et~al.}(2012)\citenamefont {Zilz}, \citenamefont {Poole}, \citenamefont {Alves}, \citenamefont {Bartolo}, \citenamefont {Levaché},\ and\ \citenamefont {Lindner}}]{zilz_geometric_2012}%
  \BibitemOpen
  \bibfield  {author} {\bibinfo {author} {\bibfnamefont {J.}~\bibnamefont {Zilz}}, \bibinfo {author} {\bibfnamefont {R.~J.}\ \bibnamefont {Poole}}, \bibinfo {author} {\bibfnamefont {M.~A.}\ \bibnamefont {Alves}}, \bibinfo {author} {\bibfnamefont {D.}~\bibnamefont {Bartolo}}, \bibinfo {author} {\bibfnamefont {B.}~\bibnamefont {Levaché}},\ and\ \bibinfo {author} {\bibfnamefont {A.}~\bibnamefont {Lindner}},\ }\bibfield  {title} {\bibinfo {title} {Geometric scaling of a purely elastic flow instability in serpentine channels},\ }\href {https://www.cambridge.org/core/product/identifier/S0022112012004119/type/journal_article} {\bibfield  {journal} {\bibinfo  {journal} {J. Fluid Mech.}\ }\textbf {\bibinfo {volume} {712}},\ \bibinfo {pages} {203} (\bibinfo {year} {2012})}\BibitemShut {NoStop}%
\bibitem [{\citenamefont {Casanellas}\ \emph {et~al.}(2016)\citenamefont {Casanellas}, \citenamefont {Alves}, \citenamefont {Poole}, \citenamefont {Lerouge},\ and\ \citenamefont {Lindner}}]{casanellas_stabilizing_2016}%
  \BibitemOpen
  \bibfield  {author} {\bibinfo {author} {\bibfnamefont {L.}~\bibnamefont {Casanellas}}, \bibinfo {author} {\bibfnamefont {M.~A.}\ \bibnamefont {Alves}}, \bibinfo {author} {\bibfnamefont {R.~J.}\ \bibnamefont {Poole}}, \bibinfo {author} {\bibfnamefont {S.}~\bibnamefont {Lerouge}},\ and\ \bibinfo {author} {\bibfnamefont {A.}~\bibnamefont {Lindner}},\ }\bibfield  {title} {\bibinfo {title} {The stabilizing effect of shear thinning on the onset of purely elastic instabilities in serpentine microflows},\ }\href@noop {} {\bibfield  {journal} {\bibinfo  {journal} {Soft Matter}\ }\textbf {\bibinfo {volume} {12}},\ \bibinfo {pages} {6167} (\bibinfo {year} {2016})}\BibitemShut {NoStop}%
\bibitem [{\citenamefont {Qin}\ and\ \citenamefont {Arratia}(2017)}]{qin_characterizing_2017}%
  \BibitemOpen
  \bibfield  {author} {\bibinfo {author} {\bibfnamefont {B.}~\bibnamefont {Qin}}\ and\ \bibinfo {author} {\bibfnamefont {P.~E.}\ \bibnamefont {Arratia}},\ }\bibfield  {title} {\bibinfo {title} {Characterizing elastic turbulence in channel flows at low {Reynolds} number},\ }\href {http://link.aps.org/doi/10.1103/PhysRevFluids.2.083302} {\bibfield  {journal} {\bibinfo  {journal} {Phys. Rev. Fluids}\ }\textbf {\bibinfo {volume} {2}},\ \bibinfo {pages} {083302} (\bibinfo {year} {2017})}\BibitemShut {NoStop}%
\bibitem [{\citenamefont {van Buel}\ and\ \citenamefont {Stark}(2022)}]{van_buel_characterizing_2022}%
  \BibitemOpen
  \bibfield  {author} {\bibinfo {author} {\bibfnamefont {R.}~\bibnamefont {van Buel}}\ and\ \bibinfo {author} {\bibfnamefont {H.}~\bibnamefont {Stark}},\ }\bibfield  {title} {\bibinfo {title} {Characterizing elastic turbulence in the three-dimensional von {Kármán} swirling flow using the {Oldroyd}-{B} model},\ }\href {https://aip.scitation.org/doi/10.1063/5.0079655} {\bibfield  {journal} {\bibinfo  {journal} {Phys. Fluids}\ }\textbf {\bibinfo {volume} {34}},\ \bibinfo {pages} {043112} (\bibinfo {year} {2022})}\BibitemShut {NoStop}%
\bibitem [{\citenamefont {Öztekin}\ \emph {et~al.}(1997)\citenamefont {Öztekin}, \citenamefont {Alakus},\ and\ \citenamefont {McKinley}}]{oztekin1997stability}%
  \BibitemOpen
  \bibfield  {author} {\bibinfo {author} {\bibfnamefont {A.}~\bibnamefont {Öztekin}}, \bibinfo {author} {\bibfnamefont {B.}~\bibnamefont {Alakus}},\ and\ \bibinfo {author} {\bibfnamefont {G.~H.}\ \bibnamefont {McKinley}},\ }\bibfield  {title} {\bibinfo {title} {Stability of planar stagnation flow of a highly viscoelastic fluid},\ }\href@noop {} {\bibfield  {journal} {\bibinfo  {journal} {J. Non-Newton. Fluid Mech.}\ }\textbf {\bibinfo {volume} {72}},\ \bibinfo {pages} {1} (\bibinfo {year} {1997})}\BibitemShut {NoStop}%
\bibitem [{\citenamefont {Arratia}\ \emph {et~al.}(2006)\citenamefont {Arratia}, \citenamefont {Thomas}, \citenamefont {Diorio},\ and\ \citenamefont {Gollub}}]{arratia_elastic_2006}%
  \BibitemOpen
  \bibfield  {author} {\bibinfo {author} {\bibfnamefont {P.~E.}\ \bibnamefont {Arratia}}, \bibinfo {author} {\bibfnamefont {C.~C.}\ \bibnamefont {Thomas}}, \bibinfo {author} {\bibfnamefont {J.}~\bibnamefont {Diorio}},\ and\ \bibinfo {author} {\bibfnamefont {J.~P.}\ \bibnamefont {Gollub}},\ }\bibfield  {title} {\bibinfo {title} {Elastic {Instabilities} of {Polymer} {Solutions} in {Cross}-{Channel} {Flow}},\ }\href {https://link.aps.org/doi/10.1103/PhysRevLett.96.144502} {\bibfield  {journal} {\bibinfo  {journal} {Phys. Rev. Lett.}\ }\textbf {\bibinfo {volume} {96}},\ \bibinfo {pages} {144502} (\bibinfo {year} {2006})}\BibitemShut {NoStop}%
\bibitem [{\citenamefont {Poole}\ \emph {et~al.}(2007)\citenamefont {Poole}, \citenamefont {Alves},\ and\ \citenamefont {Oliveira}}]{poole2007purely}%
  \BibitemOpen
  \bibfield  {author} {\bibinfo {author} {\bibfnamefont {R.~J.}\ \bibnamefont {Poole}}, \bibinfo {author} {\bibfnamefont {M.~A.}\ \bibnamefont {Alves}},\ and\ \bibinfo {author} {\bibfnamefont {P.~J.}\ \bibnamefont {Oliveira}},\ }\bibfield  {title} {\bibinfo {title} {Purely elastic flow asymmetries},\ }\href@noop {} {\bibfield  {journal} {\bibinfo  {journal} {Phys. Rev. Lett.}\ }\textbf {\bibinfo {volume} {99}},\ \bibinfo {pages} {164503} (\bibinfo {year} {2007})}\BibitemShut {NoStop}%
\bibitem [{\citenamefont {Shi}\ and\ \citenamefont {Christopher}(2016)}]{shi_growth_2016}%
  \BibitemOpen
  \bibfield  {author} {\bibinfo {author} {\bibfnamefont {X.}~\bibnamefont {Shi}}\ and\ \bibinfo {author} {\bibfnamefont {G.}~\bibnamefont {Christopher}},\ }\bibfield  {title} {\bibinfo {title} {Growth of viscoelastic instabilities around linear cylinder arrays},\ }\href {http://aip.scitation.org/doi/10.1063/1.4968221} {\bibfield  {journal} {\bibinfo  {journal} {Phys. Fluids}\ }\textbf {\bibinfo {volume} {28}},\ \bibinfo {pages} {124102} (\bibinfo {year} {2016})}\BibitemShut {NoStop}%
\bibitem [{\citenamefont {Zhao}\ \emph {et~al.}(2016)\citenamefont {Zhao}, \citenamefont {Shen},\ and\ \citenamefont {Haward}}]{zhao_flow_2016}%
  \BibitemOpen
  \bibfield  {author} {\bibinfo {author} {\bibfnamefont {Y.}~\bibnamefont {Zhao}}, \bibinfo {author} {\bibfnamefont {A.~Q.}\ \bibnamefont {Shen}},\ and\ \bibinfo {author} {\bibfnamefont {S.~J.}\ \bibnamefont {Haward}},\ }\bibfield  {title} {\bibinfo {title} {Flow of wormlike micellar solutions around confined microfluidic cylinders},\ }\href {http://xlink.rsc.org/?DOI=C6SM01597B} {\bibfield  {journal} {\bibinfo  {journal} {Soft Matter}\ }\textbf {\bibinfo {volume} {12}},\ \bibinfo {pages} {8666} (\bibinfo {year} {2016})}\BibitemShut {NoStop}%
\bibitem [{\citenamefont {Qin}\ \emph {et~al.}(2019{\natexlab{b}})\citenamefont {Qin}, \citenamefont {Salipante}, \citenamefont {Hudson},\ and\ \citenamefont {Arratia}}]{qin_upstream_2019}%
  \BibitemOpen
  \bibfield  {author} {\bibinfo {author} {\bibfnamefont {B.}~\bibnamefont {Qin}}, \bibinfo {author} {\bibfnamefont {P.~F.}\ \bibnamefont {Salipante}}, \bibinfo {author} {\bibfnamefont {S.~D.}\ \bibnamefont {Hudson}},\ and\ \bibinfo {author} {\bibfnamefont {P.~E.}\ \bibnamefont {Arratia}},\ }\bibfield  {title} {\bibinfo {title} {Upstream vortex and elastic wave in the viscoelastic flow around a confined cylinder},\ }\href {https://www.cambridge.org/core/product/identifier/S0022112019000739/type/journal_article} {\bibfield  {journal} {\bibinfo  {journal} {J. Fluid Mech.}\ }\textbf {\bibinfo {volume} {864}},\ \bibinfo {pages} {R2} (\bibinfo {year} {2019}{\natexlab{b}})}\BibitemShut {NoStop}%
\bibitem [{\citenamefont {Haward}\ \emph {et~al.}(2019)\citenamefont {Haward}, \citenamefont {Kitajima}, \citenamefont {Toda-Peters}, \citenamefont {Takahashi},\ and\ \citenamefont {Shen}}]{haward2019flow}%
  \BibitemOpen
  \bibfield  {author} {\bibinfo {author} {\bibfnamefont {S.~J.}\ \bibnamefont {Haward}}, \bibinfo {author} {\bibfnamefont {N.}~\bibnamefont {Kitajima}}, \bibinfo {author} {\bibfnamefont {K.}~\bibnamefont {Toda-Peters}}, \bibinfo {author} {\bibfnamefont {T.}~\bibnamefont {Takahashi}},\ and\ \bibinfo {author} {\bibfnamefont {A.~Q.}\ \bibnamefont {Shen}},\ }\bibfield  {title} {\bibinfo {title} {Flow of wormlike micellar solutions around microfluidic cylinders with high aspect ratio and low blockage ratio},\ }\href@noop {} {\bibfield  {journal} {\bibinfo  {journal} {Soft Matter}\ }\textbf {\bibinfo {volume} {15}},\ \bibinfo {pages} {1927} (\bibinfo {year} {2019})}\BibitemShut {NoStop}%
\bibitem [{\citenamefont {Haward}\ \emph {et~al.}(2020)\citenamefont {Haward}, \citenamefont {Hopkins},\ and\ \citenamefont {Shen}}]{haward_asymmetric_2020}%
  \BibitemOpen
  \bibfield  {author} {\bibinfo {author} {\bibfnamefont {S.~J.}\ \bibnamefont {Haward}}, \bibinfo {author} {\bibfnamefont {C.~C.}\ \bibnamefont {Hopkins}},\ and\ \bibinfo {author} {\bibfnamefont {A.~Q.}\ \bibnamefont {Shen}},\ }\bibfield  {title} {\bibinfo {title} {Asymmetric flow of polymer solutions around microfluidic cylinders: {Interaction} between shear-thinning and viscoelasticity},\ }\href {https://linkinghub.elsevier.com/retrieve/pii/S0377025720300185} {\bibfield  {journal} {\bibinfo  {journal} {J. Non-Newton. Fluid Mech.}\ }\textbf {\bibinfo {volume} {278}},\ \bibinfo {pages} {104250} (\bibinfo {year} {2020})}\BibitemShut {NoStop}%
\bibitem [{\citenamefont {Haward}\ \emph {et~al.}(2021)\citenamefont {Haward}, \citenamefont {Hopkins},\ and\ \citenamefont {Shen}}]{haward_stagnation_2021}%
  \BibitemOpen
  \bibfield  {author} {\bibinfo {author} {\bibfnamefont {S.~J.}\ \bibnamefont {Haward}}, \bibinfo {author} {\bibfnamefont {C.~C.}\ \bibnamefont {Hopkins}},\ and\ \bibinfo {author} {\bibfnamefont {A.~Q.}\ \bibnamefont {Shen}},\ }\bibfield  {title} {\bibinfo {title} {Stagnation points control chaotic fluctuations in viscoelastic porous media flow},\ }\href@noop {} {\bibfield  {journal} {\bibinfo  {journal} {Proc Natl Acad Sci USA}\ }\textbf {\bibinfo {volume} {118}},\ \bibinfo {pages} {e2111651118} (\bibinfo {year} {2021})}\BibitemShut {NoStop}%
\bibitem [{\citenamefont {Hopkins}\ \emph {et~al.}(2022{\natexlab{a}})\citenamefont {Hopkins}, \citenamefont {Haward},\ and\ \citenamefont {Shen}}]{hopkins_upstream_2022}%
  \BibitemOpen
  \bibfield  {author} {\bibinfo {author} {\bibfnamefont {C.~C.}\ \bibnamefont {Hopkins}}, \bibinfo {author} {\bibfnamefont {S.~J.}\ \bibnamefont {Haward}},\ and\ \bibinfo {author} {\bibfnamefont {A.~Q.}\ \bibnamefont {Shen}},\ }\bibfield  {title} {\bibinfo {title} {Upstream wall vortices in viscoelastic flow past a cylinder},\ }\href {http://xlink.rsc.org/?DOI=D2SM00418F} {\bibfield  {journal} {\bibinfo  {journal} {Soft Matter}\ }\textbf {\bibinfo {volume} {18}},\ \bibinfo {pages} {4868} (\bibinfo {year} {2022}{\natexlab{a}})}\BibitemShut {NoStop}%
\bibitem [{\citenamefont {Hopkins}\ \emph {et~al.}(2022{\natexlab{b}})\citenamefont {Hopkins}, \citenamefont {Shen},\ and\ \citenamefont {Haward}}]{hopkins_effect_2022}%
  \BibitemOpen
  \bibfield  {author} {\bibinfo {author} {\bibfnamefont {C.~C.}\ \bibnamefont {Hopkins}}, \bibinfo {author} {\bibfnamefont {A.~Q.}\ \bibnamefont {Shen}},\ and\ \bibinfo {author} {\bibfnamefont {S.~J.}\ \bibnamefont {Haward}},\ }\bibfield  {title} {\bibinfo {title} {Effect of blockage ratio on flow of a viscoelastic wormlike micellar solution past a cylinder in a microchannel},\ }\href {http://xlink.rsc.org/?DOI=D2SM01162J} {\bibfield  {journal} {\bibinfo  {journal} {Soft Matter}\ }\textbf {\bibinfo {volume} {18}},\ \bibinfo {pages} {8856} (\bibinfo {year} {2022}{\natexlab{b}})}\BibitemShut {NoStop}%
\bibitem [{\citenamefont {Chen}\ \emph {et~al.}(2026)\citenamefont {Chen}, \citenamefont {Browne}, \citenamefont {Haward}, \citenamefont {Shen},\ and\ \citenamefont {Datta}}]{chen_stagnation_2024}%
  \BibitemOpen
  \bibfield  {author} {\bibinfo {author} {\bibfnamefont {E.~Y.}\ \bibnamefont {Chen}}, \bibinfo {author} {\bibfnamefont {C.~A.}\ \bibnamefont {Browne}}, \bibinfo {author} {\bibfnamefont {S.~J.}\ \bibnamefont {Haward}}, \bibinfo {author} {\bibfnamefont {A.~Q.}\ \bibnamefont {Shen}},\ and\ \bibinfo {author} {\bibfnamefont {S.~S.}\ \bibnamefont {Datta}},\ }\bibfield  {title} {\bibinfo {title} {Stagnation points at grain contacts generate an elastic flow instability in {3D} porous media},\ }\href@noop {} {\bibfield  {journal} {\bibinfo  {journal} {J. Fluid. Mech.}\ }\textbf {\bibinfo {volume} {1034}},\ \bibinfo {pages} {A4} (\bibinfo {year} {2026})},\ \bibinfo {note} {arXiv:2412.03510 [physics]}\BibitemShut {NoStop}%
\bibitem [{\citenamefont {Haward}\ and\ \citenamefont {Shen}(2026)}]{haward2026effects}%
  \BibitemOpen
  \bibfield  {author} {\bibinfo {author} {\bibfnamefont {S.~J.}\ \bibnamefont {Haward}}\ and\ \bibinfo {author} {\bibfnamefont {A.~Q.}\ \bibnamefont {Shen}},\ }\bibfield  {title} {\bibinfo {title} {Effects of fluid rheology and geometric disorder on the enhanced resistance of viscoelastic flows through porous media},\ }\href@noop {} {\bibfield  {journal} {\bibinfo  {journal} {J. Non-Newton. Fluid Mech.}\ }\textbf {\bibinfo {volume} {349}},\ \bibinfo {pages} {105619} (\bibinfo {year} {2026})}\BibitemShut {NoStop}%
\bibitem [{\citenamefont {Snoeijer}\ \emph {et~al.}(2020)\citenamefont {Snoeijer}, \citenamefont {Pandey}, \citenamefont {Herrada},\ and\ \citenamefont {Eggers}}]{snoeijer_relationship_2020}%
  \BibitemOpen
  \bibfield  {author} {\bibinfo {author} {\bibfnamefont {J.~H.}\ \bibnamefont {Snoeijer}}, \bibinfo {author} {\bibfnamefont {A.}~\bibnamefont {Pandey}}, \bibinfo {author} {\bibfnamefont {M.~A.}\ \bibnamefont {Herrada}},\ and\ \bibinfo {author} {\bibfnamefont {J.}~\bibnamefont {Eggers}},\ }\bibfield  {title} {\bibinfo {title} {The relationship between viscoelasticity and elasticity},\ }\href@noop {} {\bibfield  {journal} {\bibinfo  {journal} {Proc. R. Soc. A.}\ }\textbf {\bibinfo {volume} {476}},\ \bibinfo {pages} {20200419} (\bibinfo {year} {2020})}\BibitemShut {NoStop}%
\bibitem [{sup()}]{supplementary_ref}%
  \BibitemOpen
  \href@noop {} {\bibinfo {title} {See supplemental material at [url will be inserted by publisher] for experimental methods, derivation of theory, and additional analysis.}}\BibitemShut {Stop}%
\bibitem [{\citenamefont {Trouton}(1906)}]{trouton_coefficient_1906}%
  \BibitemOpen
  \bibfield  {author} {\bibinfo {author} {\bibfnamefont {F.~T.}\ \bibnamefont {Trouton}},\ }\bibfield  {title} {\bibinfo {title} {On the coefficient of viscous traction and its relation to that of viscosity.},\ }\href@noop {} {\bibfield  {journal} {\bibinfo  {journal} {Proc. R. Soc. Lond. A}\ }\textbf {\bibinfo {volume} {77}},\ \bibinfo {pages} {426} (\bibinfo {year} {1906})}\BibitemShut {NoStop}%
\bibitem [{\citenamefont {Anna}\ and\ \citenamefont {McKinley}(2001)}]{anna_elasto-capillary_2001}%
  \BibitemOpen
  \bibfield  {author} {\bibinfo {author} {\bibfnamefont {S.~L.}\ \bibnamefont {Anna}}\ and\ \bibinfo {author} {\bibfnamefont {G.~H.}\ \bibnamefont {McKinley}},\ }\bibfield  {title} {\bibinfo {title} {Elasto-capillary thinning and breakup of model elastic liquids},\ }\href@noop {} {\bibfield  {journal} {\bibinfo  {journal} {J. Rheol.}\ }\textbf {\bibinfo {volume} {45}},\ \bibinfo {pages} {115} (\bibinfo {year} {2001})}\BibitemShut {NoStop}%
\bibitem [{\citenamefont {Datta}\ \emph {et~al.}(2013)\citenamefont {Datta}, \citenamefont {Chiang}, \citenamefont {Ramakrishnan},\ and\ \citenamefont {Weitz}}]{datta2013spatial}%
  \BibitemOpen
  \bibfield  {author} {\bibinfo {author} {\bibfnamefont {S.~S.}\ \bibnamefont {Datta}}, \bibinfo {author} {\bibfnamefont {H.}~\bibnamefont {Chiang}}, \bibinfo {author} {\bibfnamefont {T.}~\bibnamefont {Ramakrishnan}},\ and\ \bibinfo {author} {\bibfnamefont {D.~A.}\ \bibnamefont {Weitz}},\ }\bibfield  {title} {\bibinfo {title} {Spatial fluctuations of fluid velocities in flow through a three-dimensional porous medium},\ }\href@noop {} {\bibfield  {journal} {\bibinfo  {journal} {Phys. Rev. Lett.}\ }\textbf {\bibinfo {volume} {111}},\ \bibinfo {pages} {064501} (\bibinfo {year} {2013})}\BibitemShut {NoStop}%
\bibitem [{\citenamefont {Thielicke}\ and\ \citenamefont {Stamhuis}(2014)}]{thielicke_pivlab_2014}%
  \BibitemOpen
  \bibfield  {author} {\bibinfo {author} {\bibfnamefont {W.}~\bibnamefont {Thielicke}}\ and\ \bibinfo {author} {\bibfnamefont {E.~J.}\ \bibnamefont {Stamhuis}},\ }\bibfield  {title} {\bibinfo {title} {{PIVlab} – {Towards} {User}-friendly, {Affordable} and {Accurate} {Digital} {Particle} {Image} {Velocimetry} in {MATLAB}},\ }\href@noop {} {\bibfield  {journal} {\bibinfo  {journal} {J. Open Res. Softw.}\ }\textbf {\bibinfo {volume} {2}},\ \bibinfo {pages} {e30} (\bibinfo {year} {2014})}\BibitemShut {NoStop}%
\bibitem [{\citenamefont {De~Gennes}(1974)}]{de_gennes_coilstretch_1974}%
  \BibitemOpen
  \bibfield  {author} {\bibinfo {author} {\bibfnamefont {P.~G.}\ \bibnamefont {De~Gennes}},\ }\bibfield  {title} {\bibinfo {title} {Coil‐stretch transition of dilute flexible polymers under ultrahigh velocity gradients},\ }\href@noop {} {\bibfield  {journal} {\bibinfo  {journal} {J. Chem. Phys.}\ }\textbf {\bibinfo {volume} {60}},\ \bibinfo {pages} {14} (\bibinfo {year} {1974})}\BibitemShut {NoStop}%
\bibitem [{\citenamefont {Perkins}\ \emph {et~al.}(1997)\citenamefont {Perkins}, \citenamefont {Smith},\ and\ \citenamefont {Chu}}]{perkins_single_1997}%
  \BibitemOpen
  \bibfield  {author} {\bibinfo {author} {\bibfnamefont {T.~T.}\ \bibnamefont {Perkins}}, \bibinfo {author} {\bibfnamefont {D.~E.}\ \bibnamefont {Smith}},\ and\ \bibinfo {author} {\bibfnamefont {S.}~\bibnamefont {Chu}},\ }\bibfield  {title} {\bibinfo {title} {Single {Polymer} {Dynamics} in an {Elongational} {Flow}},\ }\href@noop {} {\bibfield  {journal} {\bibinfo  {journal} {Science}\ }\textbf {\bibinfo {volume} {276}},\ \bibinfo {pages} {2016} (\bibinfo {year} {1997})}\BibitemShut {NoStop}%
\bibitem [{\citenamefont {Carrington}\ and\ \citenamefont {Odell}(1996)}]{carrington_how_1996}%
  \BibitemOpen
  \bibfield  {author} {\bibinfo {author} {\bibfnamefont {S.~P.}\ \bibnamefont {Carrington}}\ and\ \bibinfo {author} {\bibfnamefont {J.~A.}\ \bibnamefont {Odell}},\ }\bibfield  {title} {\bibinfo {title} {How do polymers stretch in stagnation point extensional flow-fields?},\ }\href@noop {} {\bibfield  {journal} {\bibinfo  {journal} {J. Non-Newton. Fluid Mech.}\ }\textbf {\bibinfo {volume} {67}},\ \bibinfo {pages} {269} (\bibinfo {year} {1996})}\BibitemShut {NoStop}%
\bibitem [{\citenamefont {Haward}\ and\ \citenamefont {McKinley}(2013)}]{haward_instabilities_2013}%
  \BibitemOpen
  \bibfield  {author} {\bibinfo {author} {\bibfnamefont {S.~J.}\ \bibnamefont {Haward}}\ and\ \bibinfo {author} {\bibfnamefont {G.~H.}\ \bibnamefont {McKinley}},\ }\bibfield  {title} {\bibinfo {title} {Instabilities in stagnation point flows of polymer solutions},\ }\href@noop {} {\bibfield  {journal} {\bibinfo  {journal} {Phys. Fluids}\ }\textbf {\bibinfo {volume} {25}},\ \bibinfo {pages} {083104} (\bibinfo {year} {2013})}\BibitemShut {NoStop}%
\bibitem [{\citenamefont {Haward}(2016)}]{haward_microfluidic_2016}%
  \BibitemOpen
  \bibfield  {author} {\bibinfo {author} {\bibfnamefont {S.~J.}\ \bibnamefont {Haward}},\ }\bibfield  {title} {\bibinfo {title} {Microfluidic extensional rheometry using stagnation point flow},\ }\href@noop {} {\bibfield  {journal} {\bibinfo  {journal} {Biomicrofluidics}\ }\textbf {\bibinfo {volume} {10}},\ \bibinfo {pages} {043401} (\bibinfo {year} {2016})}\BibitemShut {NoStop}%
\bibitem [{\citenamefont {Haward}\ \emph {et~al.}(2012)\citenamefont {Haward}, \citenamefont {Oliveira}, \citenamefont {Alves},\ and\ \citenamefont {McKinley}}]{haward_optimized_2012}%
  \BibitemOpen
  \bibfield  {author} {\bibinfo {author} {\bibfnamefont {S.~J.}\ \bibnamefont {Haward}}, \bibinfo {author} {\bibfnamefont {M.~S.~N.}\ \bibnamefont {Oliveira}}, \bibinfo {author} {\bibfnamefont {M.~A.}\ \bibnamefont {Alves}},\ and\ \bibinfo {author} {\bibfnamefont {G.~H.}\ \bibnamefont {McKinley}},\ }\bibfield  {title} {\bibinfo {title} {Optimized {Cross}-{Slot} {Flow} {Geometry} for {Microfluidic} {Extensional} {Rheometry}},\ }\href@noop {} {\bibfield  {journal} {\bibinfo  {journal} {Phys. Rev. Lett.}\ }\textbf {\bibinfo {volume} {109}},\ \bibinfo {pages} {128301} (\bibinfo {year} {2012})}\BibitemShut {NoStop}%
\bibitem [{\citenamefont {Galindo-Rosales}\ \emph {et~al.}(2014)\citenamefont {Galindo-Rosales}, \citenamefont {Oliveira},\ and\ \citenamefont {Alves}}]{galindo-rosales_optimized_2014}%
  \BibitemOpen
  \bibfield  {author} {\bibinfo {author} {\bibfnamefont {F.~J.}\ \bibnamefont {Galindo-Rosales}}, \bibinfo {author} {\bibfnamefont {M.~S.~N.}\ \bibnamefont {Oliveira}},\ and\ \bibinfo {author} {\bibfnamefont {M.~A.}\ \bibnamefont {Alves}},\ }\bibfield  {title} {\bibinfo {title} {Optimized cross-slot microdevices for homogeneous extension},\ }\href@noop {} {\bibfield  {journal} {\bibinfo  {journal} {RSC Adv.}\ }\textbf {\bibinfo {volume} {4}},\ \bibinfo {pages} {7799} (\bibinfo {year} {2014})}\BibitemShut {NoStop}%
\bibitem [{\citenamefont {Al-Raoush}\ and\ \citenamefont {Willson}(2005)}]{al-raoush_extraction_2005}%
  \BibitemOpen
  \bibfield  {author} {\bibinfo {author} {\bibfnamefont {R.~I.}\ \bibnamefont {Al-Raoush}}\ and\ \bibinfo {author} {\bibfnamefont {C.~S.}\ \bibnamefont {Willson}},\ }\bibfield  {title} {\bibinfo {title} {Extraction of physically realistic pore network properties from three-dimensional synchrotron {X}-ray microtomography images of unconsolidated porous media systems},\ }\href@noop {} {\bibfield  {journal} {\bibinfo  {journal} {J. Hydrol.}\ }\textbf {\bibinfo {volume} {300}},\ \bibinfo {pages} {44} (\bibinfo {year} {2005})}\BibitemShut {NoStop}%
\bibitem [{\citenamefont {Thompson}\ \emph {et~al.}(2005)\citenamefont {Thompson}, \citenamefont {Willson}, \citenamefont {White}, \citenamefont {Nyman}, \citenamefont {Bhattacharya},\ and\ \citenamefont {Reed}}]{thompson_application_2008}%
  \BibitemOpen
  \bibfield  {author} {\bibinfo {author} {\bibfnamefont {K.~E.}\ \bibnamefont {Thompson}}, \bibinfo {author} {\bibfnamefont {C.~S.}\ \bibnamefont {Willson}}, \bibinfo {author} {\bibfnamefont {C.~D.}\ \bibnamefont {White}}, \bibinfo {author} {\bibfnamefont {S.}~\bibnamefont {Nyman}}, \bibinfo {author} {\bibfnamefont {J.}~\bibnamefont {Bhattacharya}},\ and\ \bibinfo {author} {\bibfnamefont {A.~H.}\ \bibnamefont {Reed}},\ }\bibfield  {title} {\bibinfo {title} {Application of a new grain-based reconstruction algorithm to microtomography images for quantitative characterization and flow modeling},\ }in\ \href {https://doi.org/10.2118/95887-MS} {\emph {\bibinfo {booktitle} {SPE Annual Technical Conference and Exhibition}}}\ (\bibinfo {year} {2005})\ pp.\ \bibinfo {pages} {SPE--95887--MS}\BibitemShut {NoStop}%
\bibitem [{\citenamefont {Bird}\ \emph {et~al.}(1987)\citenamefont {Bird}, \citenamefont {Armstrong},\ and\ \citenamefont {Hassager}}]{bird_dynamics_1987}%
  \BibitemOpen
  \bibfield  {author} {\bibinfo {author} {\bibfnamefont {R.~B.}\ \bibnamefont {Bird}}, \bibinfo {author} {\bibfnamefont {R.~C.}\ \bibnamefont {Armstrong}},\ and\ \bibinfo {author} {\bibfnamefont {O.}~\bibnamefont {Hassager}},\ }\href@noop {} {\emph {\bibinfo {title} {Dynamics of {Polymeric} {Liquids}: {Fluid} {Mechanics}}}},\ Vol.~\bibinfo {volume} {1}\ (\bibinfo  {publisher} {Wiley Interscience},\ \bibinfo {year} {1987})\BibitemShut {NoStop}%
\bibitem [{\citenamefont {Larson}(1999)}]{larson_structure_1999}%
  \BibitemOpen
  \bibfield  {author} {\bibinfo {author} {\bibfnamefont {R.~G.}\ \bibnamefont {Larson}},\ }\href@noop {} {\emph {\bibinfo {title} {The {Structure} and {Rheology} of {Complex} {Fluids}}}}\ (\bibinfo  {publisher} {Oxford University Press},\ \bibinfo {year} {1999})\BibitemShut {NoStop}%
\bibitem [{\citenamefont {Lodge}(1956)}]{lodge_network_1956}%
  \BibitemOpen
  \bibfield  {author} {\bibinfo {author} {\bibfnamefont {A.~S.}\ \bibnamefont {Lodge}},\ }\bibfield  {title} {\bibinfo {title} {A network theory of flow birefringence and stress in concentrated polymer solutions},\ }\href@noop {} {\bibfield  {journal} {\bibinfo  {journal} {Trans. Faraday Soc.}\ }\textbf {\bibinfo {volume} {52}},\ \bibinfo {pages} {120} (\bibinfo {year} {1956})}\BibitemShut {NoStop}%
\bibitem [{\citenamefont {Janeschitz-Kriegl}(1969)}]{cantow_flow_1969}%
  \BibitemOpen
  \bibfield  {author} {\bibinfo {author} {\bibfnamefont {H.}~\bibnamefont {Janeschitz-Kriegl}},\ }\bibfield  {title} {\bibinfo {title} {Flow birefringence of elastico-viscous polymer systems},\ }in\ \href@noop {} {\emph {\bibinfo {booktitle} {Fortschritte der {Hochpolymeren}-{Forschung}}}},\ Vol.\ \bibinfo {volume} {6/2},\ \bibinfo {editor} {edited by\ \bibinfo {editor} {\bibfnamefont {H.~J.}\ \bibnamefont {Cantow}}, \bibinfo {editor} {\bibfnamefont {G.}~\bibnamefont {Dall'Asta}}, \bibinfo {editor} {\bibfnamefont {J.~D.}\ \bibnamefont {Ferry}}, \bibinfo {editor} {\bibfnamefont {H.}~\bibnamefont {Fujita}}, \bibinfo {editor} {\bibfnamefont {W.}~\bibnamefont {Kern}}, \bibinfo {editor} {\bibfnamefont {G.}~\bibnamefont {Natta}}, \bibinfo {editor} {\bibfnamefont {S.}~\bibnamefont {Okamura}}, \bibinfo {editor} {\bibfnamefont {C.~G.}\ \bibnamefont {Overberger}}, \bibinfo {editor} {\bibfnamefont {W.}~\bibnamefont {Prins}}, \bibinfo {editor} {\bibfnamefont {G.~V.}\ \bibnamefont {Schulz}}, \bibinfo {editor}
  {\bibfnamefont {W.~P.}\ \bibnamefont {Slichter}}, \bibinfo {editor} {\bibfnamefont {A.~J.}\ \bibnamefont {Staverman}}, \bibinfo {editor} {\bibfnamefont {J.~K.}\ \bibnamefont {Stille}},\ and\ \bibinfo {editor} {\bibfnamefont {H.~A.}\ \bibnamefont {Stuart}}}\ (\bibinfo  {publisher} {Springer Berlin Heidelberg},\ \bibinfo {address} {Berlin, Heidelberg},\ \bibinfo {year} {1969})\ pp.\ \bibinfo {pages} {170--318},\ \bibinfo {note} {series Title: Advances in Polymer Science}\BibitemShut {NoStop}%
\bibitem [{\citenamefont {Sridhar}\ \emph {et~al.}(2000)\citenamefont {Sridhar}, \citenamefont {Nguyen},\ and\ \citenamefont {Fuller}}]{sridhar_birefringence_2000}%
  \BibitemOpen
  \bibfield  {author} {\bibinfo {author} {\bibfnamefont {T.}~\bibnamefont {Sridhar}}, \bibinfo {author} {\bibfnamefont {D.~A.}\ \bibnamefont {Nguyen}},\ and\ \bibinfo {author} {\bibfnamefont {G.~G.}\ \bibnamefont {Fuller}},\ }\bibfield  {title} {\bibinfo {title} {Birefringence and stress growth in uniaxial extension of polymer solutions},\ }\href@noop {} {\bibfield  {journal} {\bibinfo  {journal} {J. Non-Newton. Fluid Mech.}\ }\textbf {\bibinfo {volume} {90}},\ \bibinfo {pages} {299} (\bibinfo {year} {2000})}\BibitemShut {NoStop}%
\bibitem [{\citenamefont {Rothstein}\ and\ \citenamefont {McKinley}(2002)}]{rothstein_comparison_2002}%
  \BibitemOpen
  \bibfield  {author} {\bibinfo {author} {\bibfnamefont {J.~P.}\ \bibnamefont {Rothstein}}\ and\ \bibinfo {author} {\bibfnamefont {G.~H.}\ \bibnamefont {McKinley}},\ }\bibfield  {title} {\bibinfo {title} {A comparison of the stress and birefringence growth of dilute, semi-dilute and concentrated polymer solutions in uniaxial extensional flows},\ }\href@noop {} {\bibfield  {journal} {\bibinfo  {journal} {J. Non-Newton. Fluid Mech.}\ }\textbf {\bibinfo {volume} {108}},\ \bibinfo {pages} {275} (\bibinfo {year} {2002})}\BibitemShut {NoStop}%
\bibitem [{\citenamefont {Corona}\ \emph {et~al.}(2022)\citenamefont {Corona}, \citenamefont {Berke}, \citenamefont {Guizar-Sicairos}, \citenamefont {Leal}, \citenamefont {Liebi},\ and\ \citenamefont {Helgeson}}]{corona_fingerprinting_2022}%
  \BibitemOpen
  \bibfield  {author} {\bibinfo {author} {\bibfnamefont {P.~T.}\ \bibnamefont {Corona}}, \bibinfo {author} {\bibfnamefont {B.}~\bibnamefont {Berke}}, \bibinfo {author} {\bibfnamefont {M.}~\bibnamefont {Guizar-Sicairos}}, \bibinfo {author} {\bibfnamefont {L.~G.}\ \bibnamefont {Leal}}, \bibinfo {author} {\bibfnamefont {M.}~\bibnamefont {Liebi}},\ and\ \bibinfo {author} {\bibfnamefont {M.~E.}\ \bibnamefont {Helgeson}},\ }\bibfield  {title} {\bibinfo {title} {Fingerprinting soft material nanostructure response to complex flow histories},\ }\href@noop {} {\bibfield  {journal} {\bibinfo  {journal} {Phys. Rev. Materials}\ }\textbf {\bibinfo {volume} {6}},\ \bibinfo {pages} {045603} (\bibinfo {year} {2022})}\BibitemShut {NoStop}%
\bibitem [{\citenamefont {Kumar}\ \emph {et~al.}(2023{\natexlab{a}})\citenamefont {Kumar}, \citenamefont {Guasto},\ and\ \citenamefont {Ardekani}}]{kumar_lagrangian_2023}%
  \BibitemOpen
  \bibfield  {author} {\bibinfo {author} {\bibfnamefont {M.}~\bibnamefont {Kumar}}, \bibinfo {author} {\bibfnamefont {J.~S.}\ \bibnamefont {Guasto}},\ and\ \bibinfo {author} {\bibfnamefont {A.~M.}\ \bibnamefont {Ardekani}},\ }\bibfield  {title} {\bibinfo {title} {Lagrangian stretching reveals stress topology in viscoelastic flows},\ }\href@noop {} {\bibfield  {journal} {\bibinfo  {journal} {Proc. Natl. Acad. Sci. U.S.A.}\ }\textbf {\bibinfo {volume} {120}},\ \bibinfo {pages} {e2211347120} (\bibinfo {year} {2023}{\natexlab{a}})}\BibitemShut {NoStop}%
\bibitem [{\citenamefont {Kumar}\ \emph {et~al.}(2023{\natexlab{b}})\citenamefont {Kumar}, \citenamefont {Walkama}, \citenamefont {Ardekani},\ and\ \citenamefont {Guasto}}]{kumar_stress_2023}%
  \BibitemOpen
  \bibfield  {author} {\bibinfo {author} {\bibfnamefont {M.}~\bibnamefont {Kumar}}, \bibinfo {author} {\bibfnamefont {D.~M.}\ \bibnamefont {Walkama}}, \bibinfo {author} {\bibfnamefont {A.~M.}\ \bibnamefont {Ardekani}},\ and\ \bibinfo {author} {\bibfnamefont {J.~S.}\ \bibnamefont {Guasto}},\ }\bibfield  {title} {\bibinfo {title} {Stress and stretching regulate dispersion in viscoelastic porous media flows},\ }\href@noop {} {\bibfield  {journal} {\bibinfo  {journal} {Soft Matter}\ }\textbf {\bibinfo {volume} {19}},\ \bibinfo {pages} {6761} (\bibinfo {year} {2023}{\natexlab{b}})}\BibitemShut {NoStop}%
\bibitem [{\citenamefont {Ewoldt}\ and\ \citenamefont {Saengow}(2022)}]{ewoldt_designing_2022}%
  \BibitemOpen
  \bibfield  {author} {\bibinfo {author} {\bibfnamefont {R.~H.}\ \bibnamefont {Ewoldt}}\ and\ \bibinfo {author} {\bibfnamefont {C.}~\bibnamefont {Saengow}},\ }\bibfield  {title} {\bibinfo {title} {Designing {Complex} {Fluids}},\ }\href@noop {} {\bibfield  {journal} {\bibinfo  {journal} {Annu. Rev. Fluid Mech.}\ }\textbf {\bibinfo {volume} {54}},\ \bibinfo {pages} {413} (\bibinfo {year} {2022})}\BibitemShut {NoStop}%
\bibitem [{\citenamefont {Richards}\ \emph {et~al.}(2024)\citenamefont {Richards}, \citenamefont {Hodgson}, \citenamefont {O’Neill}, \citenamefont {DeRosa},\ and\ \citenamefont {Poon}}]{richards_optimizing_2024}%
  \BibitemOpen
  \bibfield  {author} {\bibinfo {author} {\bibfnamefont {J.~A.}\ \bibnamefont {Richards}}, \bibinfo {author} {\bibfnamefont {D.~J.~M.}\ \bibnamefont {Hodgson}}, \bibinfo {author} {\bibfnamefont {R.~E.}\ \bibnamefont {O’Neill}}, \bibinfo {author} {\bibfnamefont {M.~E.}\ \bibnamefont {DeRosa}},\ and\ \bibinfo {author} {\bibfnamefont {W.~C.~K.}\ \bibnamefont {Poon}},\ }\bibfield  {title} {\bibinfo {title} {Optimizing non-{Newtonian} fluids for impact protection of laminates},\ }\href@noop {} {\bibfield  {journal} {\bibinfo  {journal} {Proc. Natl. Acad. Sci. U.S.A.}\ }\textbf {\bibinfo {volume} {121}},\ \bibinfo {pages} {e2317832121} (\bibinfo {year} {2024})}\BibitemShut {NoStop}%
\bibitem [{\citenamefont {Stone}\ \emph {et~al.}(2004)\citenamefont {Stone}, \citenamefont {Stroock},\ and\ \citenamefont {Ajdari}}]{stone_engineering_2004}%
  \BibitemOpen
  \bibfield  {author} {\bibinfo {author} {\bibfnamefont {H.~A.}\ \bibnamefont {Stone}}, \bibinfo {author} {\bibfnamefont {A.~D.}\ \bibnamefont {Stroock}},\ and\ \bibinfo {author} {\bibfnamefont {A.}~\bibnamefont {Ajdari}},\ }\bibfield  {title} {\bibinfo {title} {Engineering {Flows} in {Small} {Devices}: {Microfluidics} {Toward} a {Lab}-on-a-{Chip}},\ }\href@noop {} {\bibfield  {journal} {\bibinfo  {journal} {Annu. Rev. Fluid Mech.}\ }\textbf {\bibinfo {volume} {36}},\ \bibinfo {pages} {381} (\bibinfo {year} {2004})}\BibitemShut {NoStop}%
\bibitem [{\citenamefont {Browne}\ \emph {et~al.}(2020)\citenamefont {Browne}, \citenamefont {Shih},\ and\ \citenamefont {Datta}}]{browne_porescale_2020}%
  \BibitemOpen
  \bibfield  {author} {\bibinfo {author} {\bibfnamefont {C.~A.}\ \bibnamefont {Browne}}, \bibinfo {author} {\bibfnamefont {A.}~\bibnamefont {Shih}},\ and\ \bibinfo {author} {\bibfnamefont {S.~S.}\ \bibnamefont {Datta}},\ }\bibfield  {title} {\bibinfo {title} {Pore‐{Scale} {Flow} {Characterization} of {Polymer} {Solutions} in {Microfluidic} {Porous} {Media}},\ }\href@noop {} {\bibfield  {journal} {\bibinfo  {journal} {Small}\ }\textbf {\bibinfo {volume} {16}},\ \bibinfo {pages} {1903944} (\bibinfo {year} {2020})}\BibitemShut {NoStop}%
\end{thebibliography}

\begin{thebibliography}{53}%
\makeatletter
\providecommand \@ifxundefined [1]{%
 \@ifx{#1\undefined}
}%
\providecommand \@ifnum [1]{%
 \ifnum #1\expandafter \@firstoftwo
 \else \expandafter \@secondoftwo
 \fi
}%
\providecommand \@ifx [1]{%
 \ifx #1\expandafter \@firstoftwo
 \else \expandafter \@secondoftwo
 \fi
}%
\providecommand \natexlab [1]{#1}%
\providecommand \enquote  [1]{``#1''}%
\providecommand \bibnamefont  [1]{#1}%
\providecommand \bibfnamefont [1]{#1}%
\providecommand \citenamefont [1]{#1}%
\providecommand \href@noop [0]{\@secondoftwo}%
\providecommand \href [0]{\begingroup \@sanitize@url \@href}%
\providecommand \@href[1]{\@@startlink{#1}\@@href}%
\providecommand \@@href[1]{\endgroup#1\@@endlink}%
\providecommand \@sanitize@url [0]{\catcode `\\12\catcode `\$12\catcode `\&12\catcode `\#12\catcode `\^12\catcode `\_12\catcode `\%12\relax}%
\providecommand \@@startlink[1]{}%
\providecommand \@@endlink[0]{}%
\providecommand \url  [0]{\begingroup\@sanitize@url \@url }%
\providecommand \@url [1]{\endgroup\@href {#1}{\urlprefix }}%
\providecommand \urlprefix  [0]{URL }%
\providecommand \Eprint [0]{\href }%
\providecommand \doibase [0]{https://doi.org/}%
\providecommand \selectlanguage [0]{\@gobble}%
\providecommand \bibinfo  [0]{\@secondoftwo}%
\providecommand \bibfield  [0]{\@secondoftwo}%
\providecommand \translation [1]{[#1]}%
\providecommand \BibitemOpen [0]{}%
\providecommand \bibitemStop [0]{}%
\providecommand \bibitemNoStop [0]{.\EOS\space}%
\providecommand \EOS [0]{\spacefactor3000\relax}%
\providecommand \BibitemShut  [1]{\csname bibitem#1\endcsname}%
\let\auto@bib@innerbib\@empty
\bibitem [{\citenamefont {Dobrynin}\ \emph {et~al.}(1995)\citenamefont {Dobrynin}, \citenamefont {Colby},\ and\ \citenamefont {Rubinstein}}]{dobrynin_scaling_1995}%
  \BibitemOpen
  \bibfield  {author} {\bibinfo {author} {\bibfnamefont {A.~V.}\ \bibnamefont {Dobrynin}}, \bibinfo {author} {\bibfnamefont {R.~H.}\ \bibnamefont {Colby}},\ and\ \bibinfo {author} {\bibfnamefont {M.}~\bibnamefont {Rubinstein}},\ }\bibfield  {title} {\bibinfo {title} {Scaling {Theory} of {Polyelectrolyte} {Solutions}},\ }\href@noop {} {\bibfield  {journal} {\bibinfo  {journal} {Macromolecules}\ }\textbf {\bibinfo {volume} {28}},\ \bibinfo {pages} {1859} (\bibinfo {year} {1995})}\BibitemShut {NoStop}%
\bibitem [{\citenamefont {Browne}\ and\ \citenamefont {Datta}(2021)}]{browne_elastic_2021}%
  \BibitemOpen
  \bibfield  {author} {\bibinfo {author} {\bibfnamefont {C.~A.}\ \bibnamefont {Browne}}\ and\ \bibinfo {author} {\bibfnamefont {S.~S.}\ \bibnamefont {Datta}},\ }\bibfield  {title} {\bibinfo {title} {Elastic turbulence generates anomalous flow resistance in porous media},\ }\href@noop {} {\bibfield  {journal} {\bibinfo  {journal} {Sci. Adv.}\ }\textbf {\bibinfo {volume} {7}},\ \bibinfo {pages} {11} (\bibinfo {year} {2021})}\BibitemShut {NoStop}%
\bibitem [{\citenamefont {Shaqfeh}(1996)}]{shaqfeh_purely_1996}%
  \BibitemOpen
  \bibfield  {author} {\bibinfo {author} {\bibfnamefont {E.~S.~G.}\ \bibnamefont {Shaqfeh}},\ }\bibfield  {title} {\bibinfo {title} {Purely {Elastic} {Instabilities} in {Viscometric} {Flows}},\ }\href@noop {} {\bibfield  {journal} {\bibinfo  {journal} {Annu. Rev. Fluid Mech.}\ }\textbf {\bibinfo {volume} {28}},\ \bibinfo {pages} {129} (\bibinfo {year} {1996})}\BibitemShut {NoStop}%
\bibitem [{\citenamefont {Larson}(1992)}]{larson_instabilities_1992}%
  \BibitemOpen
  \bibfield  {author} {\bibinfo {author} {\bibfnamefont {R.~G.}\ \bibnamefont {Larson}},\ }\bibfield  {title} {\bibinfo {title} {Instabilities in viscoelastic flows},\ }\href@noop {} {\bibfield  {journal} {\bibinfo  {journal} {Rheol. Acta}\ }\textbf {\bibinfo {volume} {31}},\ \bibinfo {pages} {213} (\bibinfo {year} {1992})}\BibitemShut {NoStop}%
\bibitem [{\citenamefont {Howe}\ \emph {et~al.}(2015)\citenamefont {Howe}, \citenamefont {Clarke},\ and\ \citenamefont {Giernalczyk}}]{howe_flow_2015}%
  \BibitemOpen
  \bibfield  {author} {\bibinfo {author} {\bibfnamefont {A.~M.}\ \bibnamefont {Howe}}, \bibinfo {author} {\bibfnamefont {A.}~\bibnamefont {Clarke}},\ and\ \bibinfo {author} {\bibfnamefont {D.}~\bibnamefont {Giernalczyk}},\ }\bibfield  {title} {\bibinfo {title} {Flow of concentrated viscoelastic polymer solutions in porous media: effect of {MW} and concentration on elastic turbulence onset in various geometries},\ }\href@noop {} {\bibfield  {journal} {\bibinfo  {journal} {Soft Matter}\ }\textbf {\bibinfo {volume} {11}},\ \bibinfo {pages} {6419} (\bibinfo {year} {2015})}\BibitemShut {NoStop}%
\bibitem [{\citenamefont {Casanellas}\ \emph {et~al.}(2016)\citenamefont {Casanellas}, \citenamefont {Alves}, \citenamefont {Poole}, \citenamefont {Lerouge},\ and\ \citenamefont {Lindner}}]{casanellas_stabilizing_2016}%
  \BibitemOpen
  \bibfield  {author} {\bibinfo {author} {\bibfnamefont {L.}~\bibnamefont {Casanellas}}, \bibinfo {author} {\bibfnamefont {M.~A.}\ \bibnamefont {Alves}}, \bibinfo {author} {\bibfnamefont {R.~J.}\ \bibnamefont {Poole}}, \bibinfo {author} {\bibfnamefont {S.}~\bibnamefont {Lerouge}},\ and\ \bibinfo {author} {\bibfnamefont {A.}~\bibnamefont {Lindner}},\ }\bibfield  {title} {\bibinfo {title} {The stabilizing effect of shear thinning on the onset of purely elastic instabilities in serpentine microflows},\ }\href@noop {} {\bibfield  {journal} {\bibinfo  {journal} {Soft Matter}\ }\textbf {\bibinfo {volume} {12}},\ \bibinfo {pages} {6167} (\bibinfo {year} {2016})}\BibitemShut {NoStop}%
\bibitem [{\citenamefont {Kolte}\ and\ \citenamefont {Szabo}(1999)}]{kolte_capillary_1999}%
  \BibitemOpen
  \bibfield  {author} {\bibinfo {author} {\bibfnamefont {M.~I.}\ \bibnamefont {Kolte}}\ and\ \bibinfo {author} {\bibfnamefont {P.}~\bibnamefont {Szabo}},\ }\bibfield  {title} {\bibinfo {title} {Capillary thinning of polymeric filaments},\ }\href@noop {} {\bibfield  {journal} {\bibinfo  {journal} {J. Rheol.}\ }\textbf {\bibinfo {volume} {43}},\ \bibinfo {pages} {609} (\bibinfo {year} {1999})}\BibitemShut {NoStop}%
\bibitem [{\citenamefont {Anna}\ and\ \citenamefont {McKinley}(2001)}]{anna_elasto-capillary_2001}%
  \BibitemOpen
  \bibfield  {author} {\bibinfo {author} {\bibfnamefont {S.~L.}\ \bibnamefont {Anna}}\ and\ \bibinfo {author} {\bibfnamefont {G.~H.}\ \bibnamefont {McKinley}},\ }\bibfield  {title} {\bibinfo {title} {Elasto-capillary thinning and breakup of model elastic liquids},\ }\href@noop {} {\bibfield  {journal} {\bibinfo  {journal} {J. Rheol.}\ }\textbf {\bibinfo {volume} {45}},\ \bibinfo {pages} {115} (\bibinfo {year} {2001})}\BibitemShut {NoStop}%
\bibitem [{\citenamefont {Entov}\ and\ \citenamefont {Hinch}(1997)}]{entov_effect_1997}%
  \BibitemOpen
  \bibfield  {author} {\bibinfo {author} {\bibfnamefont {V.~M.}\ \bibnamefont {Entov}}\ and\ \bibinfo {author} {\bibfnamefont {E.~J.}\ \bibnamefont {Hinch}},\ }\bibfield  {title} {\bibinfo {title} {Effect of a spectrum of relaxation times on the capillary thinning of a filament of elastic liquid},\ }\href@noop {} {\bibfield  {journal} {\bibinfo  {journal} {J. Non-Newton. Fluid Mech.}\ }\textbf {\bibinfo {volume} {72}},\ \bibinfo {pages} {31} (\bibinfo {year} {1997})}\BibitemShut {NoStop}%
\bibitem [{\citenamefont {Walkama}\ \emph {et~al.}(2020)\citenamefont {Walkama}, \citenamefont {Waisbord},\ and\ \citenamefont {Guasto}}]{walkama_disorder_2020}%
  \BibitemOpen
  \bibfield  {author} {\bibinfo {author} {\bibfnamefont {D.~M.}\ \bibnamefont {Walkama}}, \bibinfo {author} {\bibfnamefont {N.}~\bibnamefont {Waisbord}},\ and\ \bibinfo {author} {\bibfnamefont {J.~S.}\ \bibnamefont {Guasto}},\ }\bibfield  {title} {\bibinfo {title} {Disorder {Suppresses} {Chaos} in {Viscoelastic} {Flows}},\ }\href@noop {} {\bibfield  {journal} {\bibinfo  {journal} {Phys. Rev. Lett.}\ }\textbf {\bibinfo {volume} {124}},\ \bibinfo {pages} {164501} (\bibinfo {year} {2020})}\BibitemShut {NoStop}%
\bibitem [{\citenamefont {Haward}\ \emph {et~al.}(2021)\citenamefont {Haward}, \citenamefont {Hopkins},\ and\ \citenamefont {Shen}}]{haward_stagnation_2021}%
  \BibitemOpen
  \bibfield  {author} {\bibinfo {author} {\bibfnamefont {S.~J.}\ \bibnamefont {Haward}}, \bibinfo {author} {\bibfnamefont {C.~C.}\ \bibnamefont {Hopkins}},\ and\ \bibinfo {author} {\bibfnamefont {A.~Q.}\ \bibnamefont {Shen}},\ }\bibfield  {title} {\bibinfo {title} {Stagnation points control chaotic fluctuations in viscoelastic porous media flow},\ }\href@noop {} {\bibfield  {journal} {\bibinfo  {journal} {Proc Natl Acad Sci USA}\ }\textbf {\bibinfo {volume} {118}},\ \bibinfo {pages} {e2111651118} (\bibinfo {year} {2021})}\BibitemShut {NoStop}%
\bibitem [{\citenamefont {Carlson}\ \emph {et~al.}(2022)\citenamefont {Carlson}, \citenamefont {Toda-Peters}, \citenamefont {Shen},\ and\ \citenamefont {Haward}}]{carlson_volumetric_2022}%
  \BibitemOpen
  \bibfield  {author} {\bibinfo {author} {\bibfnamefont {D.~W.}\ \bibnamefont {Carlson}}, \bibinfo {author} {\bibfnamefont {K.}~\bibnamefont {Toda-Peters}}, \bibinfo {author} {\bibfnamefont {A.~Q.}\ \bibnamefont {Shen}},\ and\ \bibinfo {author} {\bibfnamefont {S.~J.}\ \bibnamefont {Haward}},\ }\bibfield  {title} {\bibinfo {title} {Volumetric evolution of elastic turbulence in porous media},\ }\href@noop {} {\bibfield  {journal} {\bibinfo  {journal} {J. Fluid Mech.}\ }\textbf {\bibinfo {volume} {950}},\ \bibinfo {pages} {A36} (\bibinfo {year} {2022})}\BibitemShut {NoStop}%
\bibitem [{\citenamefont {Mitchell}\ \emph {et~al.}(2016)\citenamefont {Mitchell}, \citenamefont {Lyons}, \citenamefont {Howe},\ and\ \citenamefont {Clarke}}]{mitchell_viscoelastic_2016}%
  \BibitemOpen
  \bibfield  {author} {\bibinfo {author} {\bibfnamefont {J.}~\bibnamefont {Mitchell}}, \bibinfo {author} {\bibfnamefont {K.}~\bibnamefont {Lyons}}, \bibinfo {author} {\bibfnamefont {A.~M.}\ \bibnamefont {Howe}},\ and\ \bibinfo {author} {\bibfnamefont {A.}~\bibnamefont {Clarke}},\ }\bibfield  {title} {\bibinfo {title} {Viscoelastic polymer flows and elastic turbulence in three-dimensional porous structures},\ }\href@noop {} {\bibfield  {journal} {\bibinfo  {journal} {Soft Matter}\ }\textbf {\bibinfo {volume} {12}},\ \bibinfo {pages} {460} (\bibinfo {year} {2016})}\BibitemShut {NoStop}%
\bibitem [{\citenamefont {Zami-Pierre}\ \emph {et~al.}(2016)\citenamefont {Zami-Pierre}, \citenamefont {de~Loubens}, \citenamefont {Quintard},\ and\ \citenamefont {Davit}}]{zami-pierre_transition_2016}%
  \BibitemOpen
  \bibfield  {author} {\bibinfo {author} {\bibfnamefont {F.}~\bibnamefont {Zami-Pierre}}, \bibinfo {author} {\bibfnamefont {R.}~\bibnamefont {de~Loubens}}, \bibinfo {author} {\bibfnamefont {M.}~\bibnamefont {Quintard}},\ and\ \bibinfo {author} {\bibfnamefont {Y.}~\bibnamefont {Davit}},\ }\bibfield  {title} {\bibinfo {title} {Transition in the {Flow} of {Power}-{Law} {Fluids} through {Isotropic} {Porous} {Media}},\ }\href@noop {} {\bibfield  {journal} {\bibinfo  {journal} {Phys. Rev. Lett.}\ }\textbf {\bibinfo {volume} {117}},\ \bibinfo {pages} {074502} (\bibinfo {year} {2016})}\BibitemShut {NoStop}%
\bibitem [{\citenamefont {Berg}\ and\ \citenamefont {van Wunnik}(2017)}]{berg_shear_2017}%
  \BibitemOpen
  \bibfield  {author} {\bibinfo {author} {\bibfnamefont {S.}~\bibnamefont {Berg}}\ and\ \bibinfo {author} {\bibfnamefont {J.}~\bibnamefont {van Wunnik}},\ }\bibfield  {title} {\bibinfo {title} {Shear {Rate} {Determination} from {Pore}-{Scale} {Flow} {Fields}},\ }\href@noop {} {\bibfield  {journal} {\bibinfo  {journal} {Transp Porous Med}\ }\textbf {\bibinfo {volume} {117}},\ \bibinfo {pages} {229} (\bibinfo {year} {2017})}\BibitemShut {NoStop}%
\bibitem [{\citenamefont {Thielicke}\ and\ \citenamefont {Stamhuis}(2014)}]{thielicke_pivlab_2014}%
  \BibitemOpen
  \bibfield  {author} {\bibinfo {author} {\bibfnamefont {W.}~\bibnamefont {Thielicke}}\ and\ \bibinfo {author} {\bibfnamefont {E.~J.}\ \bibnamefont {Stamhuis}},\ }\bibfield  {title} {\bibinfo {title} {{PIVlab} – {Towards} {User}-friendly, {Affordable} and {Accurate} {Digital} {Particle} {Image} {Velocimetry} in {MATLAB}},\ }\href@noop {} {\bibfield  {journal} {\bibinfo  {journal} {J. Open Res. Softw.}\ }\textbf {\bibinfo {volume} {2}},\ \bibinfo {pages} {e30} (\bibinfo {year} {2014})}\BibitemShut {NoStop}%
\bibitem [{\citenamefont {Trouton}(1906)}]{trouton_coefficient_1906}%
  \BibitemOpen
  \bibfield  {author} {\bibinfo {author} {\bibfnamefont {F.~T.}\ \bibnamefont {Trouton}},\ }\bibfield  {title} {\bibinfo {title} {On the coefficient of viscous traction and its relation to that of viscosity.},\ }\href@noop {} {\bibfield  {journal} {\bibinfo  {journal} {Proc. R. Soc. Lond. A}\ }\textbf {\bibinfo {volume} {77}},\ \bibinfo {pages} {426} (\bibinfo {year} {1906})}\BibitemShut {NoStop}%
\bibitem [{\citenamefont {Petrie}(2006)}]{petrie_extensional_2006}%
  \BibitemOpen
  \bibfield  {author} {\bibinfo {author} {\bibfnamefont {C.~J.~S.}\ \bibnamefont {Petrie}},\ }\bibfield  {title} {\bibinfo {title} {Extensional viscosity: {A} critical discussion},\ }\href@noop {} {\bibfield  {journal} {\bibinfo  {journal} {J. Non-Newton. Fluid Mech.}\ }\textbf {\bibinfo {volume} {137}},\ \bibinfo {pages} {15} (\bibinfo {year} {2006})}\BibitemShut {NoStop}%
\bibitem [{\citenamefont {Haward}\ \emph {et~al.}(2023)\citenamefont {Haward}, \citenamefont {Varchanis}, \citenamefont {McKinley}, \citenamefont {Alves},\ and\ \citenamefont {Shen}}]{haward_extensional_2023}%
  \BibitemOpen
  \bibfield  {author} {\bibinfo {author} {\bibfnamefont {S.~J.}\ \bibnamefont {Haward}}, \bibinfo {author} {\bibfnamefont {S.}~\bibnamefont {Varchanis}}, \bibinfo {author} {\bibfnamefont {G.~H.}\ \bibnamefont {McKinley}}, \bibinfo {author} {\bibfnamefont {M.~A.}\ \bibnamefont {Alves}},\ and\ \bibinfo {author} {\bibfnamefont {A.~Q.}\ \bibnamefont {Shen}},\ }\bibfield  {title} {\bibinfo {title} {Extensional rheometry of mobile fluids. {Part} {II}: {Comparison} between the uniaxial, planar, and biaxial extensional rheology of dilute polymer solutions using numerically optimized stagnation point microfluidic devices},\ }\href@noop {} {\bibfield  {journal} {\bibinfo  {journal} {J. Rheol.}\ }\textbf {\bibinfo {volume} {67}},\ \bibinfo {pages} {1011} (\bibinfo {year} {2023})}\BibitemShut {NoStop}%
\bibitem [{\citenamefont {Macosko}(1994)}]{macosko_rheology_1994}%
  \BibitemOpen
  \bibfield  {author} {\bibinfo {author} {\bibfnamefont {C.~W.}\ \bibnamefont {Macosko}},\ }\href@noop {} {\emph {\bibinfo {title} {Rheology: principles, measurements, and applications}}},\ Advances in interfacial engineering series\ (\bibinfo  {publisher} {VCH},\ \bibinfo {address} {New York},\ \bibinfo {year} {1994})\BibitemShut {NoStop}%
\bibitem [{\citenamefont {Larson}(1999)}]{larson_structure_1999}%
  \BibitemOpen
  \bibfield  {author} {\bibinfo {author} {\bibfnamefont {R.~G.}\ \bibnamefont {Larson}},\ }\href@noop {} {\emph {\bibinfo {title} {The {Structure} and {Rheology} of {Complex} {Fluids}}}}\ (\bibinfo  {publisher} {Oxford University Press},\ \bibinfo {year} {1999})\BibitemShut {NoStop}%
\bibitem [{\citenamefont {McKinley}\ and\ \citenamefont {Sridhar}(2002)}]{mckinley_filament-stretching_2002}%
  \BibitemOpen
  \bibfield  {author} {\bibinfo {author} {\bibfnamefont {G.~H.}\ \bibnamefont {McKinley}}\ and\ \bibinfo {author} {\bibfnamefont {T.}~\bibnamefont {Sridhar}},\ }\bibfield  {title} {\bibinfo {title} {Filament-{Stretching} {Rheometry} of {Complex} {Fluids}},\ }\href@noop {} {\bibfield  {journal} {\bibinfo  {journal} {Annu. Rev. Fluid Mech.}\ }\textbf {\bibinfo {volume} {34}},\ \bibinfo {pages} {375} (\bibinfo {year} {2002})}\BibitemShut {NoStop}%
\bibitem [{\citenamefont {Corona}\ \emph {et~al.}(2022)\citenamefont {Corona}, \citenamefont {Berke}, \citenamefont {Guizar-Sicairos}, \citenamefont {Leal}, \citenamefont {Liebi},\ and\ \citenamefont {Helgeson}}]{corona_fingerprinting_2022}%
  \BibitemOpen
  \bibfield  {author} {\bibinfo {author} {\bibfnamefont {P.~T.}\ \bibnamefont {Corona}}, \bibinfo {author} {\bibfnamefont {B.}~\bibnamefont {Berke}}, \bibinfo {author} {\bibfnamefont {M.}~\bibnamefont {Guizar-Sicairos}}, \bibinfo {author} {\bibfnamefont {L.~G.}\ \bibnamefont {Leal}}, \bibinfo {author} {\bibfnamefont {M.}~\bibnamefont {Liebi}},\ and\ \bibinfo {author} {\bibfnamefont {M.~E.}\ \bibnamefont {Helgeson}},\ }\bibfield  {title} {\bibinfo {title} {Fingerprinting soft material nanostructure response to complex flow histories},\ }\href@noop {} {\bibfield  {journal} {\bibinfo  {journal} {Phys. Rev. Materials}\ }\textbf {\bibinfo {volume} {6}},\ \bibinfo {pages} {045603} (\bibinfo {year} {2022})}\BibitemShut {NoStop}%
\bibitem [{\citenamefont {Kumar}\ \emph {et~al.}(2023{\natexlab{a}})\citenamefont {Kumar}, \citenamefont {Guasto},\ and\ \citenamefont {Ardekani}}]{kumar_lagrangian_2023}%
  \BibitemOpen
  \bibfield  {author} {\bibinfo {author} {\bibfnamefont {M.}~\bibnamefont {Kumar}}, \bibinfo {author} {\bibfnamefont {J.~S.}\ \bibnamefont {Guasto}},\ and\ \bibinfo {author} {\bibfnamefont {A.~M.}\ \bibnamefont {Ardekani}},\ }\bibfield  {title} {\bibinfo {title} {Lagrangian stretching reveals stress topology in viscoelastic flows},\ }\href@noop {} {\bibfield  {journal} {\bibinfo  {journal} {Proc. Natl. Acad. Sci. U.S.A.}\ }\textbf {\bibinfo {volume} {120}},\ \bibinfo {pages} {e2211347120} (\bibinfo {year} {2023}{\natexlab{a}})}\BibitemShut {NoStop}%
\bibitem [{\citenamefont {Kumar}\ \emph {et~al.}(2023{\natexlab{b}})\citenamefont {Kumar}, \citenamefont {Walkama}, \citenamefont {Ardekani},\ and\ \citenamefont {Guasto}}]{kumar_stress_2023}%
  \BibitemOpen
  \bibfield  {author} {\bibinfo {author} {\bibfnamefont {M.}~\bibnamefont {Kumar}}, \bibinfo {author} {\bibfnamefont {D.~M.}\ \bibnamefont {Walkama}}, \bibinfo {author} {\bibfnamefont {A.~M.}\ \bibnamefont {Ardekani}},\ and\ \bibinfo {author} {\bibfnamefont {J.~S.}\ \bibnamefont {Guasto}},\ }\bibfield  {title} {\bibinfo {title} {Stress and stretching regulate dispersion in viscoelastic porous media flows},\ }\href@noop {} {\bibfield  {journal} {\bibinfo  {journal} {Soft Matter}\ }\textbf {\bibinfo {volume} {19}},\ \bibinfo {pages} {6761} (\bibinfo {year} {2023}{\natexlab{b}})}\BibitemShut {NoStop}%
\bibitem [{\citenamefont {Bird}\ \emph {et~al.}(1987)\citenamefont {Bird}, \citenamefont {Armstrong},\ and\ \citenamefont {Hassager}}]{bird_dynamics_1987}%
  \BibitemOpen
  \bibfield  {author} {\bibinfo {author} {\bibfnamefont {R.~B.}\ \bibnamefont {Bird}}, \bibinfo {author} {\bibfnamefont {R.~C.}\ \bibnamefont {Armstrong}},\ and\ \bibinfo {author} {\bibfnamefont {O.}~\bibnamefont {Hassager}},\ }\href@noop {} {\emph {\bibinfo {title} {Dynamics of {Polymeric} {Liquids}: {Fluid} {Mechanics}}}},\ Vol.~\bibinfo {volume} {1}\ (\bibinfo  {publisher} {Wiley Interscience},\ \bibinfo {year} {1987})\BibitemShut {NoStop}%
\bibitem [{\citenamefont {Maklad}\ and\ \citenamefont {Poole}(2021)}]{maklad_review_2021}%
  \BibitemOpen
  \bibfield  {author} {\bibinfo {author} {\bibfnamefont {O.}~\bibnamefont {Maklad}}\ and\ \bibinfo {author} {\bibfnamefont {R.~J.}\ \bibnamefont {Poole}},\ }\bibfield  {title} {\bibinfo {title} {A review of the second normal-stress difference; its importance in various flows, measurement techniques, results for various complex fluids and theoretical predictions},\ }\href@noop {} {\bibfield  {journal} {\bibinfo  {journal} {J. Non-Newton. Fluid Mech.}\ }\textbf {\bibinfo {volume} {292}},\ \bibinfo {pages} {104522} (\bibinfo {year} {2021})}\BibitemShut {NoStop}%
\bibitem [{\citenamefont {De~Gennes}(1974)}]{de_gennes_coilstretch_1974}%
  \BibitemOpen
  \bibfield  {author} {\bibinfo {author} {\bibfnamefont {P.~G.}\ \bibnamefont {De~Gennes}},\ }\bibfield  {title} {\bibinfo {title} {Coil‐stretch transition of dilute flexible polymers under ultrahigh velocity gradients},\ }\href@noop {} {\bibfield  {journal} {\bibinfo  {journal} {J. Chem. Phys.}\ }\textbf {\bibinfo {volume} {60}},\ \bibinfo {pages} {14} (\bibinfo {year} {1974})}\BibitemShut {NoStop}%
\bibitem [{\citenamefont {Perkins}\ \emph {et~al.}(1997)\citenamefont {Perkins}, \citenamefont {Smith},\ and\ \citenamefont {Chu}}]{perkins_single_1997}%
  \BibitemOpen
  \bibfield  {author} {\bibinfo {author} {\bibfnamefont {T.~T.}\ \bibnamefont {Perkins}}, \bibinfo {author} {\bibfnamefont {D.~E.}\ \bibnamefont {Smith}},\ and\ \bibinfo {author} {\bibfnamefont {S.}~\bibnamefont {Chu}},\ }\bibfield  {title} {\bibinfo {title} {Single {Polymer} {Dynamics} in an {Elongational} {Flow}},\ }\href@noop {} {\bibfield  {journal} {\bibinfo  {journal} {Science}\ }\textbf {\bibinfo {volume} {276}},\ \bibinfo {pages} {2016} (\bibinfo {year} {1997})}\BibitemShut {NoStop}%
\bibitem [{\citenamefont {Anna}\ \emph {et~al.}(2001)\citenamefont {Anna}, \citenamefont {McKinley}, \citenamefont {Nguyen}, \citenamefont {Sridhar}, \citenamefont {Muller}, \citenamefont {Huang},\ and\ \citenamefont {James}}]{anna_interlaboratory_2001}%
  \BibitemOpen
  \bibfield  {author} {\bibinfo {author} {\bibfnamefont {S.~L.}\ \bibnamefont {Anna}}, \bibinfo {author} {\bibfnamefont {G.~H.}\ \bibnamefont {McKinley}}, \bibinfo {author} {\bibfnamefont {D.~A.}\ \bibnamefont {Nguyen}}, \bibinfo {author} {\bibfnamefont {T.}~\bibnamefont {Sridhar}}, \bibinfo {author} {\bibfnamefont {S.~J.}\ \bibnamefont {Muller}}, \bibinfo {author} {\bibfnamefont {J.}~\bibnamefont {Huang}},\ and\ \bibinfo {author} {\bibfnamefont {D.~F.}\ \bibnamefont {James}},\ }\bibfield  {title} {\bibinfo {title} {An interlaboratory comparison of measurements from filament-stretching rheometers using common test fluids},\ }\href@noop {} {\bibfield  {journal} {\bibinfo  {journal} {Journal of Rheology}\ }\textbf {\bibinfo {volume} {45}},\ \bibinfo {pages} {83} (\bibinfo {year} {2001})}\BibitemShut {NoStop}%
\bibitem [{\citenamefont {Haward}\ and\ \citenamefont {Odell}(2003)}]{haward_viscosity_2003}%
  \BibitemOpen
  \bibfield  {author} {\bibinfo {author} {\bibfnamefont {S.~J.}\ \bibnamefont {Haward}}\ and\ \bibinfo {author} {\bibfnamefont {J.~A.}\ \bibnamefont {Odell}},\ }\bibfield  {title} {\bibinfo {title} {Viscosity enhancement in non-{Newtonian} flow of dilute polymer solutions through crystallographic porous media},\ }\href@noop {} {\bibfield  {journal} {\bibinfo  {journal} {Rheol. Acta}\ }\textbf {\bibinfo {volume} {42}},\ \bibinfo {pages} {516} (\bibinfo {year} {2003})}\BibitemShut {NoStop}%
\bibitem [{\citenamefont {Odell}\ and\ \citenamefont {Haward}(2006)}]{odell_viscosity_2006}%
  \BibitemOpen
  \bibfield  {author} {\bibinfo {author} {\bibfnamefont {J.~A.}\ \bibnamefont {Odell}}\ and\ \bibinfo {author} {\bibfnamefont {S.~J.}\ \bibnamefont {Haward}},\ }\bibfield  {title} {\bibinfo {title} {Viscosity enhancement in non-{Newtonian} flow of dilute aqueous polymer solutions through crystallographic and random porous media},\ }\href@noop {} {\bibfield  {journal} {\bibinfo  {journal} {Rheol. Acta}\ }\textbf {\bibinfo {volume} {45}},\ \bibinfo {pages} {853} (\bibinfo {year} {2006})}\BibitemShut {NoStop}%
\bibitem [{\citenamefont {Haas}\ and\ \citenamefont {Durst}(1982)}]{haas_viscoelastic_1982}%
  \BibitemOpen
  \bibfield  {author} {\bibinfo {author} {\bibfnamefont {R.}~\bibnamefont {Haas}}\ and\ \bibinfo {author} {\bibfnamefont {F.}~\bibnamefont {Durst}},\ }\bibfield  {title} {\bibinfo {title} {Viscoelastic flow of dilute polymer solutions in regularly packed beds},\ }\href@noop {} {\bibfield  {journal} {\bibinfo  {journal} {Rheol Acta}\ }\textbf {\bibinfo {volume} {21}},\ \bibinfo {pages} {566} (\bibinfo {year} {1982})}\BibitemShut {NoStop}%
\bibitem [{\citenamefont {Farinato}\ and\ \citenamefont {Yen}(1987)}]{farinato_polymer_1987}%
  \BibitemOpen
  \bibfield  {author} {\bibinfo {author} {\bibfnamefont {R.~S.}\ \bibnamefont {Farinato}}\ and\ \bibinfo {author} {\bibfnamefont {W.~S.}\ \bibnamefont {Yen}},\ }\bibfield  {title} {\bibinfo {title} {Polymer degradation in porous media flow},\ }\href@noop {} {\bibfield  {journal} {\bibinfo  {journal} {J. Appl. Polym. Sci.}\ }\textbf {\bibinfo {volume} {33}},\ \bibinfo {pages} {2353} (\bibinfo {year} {1987})}\BibitemShut {NoStop}%
\bibitem [{\citenamefont {Dyakonova}\ \emph {et~al.}(1996)\citenamefont {Dyakonova}, \citenamefont {Odell}, \citenamefont {Brestkin}, \citenamefont {Lyulin},\ and\ \citenamefont {Saez}}]{dyakonova_macromolecular_1996}%
  \BibitemOpen
  \bibfield  {author} {\bibinfo {author} {\bibfnamefont {N.~E.}\ \bibnamefont {Dyakonova}}, \bibinfo {author} {\bibfnamefont {J.~A.}\ \bibnamefont {Odell}}, \bibinfo {author} {\bibfnamefont {Y.~V.}\ \bibnamefont {Brestkin}}, \bibinfo {author} {\bibfnamefont {A.~V.}\ \bibnamefont {Lyulin}},\ and\ \bibinfo {author} {\bibfnamefont {A.~E.}\ \bibnamefont {Saez}},\ }\bibfield  {title} {\bibinfo {title} {Macromolecular strain in periodic models of porous media flows},\ }\href@noop {} {\bibfield  {journal} {\bibinfo  {journal} {J. Non-Newton. Fluid Mech.}\ }\textbf {\bibinfo {volume} {67}},\ \bibinfo {pages} {285} (\bibinfo {year} {1996})}\BibitemShut {NoStop}%
\bibitem [{\citenamefont {Müller}\ and\ \citenamefont {Sáez}(1999)}]{nguyen_rheology_1999}%
  \BibitemOpen
  \bibfield  {author} {\bibinfo {author} {\bibfnamefont {A.~J.}\ \bibnamefont {Müller}}\ and\ \bibinfo {author} {\bibfnamefont {A.~E.}\ \bibnamefont {Sáez}},\ }\bibfield  {title} {\bibinfo {title} {The {Rheology} of {Polymer} {Solutions} in {Porous} {Media}},\ }in\ \href@noop {} {\emph {\bibinfo {booktitle} {Flexible {Polymer} {Chains} in {Elongational} {Flow}}}},\ \bibinfo {editor} {edited by\ \bibinfo {editor} {\bibfnamefont {T.~Q.}\ \bibnamefont {Nguyen}}\ and\ \bibinfo {editor} {\bibfnamefont {H.-H.}\ \bibnamefont {Kausch}}}\ (\bibinfo  {publisher} {Springer Berlin Heidelberg},\ \bibinfo {address} {Berlin, Heidelberg},\ \bibinfo {year} {1999})\ pp.\ \bibinfo {pages} {335--393}\BibitemShut {NoStop}%
\bibitem [{\citenamefont {Chmielewski}\ and\ \citenamefont {Jayaraman}(1992)}]{chmielewski_effect_1992}%
  \BibitemOpen
  \bibfield  {author} {\bibinfo {author} {\bibfnamefont {C.}~\bibnamefont {Chmielewski}}\ and\ \bibinfo {author} {\bibfnamefont {K.}~\bibnamefont {Jayaraman}},\ }\bibfield  {title} {\bibinfo {title} {The effect of polymer extensibility on crossflow of polymer solutions through cylinder arrays},\ }\href@noop {} {\bibfield  {journal} {\bibinfo  {journal} {J. Rheol.}\ }\textbf {\bibinfo {volume} {36}},\ \bibinfo {pages} {1105} (\bibinfo {year} {1992})}\BibitemShut {NoStop}%
\bibitem [{\citenamefont {Chmielewski}\ and\ \citenamefont {Jayaraman}(1993)}]{chmielewski_elastic_1993}%
  \BibitemOpen
  \bibfield  {author} {\bibinfo {author} {\bibfnamefont {C.}~\bibnamefont {Chmielewski}}\ and\ \bibinfo {author} {\bibfnamefont {K.}~\bibnamefont {Jayaraman}},\ }\bibfield  {title} {\bibinfo {title} {Elastic instability in crossflow of polymer solutions through periodic arrays of cylinders},\ }\href@noop {} {\bibfield  {journal} {\bibinfo  {journal} {J. Non-Newton. Fluid Mech.}\ }\textbf {\bibinfo {volume} {48}},\ \bibinfo {pages} {285} (\bibinfo {year} {1993})}\BibitemShut {NoStop}%
\bibitem [{\citenamefont {James}\ \emph {et~al.}(2012)\citenamefont {James}, \citenamefont {Yip},\ and\ \citenamefont {Currie}}]{james_slow_2012}%
  \BibitemOpen
  \bibfield  {author} {\bibinfo {author} {\bibfnamefont {D.~F.}\ \bibnamefont {James}}, \bibinfo {author} {\bibfnamefont {R.}~\bibnamefont {Yip}},\ and\ \bibinfo {author} {\bibfnamefont {I.~G.}\ \bibnamefont {Currie}},\ }\bibfield  {title} {\bibinfo {title} {Slow flow of {Boger} fluids through model fibrous porous media},\ }\href@noop {} {\bibfield  {journal} {\bibinfo  {journal} {J. Rheol.}\ }\textbf {\bibinfo {volume} {56}},\ \bibinfo {pages} {1249} (\bibinfo {year} {2012})}\BibitemShut {NoStop}%
\bibitem [{\citenamefont {Chen}\ \emph {et~al.}(2026)\citenamefont {Chen}, \citenamefont {Browne}, \citenamefont {Haward}, \citenamefont {Shen},\ and\ \citenamefont {Datta}}]{chen_stagnation_2024}%
  \BibitemOpen
  \bibfield  {author} {\bibinfo {author} {\bibfnamefont {E.~Y.}\ \bibnamefont {Chen}}, \bibinfo {author} {\bibfnamefont {C.~A.}\ \bibnamefont {Browne}}, \bibinfo {author} {\bibfnamefont {S.~J.}\ \bibnamefont {Haward}}, \bibinfo {author} {\bibfnamefont {A.~Q.}\ \bibnamefont {Shen}},\ and\ \bibinfo {author} {\bibfnamefont {S.~S.}\ \bibnamefont {Datta}},\ }\bibfield  {title} {\bibinfo {title} {Stagnation points at grain contacts generate an elastic flow instability in {3D} porous media},\ }\href@noop {} {\bibfield  {journal} {\bibinfo  {journal} {J. Fluid. Mech.}\ }\textbf {\bibinfo {volume} {1034}},\ \bibinfo {pages} {A4} (\bibinfo {year} {2026})},\ \bibinfo {note} {arXiv:2412.03510 [physics]}\BibitemShut {NoStop}%
\bibitem [{\citenamefont {Haward}\ \emph {et~al.}(2012)\citenamefont {Haward}, \citenamefont {Oliveira}, \citenamefont {Alves},\ and\ \citenamefont {McKinley}}]{haward_optimized_2012}%
  \BibitemOpen
  \bibfield  {author} {\bibinfo {author} {\bibfnamefont {S.~J.}\ \bibnamefont {Haward}}, \bibinfo {author} {\bibfnamefont {M.~S.~N.}\ \bibnamefont {Oliveira}}, \bibinfo {author} {\bibfnamefont {M.~A.}\ \bibnamefont {Alves}},\ and\ \bibinfo {author} {\bibfnamefont {G.~H.}\ \bibnamefont {McKinley}},\ }\bibfield  {title} {\bibinfo {title} {Optimized {Cross}-{Slot} {Flow} {Geometry} for {Microfluidic} {Extensional} {Rheometry}},\ }\href@noop {} {\bibfield  {journal} {\bibinfo  {journal} {Phys. Rev. Lett.}\ }\textbf {\bibinfo {volume} {109}},\ \bibinfo {pages} {128301} (\bibinfo {year} {2012})}\BibitemShut {NoStop}%
\bibitem [{\citenamefont {Galindo-Rosales}\ \emph {et~al.}(2014)\citenamefont {Galindo-Rosales}, \citenamefont {Oliveira},\ and\ \citenamefont {Alves}}]{galindo-rosales_optimized_2014}%
  \BibitemOpen
  \bibfield  {author} {\bibinfo {author} {\bibfnamefont {F.~J.}\ \bibnamefont {Galindo-Rosales}}, \bibinfo {author} {\bibfnamefont {M.~S.~N.}\ \bibnamefont {Oliveira}},\ and\ \bibinfo {author} {\bibfnamefont {M.~A.}\ \bibnamefont {Alves}},\ }\bibfield  {title} {\bibinfo {title} {Optimized cross-slot microdevices for homogeneous extension},\ }\href@noop {} {\bibfield  {journal} {\bibinfo  {journal} {RSC Adv.}\ }\textbf {\bibinfo {volume} {4}},\ \bibinfo {pages} {7799} (\bibinfo {year} {2014})}\BibitemShut {NoStop}%
\bibitem [{\citenamefont {Arratia}\ \emph {et~al.}(2006)\citenamefont {Arratia}, \citenamefont {Thomas}, \citenamefont {Diorio},\ and\ \citenamefont {Gollub}}]{arratia_elastic_2006}%
  \BibitemOpen
  \bibfield  {author} {\bibinfo {author} {\bibfnamefont {P.~E.}\ \bibnamefont {Arratia}}, \bibinfo {author} {\bibfnamefont {C.~C.}\ \bibnamefont {Thomas}}, \bibinfo {author} {\bibfnamefont {J.}~\bibnamefont {Diorio}},\ and\ \bibinfo {author} {\bibfnamefont {J.~P.}\ \bibnamefont {Gollub}},\ }\bibfield  {title} {\bibinfo {title} {Elastic {Instabilities} of {Polymer} {Solutions} in {Cross}-{Channel} {Flow}},\ }\href {https://link.aps.org/doi/10.1103/PhysRevLett.96.144502} {\bibfield  {journal} {\bibinfo  {journal} {Phys. Rev. Lett.}\ }\textbf {\bibinfo {volume} {96}},\ \bibinfo {pages} {144502} (\bibinfo {year} {2006})}\BibitemShut {NoStop}%
\bibitem [{\citenamefont {Sousa}\ \emph {et~al.}(2018)\citenamefont {Sousa}, \citenamefont {Pinho},\ and\ \citenamefont {Alves}}]{sousa_purely-elastic_2018}%
  \BibitemOpen
  \bibfield  {author} {\bibinfo {author} {\bibfnamefont {P.~C.}\ \bibnamefont {Sousa}}, \bibinfo {author} {\bibfnamefont {F.~T.}\ \bibnamefont {Pinho}},\ and\ \bibinfo {author} {\bibfnamefont {M.~A.}\ \bibnamefont {Alves}},\ }\bibfield  {title} {\bibinfo {title} {Purely-elastic flow instabilities and elastic turbulence in microfluidic cross-slot devices},\ }\href@noop {} {\bibfield  {journal} {\bibinfo  {journal} {Soft Matter}\ }\textbf {\bibinfo {volume} {14}},\ \bibinfo {pages} {1344} (\bibinfo {year} {2018})}\BibitemShut {NoStop}%
\bibitem [{\citenamefont {Davoodi}\ \emph {et~al.}(2019)\citenamefont {Davoodi}, \citenamefont {Domingues},\ and\ \citenamefont {Poole}}]{davoodi_control_2019}%
  \BibitemOpen
  \bibfield  {author} {\bibinfo {author} {\bibfnamefont {M.}~\bibnamefont {Davoodi}}, \bibinfo {author} {\bibfnamefont {A.~F.}\ \bibnamefont {Domingues}},\ and\ \bibinfo {author} {\bibfnamefont {R.~J.}\ \bibnamefont {Poole}},\ }\bibfield  {title} {\bibinfo {title} {Control of a purely elastic symmetry-breaking flow instability in cross-slot geometries},\ }\href@noop {} {\bibfield  {journal} {\bibinfo  {journal} {J. Fluid Mech.}\ }\textbf {\bibinfo {volume} {881}},\ \bibinfo {pages} {1123} (\bibinfo {year} {2019})}\BibitemShut {NoStop}%
\bibitem [{\citenamefont {Yokokoji}\ \emph {et~al.}(2023)\citenamefont {Yokokoji}, \citenamefont {Varchanis}, \citenamefont {Shen},\ and\ \citenamefont {Haward}}]{yokokoji_rheological_2023}%
  \BibitemOpen
  \bibfield  {author} {\bibinfo {author} {\bibfnamefont {A.}~\bibnamefont {Yokokoji}}, \bibinfo {author} {\bibfnamefont {S.}~\bibnamefont {Varchanis}}, \bibinfo {author} {\bibfnamefont {A.~Q.}\ \bibnamefont {Shen}},\ and\ \bibinfo {author} {\bibfnamefont {S.~J.}\ \bibnamefont {Haward}},\ }\bibfield  {title} {\bibinfo {title} {Rheological effects on purely-elastic flow asymmetries in the cross-slot geometry},\ }\href@noop {} {\bibfield  {journal} {\bibinfo  {journal} {Soft Matter}\ }\textbf {\bibinfo {volume} {20}},\ \bibinfo {pages} {152} (\bibinfo {year} {2023})}\BibitemShut {NoStop}%
\bibitem [{\citenamefont {Browne}\ \emph {et~al.}(2023)\citenamefont {Browne}, \citenamefont {Huang}, \citenamefont {Zheng},\ and\ \citenamefont {Datta}}]{browne_homogenizing_2023}%
  \BibitemOpen
  \bibfield  {author} {\bibinfo {author} {\bibfnamefont {C.~A.}\ \bibnamefont {Browne}}, \bibinfo {author} {\bibfnamefont {R.~B.}\ \bibnamefont {Huang}}, \bibinfo {author} {\bibfnamefont {C.~W.}\ \bibnamefont {Zheng}},\ and\ \bibinfo {author} {\bibfnamefont {S.~S.}\ \bibnamefont {Datta}},\ }\bibfield  {title} {\bibinfo {title} {Homogenizing fluid transport in stratified porous media using an elastic flow instability},\ }\href@noop {} {\bibfield  {journal} {\bibinfo  {journal} {J. Fluid Mech.}\ }\textbf {\bibinfo {volume} {963}},\ \bibinfo {pages} {A30} (\bibinfo {year} {2023})}\BibitemShut {NoStop}%
\bibitem [{\citenamefont {Browne}\ and\ \citenamefont {Datta}(2024)}]{browne_harnessing_2024}%
  \BibitemOpen
  \bibfield  {author} {\bibinfo {author} {\bibfnamefont {C.~A.}\ \bibnamefont {Browne}}\ and\ \bibinfo {author} {\bibfnamefont {S.~S.}\ \bibnamefont {Datta}},\ }\bibfield  {title} {\bibinfo {title} {Harnessing elastic instabilities for enhanced mixing and reaction kinetics in porous media},\ }\href@noop {} {\bibfield  {journal} {\bibinfo  {journal} {Proc. Natl. Acad. Sci. U.S.A.}\ }\textbf {\bibinfo {volume} {121}},\ \bibinfo {pages} {e2320962121} (\bibinfo {year} {2024})}\BibitemShut {NoStop}%
\bibitem [{\citenamefont {Al-Raoush}\ and\ \citenamefont {Willson}(2005)}]{al-raoush_extraction_2005}%
  \BibitemOpen
  \bibfield  {author} {\bibinfo {author} {\bibfnamefont {R.~I.}\ \bibnamefont {Al-Raoush}}\ and\ \bibinfo {author} {\bibfnamefont {C.~S.}\ \bibnamefont {Willson}},\ }\bibfield  {title} {\bibinfo {title} {Extraction of physically realistic pore network properties from three-dimensional synchrotron {X}-ray microtomography images of unconsolidated porous media systems},\ }\href@noop {} {\bibfield  {journal} {\bibinfo  {journal} {J. Hydrol.}\ }\textbf {\bibinfo {volume} {300}},\ \bibinfo {pages} {44} (\bibinfo {year} {2005})}\BibitemShut {NoStop}%
\bibitem [{\citenamefont {Thompson}\ \emph {et~al.}(2005)\citenamefont {Thompson}, \citenamefont {Willson}, \citenamefont {White}, \citenamefont {Nyman}, \citenamefont {Bhattacharya},\ and\ \citenamefont {Reed}}]{thompson_application_2008}%
  \BibitemOpen
  \bibfield  {author} {\bibinfo {author} {\bibfnamefont {K.~E.}\ \bibnamefont {Thompson}}, \bibinfo {author} {\bibfnamefont {C.~S.}\ \bibnamefont {Willson}}, \bibinfo {author} {\bibfnamefont {C.~D.}\ \bibnamefont {White}}, \bibinfo {author} {\bibfnamefont {S.}~\bibnamefont {Nyman}}, \bibinfo {author} {\bibfnamefont {J.}~\bibnamefont {Bhattacharya}},\ and\ \bibinfo {author} {\bibfnamefont {A.~H.}\ \bibnamefont {Reed}},\ }\bibfield  {title} {\bibinfo {title} {Application of a new grain-based reconstruction algorithm to microtomography images for quantitative characterization and flow modeling},\ }in\ \href {https://doi.org/10.2118/95887-MS} {\emph {\bibinfo {booktitle} {SPE Annual Technical Conference and Exhibition}}}\ (\bibinfo {year} {2005})\ pp.\ \bibinfo {pages} {SPE--95887--MS}\BibitemShut {NoStop}%
\bibitem [{\citenamefont {De}\ \emph {et~al.}(2017{\natexlab{a}})\citenamefont {De}, \citenamefont {Kuipers}, \citenamefont {Peters},\ and\ \citenamefont {Padding}}]{de_viscoelastic_2017}%
  \BibitemOpen
  \bibfield  {author} {\bibinfo {author} {\bibfnamefont {S.}~\bibnamefont {De}}, \bibinfo {author} {\bibfnamefont {J.~A.~M.}\ \bibnamefont {Kuipers}}, \bibinfo {author} {\bibfnamefont {E.~A. J.~F.}\ \bibnamefont {Peters}},\ and\ \bibinfo {author} {\bibfnamefont {J.~T.}\ \bibnamefont {Padding}},\ }\bibfield  {title} {\bibinfo {title} {Viscoelastic flow simulations in random porous media},\ }\href@noop {} {\bibfield  {journal} {\bibinfo  {journal} {J. Non-Newton. Fluid Mech.}\ }\textbf {\bibinfo {volume} {248}},\ \bibinfo {pages} {50} (\bibinfo {year} {2017}{\natexlab{a}})}\BibitemShut {NoStop}%
\bibitem [{\citenamefont {De}\ \emph {et~al.}(2017{\natexlab{b}})\citenamefont {De}, \citenamefont {Kuipers}, \citenamefont {Peters},\ and\ \citenamefont {Padding}}]{de_viscoelastic_2017-1}%
  \BibitemOpen
  \bibfield  {author} {\bibinfo {author} {\bibfnamefont {S.}~\bibnamefont {De}}, \bibinfo {author} {\bibfnamefont {J.~A.~M.}\ \bibnamefont {Kuipers}}, \bibinfo {author} {\bibfnamefont {E.~A. J.~F.}\ \bibnamefont {Peters}},\ and\ \bibinfo {author} {\bibfnamefont {J.~T.}\ \bibnamefont {Padding}},\ }\bibfield  {title} {\bibinfo {title} {Viscoelastic flow simulations in model porous media},\ }\href@noop {} {\bibfield  {journal} {\bibinfo  {journal} {Phys. Rev. Fluids}\ }\textbf {\bibinfo {volume} {2}},\ \bibinfo {pages} {053303} (\bibinfo {year} {2017}{\natexlab{b}})}\BibitemShut {NoStop}%
\bibitem [{\citenamefont {James}(2016)}]{james_n1_2016}%
  \BibitemOpen
  \bibfield  {author} {\bibinfo {author} {\bibfnamefont {D.~F.}\ \bibnamefont {James}},\ }\bibfield  {title} {\bibinfo {title} {N1 stresses in extensional flows},\ }\href@noop {} {\bibfield  {journal} {\bibinfo  {journal} {J. Non-Newton. Fluid Mech.}\ }\textbf {\bibinfo {volume} {232}},\ \bibinfo {pages} {33} (\bibinfo {year} {2016})}\BibitemShut {NoStop}%
\end{thebibliography}

%

\end{document}